\documentclass[11pt]{article} 
\usepackage{geometry}
\usepackage{authblk}
\usepackage{amsfonts, amsmath, amsthm, enumerate, rotate}
\usepackage{amssymb, bm, bbm, latexsym, mathrsfs}
\usepackage{graphicx}
\usepackage{color, url}
\usepackage{natbib}
\usepackage[title]{appendix}
\newenvironment{changemargin}[2]{%
\begin{list}{}{%
\setlength{\topsep}{0pt}%
\setlength{\leftmargin}{#1}%
\setlength{\rightmargin}{#2}%
\setlength{\listparindent}{\parindent}%
\setlength{\itemindent}{\parindent}%
\setlength{\parsep}{\parskip}%
}%
\item[]}{\end{list}}

\title{Copula Approximate Bayesian Computation  \\
Using Distribution Random Forests}
\author[1]{George Karabatsos}
\affil[1]{Departments of Mathematics, Statistics, and Computer Sciences\protect\\
and Educational Psychology in Statistics and Measurement \protect
\\ University of Illinois-Chicago \protect\\
e-mail: georgek@uic.edu, gkarabatsos1@gmail.com}
\date{\today}

\begin{document}
\maketitle
\begin{abstract}
Ongoing modern computational advancements continue to make it easier to collect increasingly large and complex datasets, which can often only be realistically analyzed using models defined by intractable likelihood functions. This invited feature article introduces and provides an extensive simulation study of a new Approximate Bayesian Computation (ABC) framework for estimating the posterior distribution and the maximum likelihood estimate (MLE) of the parameters of models defined by intractable likelihoods, which unifies and extends previous ABC methods proposed separately. This framework, \texttt{copulaABcdrf}, aims to accurately estimate and describe the possibly skewed and high dimensional posterior distribution by a novel multivariate copula-based meta-\textit{t} distribution, based on univariate marginal posterior distributions which can be accurately estimated by Distribution Random Forests (\texttt{drf}), while performing automatic summary statistics (covariates) selection, and based on robustly-estimated copula dependence parameters. The \texttt{copulaABcdrf} framework also provides a novel multivariate mode estimator to perform MLE and posterior mode estimation, and an optional step to perform model selection from a given set of models using posterior probabilities estimated by \texttt{drf}. The posterior distribution estimation accuracy of the ABC framework is illustrated and compared with previous standard ABC methods, through several simulation studies involving low- and high-dimensional models with computable posterior distributions, which are either unimodal, skewed, or multimodal; and exponential random graph and mechanistic network models, each defined by an intractable likelihood from which it is costly to simulate large network datasets. This paper also proposes and studies a new solution to the simulation cost problem in ABC, involving posterior estimation of parameters from datasets simulated from the given model that are smaller compared to the potentially large size of the dataset being analyzed. This proposal is motivated by the fact that for many models defined by intractable likelihoods, such as the network models when they are applied to analyze massive networks, the repeated simulation of large datasets (networks) for posterior-based parameter estimation can be too computationally costly, and vastly slow down or prohibit the use standard \texttt{ABC} methods. The \texttt{copulaABcdrf} framework and standard \texttt{ABC} methods are further illustrated through analyses of large real-life networks of sizes ranging between 28,000 to 65.6 million nodes (between 3 million to 1.8 billion edges), including a large multilayer network with weighted directed edges. The results of the simulation studies showed that, in settings where the true posterior distribution is not highly-multimodal, \texttt{copulaABcdrf} usually produced similar point estimates from the posterior distribution for low dimensional parametric models as previous \texttt{ABC} methods, but the copula-based method can produce more accurate estimates from the posterior distribution for high-dimensional models, and in both dimensionality cases, usually produced more accurate estimates of univariate marginal posterior distributions of parameters. Also, posterior estimation accuracy was usually improved when pre-selecting the important summary statistics using \texttt{drf}, compared to \texttt{ABC} employing no pre-selection of the subset of important summaries. For all \texttt{ABC} methods studied, accurate estimation of a highly-multimodal posterior distribution was challenging. In light of the results of all the simulation studies, the article concludes by discussing how \texttt{copulaABcdrf} framework can be improved for future research.\\\\
   \textbf{Keywords:}  Bayesian analysis, Maximum Likelihood, Intractable likelihood.
\end{abstract}

\section{Introduction}

Statistical models defined by intractable likelihood functions are important for analyzing complex and large datasets from many scientific fields. The broad field of ABC provides alternative algorithms for estimating the approximate posterior distribution or MLE of the parameters of any model defined by a likelihood which may be intractable, either because it does not have an explicit form, the data set being analyzed is large, or because the model is high-dimensional (i.e., the model has many parameters).

For the model chosen to analyze a given observed dataset, represented by a set of \textit{observed data summary statistics} (or \textit{observed summaries}) which are ideally sufficient (at least approximately), the original rejection (vanilla) ABC algorithm \citep{TavareEtAl97,PritchardEtAl99}, here referred to as \texttt{rejectionABC}, obtains samples from the approximate posterior distribution of the model parameters from the subsample of many samples from the prior distribution, where for each prior sample, the summary statistics of a pseudo-dataset (or \textit{pseudo-data summaries}) drawn from the model's exact likelihood conditionally on the prior sample of the model parameters, is within a chosen small $\epsilon \geq 0$ in (e.g., Euclidean) distance to the observed data summary statistics. Then from this subsample, estimates of marginal posterior means, medians, density and marginal densities, and estimates of the density and mode and MLE of all the parameters can be obtained. A central object of ABC is the \textit{Reference Table}, consisting of many rows, where each table row contains a prior sample of the model parameters, and the corresponding summaries of a pseudo-dataset drawn from the model's exact likelihood, conditional on this prior sample. The applicability of \texttt{rejectionABC} hinges on the ability to efficiently sample from the model's likelihood. A main motivation for \texttt{rejectionABC} is that it produces exact samples from the posterior distribution, as $\epsilon \rightarrow 0$ when the summary statistics are sufficient \citep{MarinEtAl12}. To explain, consider any parametric model with parameters $\theta\in\Theta\subseteq\mathbb{R}^d$ ($1\leq d < \infty$), a prior distribution $\pi(\theta)$ over $\Theta$, and likelihood $f(x\mid\theta)$ for any given dataset $x$ with summary statistics $s(x)$. Further, if $s$ is (minimally) sufficient for $x$, then the posterior distribution $\pi(\theta\mid x)\propto f(x\mid\theta)\pi(\theta)$ satisfies $\pi(\theta\mid x)=\pi(\theta\mid s(x))$ for all $\pi(\theta)$ \citep[][\S5.1.4]{BernardoSmith94}. Then for any random dataset $Y \sim f(\cdot\mid\theta)$, the posterior $\pi(\theta\mid s(Y))$ converges in distribution to the exact posterior $\pi(\theta\mid x)$ as the distance $\delta$ between $s(Y)$ and $s(x)$ vanishes, $\delta(s(Y),s(x))=\epsilon \rightarrow 0$, as then $s(Y)$ becomes the sufficient statistics $s(x)$ of $x$. Therefore, when $\epsilon$ is small and in more typical scenarios where sufficient statistics are unavailable, the \texttt{rejectionABC} produces samples from the approximate posterior distribution and the summary statistics are approximately sufficient for $x$ at best, for the given dataset and the model applied to analyze it. Often in \texttt{rejectionABC} practice, the tolerance $\epsilon$ is chosen as a small \textit{q}\% (e.g., 1\%) quantile of the simulated distances, to define a  \textit{q}\%-nearest-neighbor \texttt{rejectionABC} \citep{BiauCerouGuyader15}. In ABC, asymptotically efficient estimation of posterior expectations relies on the number of summary statistics to equal to the number of parameters of the given model, where each parameter corresponds to a summary statistic being an approximate MLE of the model parameter \citep{LiFearnhead18}. Therefore, in scenarios where the number of summary statistics (summaries) exceed the number of parameters of the given model, semiautomatic \texttt{rejectionABC} \citep{FearnheadPrangle12} is often applied, which pre-selects the summary statistics that are most important and equal in number to the number of parameters of the given model (here, referred to as \texttt{rejectionABCselect}), using any one of several available methods to select important summaries for ABC data analysis \citep[e.g.,][]{BlumEtAl13}.

The original \texttt{rejectionABC} method (and variants) quintessentially represents ABC, but for the purposes of obtaining reliably-accurate posterior inferences, it can be computationally inefficient or prohibitive to apply easily because it can require generating a huge number of prior samples, especially for models with more than a few parameters and/or from which it is computationally expensive to simulate pseudo-datasets from. Also, while the appropriate selection of relevant summary statistics becomes important in typical scenarios where they are not sufficient. Further, the choices of distance measure and small $\epsilon$ are also important tuning parameters for the rejection (vanilla) ABC algorithm, which can impact the level of posterior inference accuracy. Therefore, more recent research has developed various Monte Carlo algorithms that can more efficiently sample the ABC posterior distribution, with possibly less dependence on the tolerance, distance measure, and summary statistics tuning parameters of \texttt{rejectionABC}. Essentially all ABC methods can each be viewed as defining a specific approximate model likelihood providing a surrogate to the exact intractable likelihood of the given model, which measures the closeness of summary statistics of the observed dataset to the same summaries of a random dataset simulated from the model's exact likelihood, conditionally on any proposed set of model parameters. The ABC field has had many reviews due to its wide theoretical scope and applicability \citep[e.g.,][]{SunnakerEtAl13,KarabatsosLeisen18abc,SissonEtAl18a,GrazianFan20,CranmerEtAl20,CraiuLevi22,Karabatsos23,PesonenEtAl23,MartinEtAl23}.

The main research objective of this paper is to introduce and study a novel ABC framework, \texttt{copulaABCdrf}, for estimating the posterior distribution and the MLE of the parameters of models defined by intractable likelihood functions, from the given generated Reference Table. The \texttt{copulaABCdrf} framework unifies and extends \texttt{copulaABC}  \citep{LiNottEtAl17,NottEtAl18handABC,ChenGutmann19,KleinEtAl24}, ABC-Random Forests (\texttt{abcrf}) \citep{RaynalEtAl18}, and ABC and other simulation-based (\texttt{AMLE}) methods for calculating the approximate MLE for intractable likelihood models \citep[][]{RubioJohansen13,KajiharaEtAl18,PicchiniAnderson17,GutmannCorander16,YildirimEtAl15,DeanEtAl14,GourierouxEtAl93,McFadden89}. Notably, \texttt{copulaABCdrf} provides a single framework to perform all the tasks of inferences from a possibly high dimensional posterior distribution (including estimation of the posterior mean, median, mode, univariate marginal posteriors), MLE estimation, automatic summary statistics selection, and model selection. In contrast, nearly all the other proposed ABC methods, typically \texttt{rejectionABC} methods, were proposed to perform only a small subset of these tasks in a non-unified manner. Therefore, a related objective of this paper is to compare the results of \texttt{copulaABCdrf} with other existing \texttt{rejectionABC} methods in terms of accuracy in estimating quantities from the posterior distribution. While some \texttt{rejectionABC} methods were already mentioned, others are reviewed later in due course.

To elaborate, \texttt{copulaABCdrf} combines \texttt{copulaABC}'s ability to approximate high dimensional posterior distributions, based on estimation of one-dimensional marginal posterior distributions (unlike \texttt{abcrf} and \texttt{AMLE} methods), while separately estimating the dependence structure of the full multivariate posterior; with with \texttt{abcrf}'s ability to estimate these one-dimensional marginal distributions (of the dependent variables) using Random Forests (\texttt{rf}), conditionally on observed data summaries, while automatically selecting relevant summary statistics (covariates) for the given model and dataset under analysis (unlike nearly all \texttt{copulaABC} and \texttt{AMLE} methods); and the ability to calculate the approximate MLE (unlike \texttt{abcrf} and \texttt{copula ABC}). In particular, \texttt{copulaABCdrf} unlike \texttt{abcrf} employs a more modern, Distribution Random Forests (\texttt{drf}) \citep[][\S4.1]{CevidEtAl22}, which more directly targets the estimation of the one-dimensional marginal distributions of the model parameters from the Reference Table. A benefit of using \texttt{drf} for ABC is that it allows the data analyst to avoid the inefficient accept-reject procedure used in the original \texttt{rejectionABC} algorithm, while not requiring the analyst to pre-select summary statistics, and the tolerance and distance measure tuning parameters of \texttt{rejectionABC}. Besides \texttt{copulaABCdrf}, other methods in the literature use generative models to perform this multivariate density regression task of estimating the full joint posterior distribution, without using copulas. They include autoregressive flow models \citep{PapamakariosEtAl19} and conditional generative adversarial models \citep{WangRockova23}, which have demonstrated to accurately estimate low-dimensional multimodal posterior distributions. Meanwhile, \texttt{copulaABC} has shown the ability to accurately estimate a skewed unimodal posterior distribution of a 250-parameter Gaussian model \citep{LiNottEtAl17,NottEtAl18handABC}.

The \texttt{copulaABCdrf} framework aims to estimate (at least approximately) the posterior distribution of model parameters of possibly-high dimension $d =$ dim$(\theta)$, using a multivariate copula-based distribution estimated from a constructed Reference Table. It does so while accounting for any skewness in the posterior, and for the fact that the posterior converges to a multivariate Gaussian distribution as the data sample size increases, even for misspecified models, under regularity conditions \citep{KleijnVanderVaart12}. For the purposes of estimating the posterior mode and MLE in practical applications, such a posterior distribution estimate should be easy to compute its probability density function (pdf) and to sample from.

According to Sklar's (1959\nocite{Sklar59}) theorem, a continuous $d$-variate random parameter vector $\theta$ can be described by a joint (posterior) cumulative distribution function (cdf), called a copula, which can be uniquely represented by its univariate marginal densities (distributions) and a copula density function describing its dependence structure. If the random vector $\theta$ has any discrete-valued parameters, then these discrete random variables can be continued using jittering methods \citep{DenuitEtAl05,MadsenFang11} which maintain a unique representation of the joint copula-based posterior cdf of $\theta$, while ensuring that no information is lost and preserving the same dependence relationship among the model parameters.

Approximating a possibly skewed high-dimensional posterior distribution requires a suitable choice of multivariate copula-based distribution family. Choices are limited because it is difficult to construct high-dimensional copulas \citep[][p.105]{Nelsen06}. Meanwhile, skewness can be introduced into a multivariate copula-based distribution, either: (1) by using skewed univariate margins \citep[e.g.,][]{HutsonEtAl15,SmithVahey16,BaillienEtAl22}; (2) by using a skewed copula density function to account for asymmetric dependencies in the multivariate distribution, such as a skewed, grouped Student-$t$, or mixture of $K$ Gaussian or $t$ copula density functions \citep
[e.g.,][]{WeiEtAl16,Yoshiba18,DemartaMcNeil05,DaulEtAl03,KosmidisKarlis16}, or using a covariate-dependent copula density function \citep{SmithKlein21, KleinEtAl24, AcarEtAl13}; or (3) by combining both skewed marginals and skewed densities, but this can unnecessarily increase the complexity of the copula model \citep{BaillienEtAl22}. In high dimensions, estimating the parameters of skewed copulas is challenging and cumbersome \citep{HintzEtAl22b}, and arguably the mixture copulas are over-parameterized, with $Kd(d-1)/2$ correlation matrix parameters, while the number of correlation parameters grows quadratically with $d$, and it is not easy to reliably estimate $K$ correlation matrices \citep{DellaportasTsionas19,QuLu21}. Besides, the \texttt{copulaABCdrf} framework also aims to estimate a multivariate posterior mode, and the MLE as a mode of a posterior-to-prior pdf ratio, while the mode is robust to the shapes of the tails of the distribution.

Therefore, all things considered, \texttt{copulaABCdrf} employs a novel multivariate meta-$t$ \citep{FangEtAl02} copula-based distribution approximation of the posterior, which allows for multivariate skewness by using skewed marginals which are flexibly modeled and estimated by \texttt{drf}s, respectively. This multivariate distribution is defined by univariate marginals, which are (covariate-dependent) \texttt{drf} regressions, along with a multivariate $t$ parametric copula that is independent of covariates, extending previous related copula models \citep{PittEtAl06, Song00}. The dependence structure of the meta-$t$ distribution is defined by the density of a multivariate $t$ distribution with (posterior) degrees of freedom and scale matrix parameters, which accounts for symmetric tail dependence and inherits the robust properties of the $t$ distribution, by inversely-weighting each observation by its Mahalanobis distance from the data distribution center \citep{LangeLittleTaylor89}.

The \texttt{drf} targets the whole distribution of the dependent variable more appropriately than the mean functions of the dependent variable achieved by the earlier \texttt{rf} and \texttt{abcrf} algorithms. The capability of \texttt{drf} to automatically select summary statistics enables the new ABC framework to avoid the potential issues involved with pre-selecting summary statistics \citep{DrovandiEtAl23}, and to avoid having to introduce an extra prior distribution to do this selection, as done in other ABC methods. The original \texttt{copulaABC} method \citep{LiNottEtAl17,NottEtAl18handABC} was based on a Gaussian copula density, defined by tails that are thinner than those of the more-robust $t$ distribution (but the correlation matrix parameters of the meta-Gaussian can be robustly estimated using robust correlations, as done in \citep{AnEtAl20}), and defined by a corresponding meta-Gaussian distribution with univariate densities estimated from the Reference Table by smooth kernel density estimation methods, conditionally on bandwidth parameters and summary statistics that need to be non-automatically preselected by the data-analyst.

Further, for scenarios where multiple models are considered to analyze the same dataset, \texttt{copulaABCdrf} employs an optional step which can be used to perform model selection, based on posterior model probabilities that are estimated using \texttt{drf} (conditionally on the observed data summaries). This provides a simple minor extension of the previously-validated \texttt{rf}-based approach to model selection in ABC \citep[][]{PudloEtAl16,MarinEtAl18}, by directly estimating the \textit{M}-category multinomial distribution of the model index dependent variable, among the \textit{M} models compared, instead of estimating \textit{M} separate classification \texttt{rf}s for binary (0 or 1 valued) indices of models.

Finally, the \texttt{copulaABCdrf} framework calculates an estimate of the multivariate posterior mode and MLE of the parameters $\theta$ of the (selected) model. The MLE is the multivariate mode of the ratio of the meta-$t$ posterior pdf estimate to the prior pdf, or the posterior mode under a proper uniform prior supporting the MLE. The new multivariate mode estimator maximizes a meta-$t$ posterior density estimate (or finds the MLE by maximizing the ratio of the posterior density estimate to the prior density) over parameter samples of points from the Reference Table (or over an extra set of parameter samples of points drawn from the meta-$t$ posterior density estimate if necessary), which avoids a possibly costly grid search involved in multivariate mode estimation by global maximization. Similarly, other  "indirect" multivariate mode estimation methods \citep{Devroye79,AbrahamEtAl03a,AbrahamEtAl04,HsuWu13,DasguptaKpotufe14} use the data sample points to estimate the mode from a nonparametric kernel, $k$-nearest-neighbor, or parametric density estimate, while respectively, relying on choice of bandwidth, or $k$, or on a successful normal transformation of the data. The \texttt{copulaABCdrf} framework employs a novel semiparametric approach to posterior mode and MLE estimation, because it is copula-based and thus applicable to high dimensions as shown by previous \texttt{copulaABC} research \citep{NottEtAl18handABC} using Gaussian copulas and not using \texttt{rf} or summary statistics selection algorithms. Other multivariate mode estimators seem only applicable to lower dimensions \citep[see][for a review]{Chacon20}. \texttt{copulaABCdrf} estimates the MLE (mode) while performing automatic summary statistics (variable) selection via \texttt{drf}s for the meta-$t$ marginals.

Next, Section \S\ref{ABCframework} describes the \texttt{copulaABCdrf} framework. Then \S\ref{NumericalExamples} investigates this framework through several simulation studies, which compare the results of \\\texttt{copulaABCdrf} with the results from existing \texttt{rejectionABC}  methods. The first three simulation studies consider models with low-dimensional, high dimensional, and multimodal joint posterior distributions, being either analytically or simulation-based computable posterior distributions, in order to allow their direct comparisons with posterior distributions estimated under different ABC methods. The subsequent simulation studies and real data analyses in \S\ref{NumericalExamples} focus on models defined by intractable likelihoods. Most models defined by intractable likelihoods have the two properties that (1) it is possible to rapidly simulate datasets from the model's likelihood, in which case ABC or synthetic likelihood methods \citep[][and references therein]{PicchiniEtAl23,PicchiniTamborrino24} can be used to estimate the (approximate) posterior distribution and MLE of the model parameters; and/or (2) the MLE or another point estimator and its sampling distribution can be computed for the model, in which case the approximate posterior distribution of the model parameters can be computed using certain ABC or bootstrap methods \citep[e.g.,][] {Karabatsos23,NewtonEtAl21,BarrientosPena20,LyddonEtAl19,ZhuMarinLeisen16}.

Therefore, the rest of \S\ref{NumericalExamples} focuses on models defined by intractable likelihoods which do not possess these two properties, at least in large data scenarios. They include Exponential Random Graph Models (ERGMs) and mechanistic models for network datasets, ubiquitous in scientific fields. For such a model, the likelihood is intractable for a large network dataset, and it is too costly or prohibitive to simulate large network datasets from the likelihood, and to compute the MLE or other point estimator of the model parameters from large networks. Along these lines, another contribution of this paper is a new solution to simulation cost problem in ABC, which involves constructing a Reference Table by simulating datasets from the exact model likelihood (given proposed model parameters, resp.) that are smaller compared to the potentially large size of the original dataset being analyzed by the model. This approach can be useful in settings where it is computationally costly to simulate data from the model, including the modeling of very large networks. Using these models, the \texttt{copulaABCdrf} framework is illustrated through the analyses of three large real-life networks, of sizes ranging between twenty thousand to over 6 million nodes, including a large multilayer network with weighted directed edges. In these large data scenarios, using \texttt{copulaABCdrf} and \texttt{rejectionABC} methods, posterior inferences for these models is achieved by employing summary statistics that are invariant to the size of the network. This allows the Reference Table to be constructed by simulating summary statistics based on networks of smaller tractable size compared to the size of the original observed large network dataset being analyzed, while summary statistics of the latter large network can be efficiently computed once using subsampling methods when necessary. As shown, the \texttt{copulaABCdrf} framework provides practical methods for analyzing widely-available complex and large datasets. Finally, \S4 ends with Conclusions with potential future research directions based on the results of this paper.

\section{Methods: The \texttt{copulaABCdrf} Framework\label{ABCframework}}

The proposed \texttt{copulaABCdrf} framework is summarized in Algorithm 1, shown below. The algorithm can be applied to any set of $M\geq1$ models. It outputs an estimate the ABC posterior distribution $\pi_{\widehat{m}}^{\text{ABC}}(\theta\mid s(x))$, a probability density function (pdf) based on corresponding cumulative distribution function (cdf) $\Pi_{\widehat{m}}^{\text{ABC}}$, the selected best predicted model, $\widehat{m}\in\{m\}_{m=1}^{M}$ among $M\geq1$ models compared, and the model's prior distribution (pdf) $\pi_{\widehat{m}}(\theta)$, and its approximate likelihood of the generic form $\int f_{\widehat{m},s}(s(x)+\varepsilon\nu\mid\theta)K(\nu)\mathrm{d}\nu$ \citep{LiFearnhead18} based on summary statistics vector $s$ and some kernel density function $\mathcal{K}$.

Optional Step 2 of Algorithm 1, used when a set of $M>1$ Bayesian models is considered by the data analyst, performs model selection based on \texttt{drf}-based \citep{CevidEtAl22} estimates of the $M$ posterior model probabilities, conditionally on observed data summaries $s(x)$. As mentioned in \S1, Step 2 provides a simple minor extension to a previously-validated Random Forest approach to model selection in ABC \citep[][]{PudloEtAl16,MarinEtAl18} by directly targeting the full \textit{M}-category multinomial distribution dependent variable, over the \textit{M} models being compared in the given model selection problem; instead of running \textit{M} separate classification \texttt{rf}s for binary (0 or 1 valued) indices of models (resp.) as the previous Random Forest approach to model selection did. It has been discussed in the literature \citep[e.g.,][]{RobertEtAl11} that summary statistics can be insufficient for model comparison. Tree-based models can alleviate this issue when enough suitably-chosen summary statistics are used for model selection \citep{PudloEtAl16}.

Step 3 of Algorithm 1 aims to estimate the posterior distribution of possibly high-dimensional model parameters at least approximately-well, using a tractable multivariate copula-based distribution, based on Sklar's (1959\nocite{Sklar59}) theorem.

To explain Sklar's theorem in terms of Algorithm 1, let
\[
\Pi^{\text{ABC}}(\theta\mid s(x))\equiv\Pi_{x}(\theta_{1},\ldots,\theta
_{d})=\Pi_{x}(\theta)
\]
be the cdf of the approximate posterior distribution (approximate because the summaries $s$ are not necessarily sufficient), with corresponding univariate posterior marginal cdfs 
\\$\Pi_{x,1}(\theta_{1}),\ldots,\Pi_{x,d}(\theta_{d})$ and pdfs $\pi_{x,1}(\theta_{1}),\ldots,\pi_{x,d}(\theta_{d})$.

Sklar's (1959\nocite{Sklar59}) theorem implies that there exists a copula $C$ (cdf) such that for all $(\theta_{1},\ldots,\theta_{d})\in\mathbb{R}^{d}$, the joint posterior cdf can be represented by
\[
\Pi_{x}(\theta_{1},\ldots,\theta_{d})=C\{\Pi_{x,1}(\theta_{1}),\ldots
,\Pi_{x,d}(\theta_{d})\}.
\]
If the $\Pi_{x,k}$ (cdfs) are continuous for all $k=1,\ldots,d$ (with corresponding pdfs $\pi_{x,k}$), then $C$ is unique, with uniform distributed $\Pi_{x,k}(\theta_{k})\sim$ $\mathcal{U}_{(0,1)}$ over the random $\theta_{k}$ (for $k=1,\ldots,d$) by the probability integral transform, and the copula $C$ is `margin-free' in the sense that it is invariant under increasing transformations of the margins \citep[][Theorem 2.4.3]{Nelsen06}. Then for an arbitrary continuous multivariate distribution \citep[e.g.,][ \S13.2]{OkhrinEtAl17}, its copula $C$ (cdf) can be determined from the transformation
\[
C(u_{1},\ldots,u_{d})=\Pi_{x}(\Pi_{x,1}^{-1}(u_{1}),\ldots,\Pi_{x,d}%
^{-1}(u_{d})),\text{ for }u_{1},\ldots,u_{d}\in\lbrack0,1],
\]
with corresponding copula density (pdf)\ $c(u_{1},\ldots,u_{d})=\tfrac
{\partial^{d}C(u_{1},\ldots,u_{d})}{\partial u_{1},\ldots,\partial u_{d}}$
(for $u_{1},\ldots,u_{d}\in\lbrack0,1]$), where the $\Pi_{x,k}^{-1}$ (for
$k=1,\ldots,d$) are inverse marginal distribution functions. Also, the
posterior distribution pdf of $\theta$ is given by 
\[
\pi_{x}(\theta_{1},\ldots,\theta_{d})=c\{\Pi_{x,1}(\theta_{1}),\ldots
,\Pi_{x,d}(\theta_{d})\}
{\textstyle\prod\nolimits_{k=1}^{d}}
\pi_{x,k}(\theta_{k})\text{ for }\theta_{1},\ldots,\theta_{d}\in\mathbb{R}\cup\{\infty,-\infty\}.
\]

As mentioned, the ABC framework (Algorithm 1) employs a novel multivariate meta-$t$ \citep{FangEtAl02} copula-based distribution approximation of the posterior, which allows for multivariate skewness by using skewed marginals which are modeled flexibly and accurately estimated by \texttt{drf}s, respectively. This multivariate distribution has marginals which are (covariate-dependent) regressions and a multivariate $t$ parametric copula that is independent of covariates, extending previous related copula models \citep{PittEtAl06, Song00} through the use of \texttt{drf}. The meta-$t$ distribution is defined by a $t$ copula cdf $C_{\nu,\rho}$ (copula density pdf, $c_{\nu,\rho}$) with degrees of freedom $\nu$ and $d\times d$ scale matrix parameters $\rho$, which is a correlation matrix if the density is that of a normal distribution.

As an aside, when the random vector $\theta$ contains any discrete variables, the joint cdf $C$ is not identifiable under Sklar's theorem \citep[e.g.,][]{Geenens20}. Therefore, to maintain direct use of Sklar's theorem using the $t$-copula, Step 1(e) of Algorithm 1 can easily apply jittering \citep{DenuitEtAl05,MadsenFang11} to continue each (of any) discrete model parameter (integer-valued without loss of generality), while ensuring that no information is lost and preserving the same dependence relationship among all model parameters. Each (of any) discrete (integer) model parameter $\theta_{k}^{\ast}\in\theta$ with posterior cdf $\Pi_{x,k}^{\ast}(\theta_{k})$ is continued (\textquotedblleft jittered\textquotedblright) into a continuous parameter $\theta_{k}=\theta_{k}^{\ast}-U_{k}$ with uniform random variable $U_{k}\sim$ $\mathcal{U}_{(0,1)}$ (with $U_{k}\perp\theta_{k}^{\ast}$ and $U_{k}\perp
U_{l}$ for $k\neq l$) and posterior cdf
\[
\Pi_{x,k}(\theta_{k})=\Pi_{x,k}^{\ast}(\lfloor\theta_{k}\rfloor)+(\theta
_{k}-\lfloor\theta_{k}\rfloor)\Pr(\vartheta_{k}^{\ast}=\lfloor\theta
_{k}+1\rfloor)
\]
and pdf $\pi_{x,k}(\theta_{k})=\Pr(\vartheta_{k}^{\ast}=\lfloor\theta_{k}+1\rfloor)$, where $\lfloor z\rfloor=$ floor$(z)$. The parameters of $\Pi_{x,k}$ and $\pi_{x,k}$ are exactly those of $\Pi_{x,k}^{\ast}$, and $\theta_{k}^{\ast}$ can be recovered from $\theta_{k}$ as $\theta_{k}^{\ast}=\lfloor\theta_{k}+1\rfloor$.

Step 3 of Algorithm 1 performs two-stage semiparametric estimation \citep{GenestEtAl95} of the meta-$t$ posterior distribution parameters, $\{\nu,\rho,\{\pi_{k},\Pi_{k}\}_{k=1}^{d}\}$. The first stage calculates nonparametric marginal posterior cdfs and pdfs, $\{\widehat{u}_{k}^{(j)}=\widehat{\Pi}_{x,k}(\theta_{k}^{(j)}),\widehat{\pi}_{x,k}(\theta_{k}^{(j)})\}_{j=1}^{N},$ conditionally on $s(x)$, from \texttt{drf}s trained on columns of the reference table, $\{\theta_{k}^{(j)},s(y^{(j)})\}_{j=1}^{N}$ for $k=1,\ldots,d$ (resp.), while performing automatic summary statistics (covariate) selection. The second stage employs an Expectation-Maximization algorithm \citep{HintzEtAl22a,HintzEtAl22b} to calculate the MLE $(\widehat{\nu},\widehat{\rho})$ of the copula density parameters of the meta-$t$ pdf based on these estimated cdfs and pdfs, for the subset of the \textit{N} rows of  $\{(\widehat{u}_{k}^{(j)}=\widehat{\Pi}_{x,k}(\theta_{k}^{(j)}|,\widehat{\pi}_{x,k}(\theta_{k}^{(j)}) \mid  $\textit{k} = 1,\ldots,\textit{d}$)\}_{j=1}^{N},$ for which $0 < \widehat{u}_{k}^{(j)} < 1$ for all $k=1,\ldots,d$, as this MLE is calculable only for this subset; which can be done using the \texttt{nvmix} R package \citep{HintzEtAl22a}. 

The \texttt{drf} is trained (estimated) on the reference table $\{\theta_{k}^{(j)},s(y^{(j)})\}_{j=1}^{N}$ (the training dataset), such that prior parameter samples $\theta_{k}^{(j)}$ (dependent variable) are regressed on the summary statistics covariate vectors $s(y^{(j)}),$ while targeting the conditional distribution of the dependent variable, for each model parameter indexed by $k=1,\ldots,d$. Then based on each of these $d$ trained \texttt{drf}s, the posterior distribution (cdfs $\Pi_{x,k}$ and pdfs $\pi_{x,k}$, for $k=1,\ldots,d$) can be accurately predicted (estimated) \citep[][]{CevidEtAl22}, conditionally on the summary statistics of the original observed dataset, $s(x)$. All this is done while performing automatic summary statistics (variable) selection from a potentially larger set of summary statistics, and accounting for the uncertainty in variable selection, without requiring the user to pre-select the subset of summary statistics relevant to the intractable-likelihood model under consideration. 

The \texttt{drf} is a weighted nearest neighbor method \citep[][]{LinJeon06} which performs locally-adaptive estimation of the conditional distribution through the aggregation of dependent variable predictions of an ensemble of randomized flexible Classification of Regression Trees (CARTs) \citep{BreimanEtAl84}, respectively, where each CART is estimated from a random subsample of training dataset (drawn without replacement), and that each level of the CART's binary tree is constructed by splitting the training data points $\{\theta_{k}^{(j)},s(y^{(j)})\}_{j=1}^{N}$ based on a covariate $s_{k}(y^{(j)})\in s(y^{(j)})$ and its split point $c$ both chosen in such a way that the distribution of the dependent responses $\theta_{k}^{(j)}$ for which $s_{k}(y^{(j)})\leq c$ differs the most compared to the distribution for which $s_{k}(y^{(j)})>c$, according to the Maximal Mean Discrepancy (MMD) \citep[][]{GrettonEtAl07a}. The MMD is a quickly-computable two-sample test statistic that can detect a wide variety of distributional changes. This way, the dependent variable distribution in each of the resulting leaf nodes is as homogeneous as

\begin{center}
\noindent
\begingroup
\begin{tabular}
[c]{lll}
\multicolumn{3}{l}{\textbf{Algorithm 1 (\texttt{copulaABCdrf})}. Estimate the posterior probability density function:}\\
\multicolumn{3}{c}{$\pi_{\widehat{m}}^{\text{ABC}}(\theta\mid s(x))\propto\pi_{\widehat{m}}(\theta){\textstyle\int}f_{\widehat{m},s}(s(x)+\varepsilon\nu\mid\theta)\mathcal{K}(\nu)\mathrm{d}\nu$,}\\
\multicolumn{3}{l}{for the selected (best) model $\widehat{m}$ among $M\geq1$ models considered.}\\\hline\hline

\multicolumn{3}{l}{Inputs: Dataset $x$ of size $n$; number of iterations $N_{M}$; $M\geq1$ models $\{m\}_{m=1}^{M}$ for}\\
\multicolumn{3}{l}{dataset $x$; each model $m$ with exact likelihood
$f_{m}(\cdot\mid\theta)$, prior probability $\pi(m)$,}\\
\multicolumn{3}{l}{prior distribution $\pi_{m}(\theta)$ for model parameters $\theta=(\theta_{k})_{k=1}^{d(m)}\in\Theta^{(m)}\subseteq\mathbb{R}^{d(m)}$,}\\
\multicolumn{3}{l}{candidate summary statistics $s(\cdot)$, size $n_{\text{sim}}>0$ of data $y$ to simulate $y \sim f_{m}(\cdot\mid\theta)$.}\\\hline\\

\multicolumn{3}{l}{\textbf{(Step 1)} Construct an initial Reference Table.}\\
\multicolumn{3}{l}{for $j=1,\ldots,N_{M}$ do \ \  (using parallel computing cores if necessary)}\\
& (a) & If $M>1$, draw model index, $m(j)\sim\pi(m)$. \ If $M=1$, set $m(j)\equiv\widehat{m}\equiv1$.\\
& (b) & Draw a parameter sample, $\theta^{(j)}\sim\pi_{m(j)}(\theta)$.\\
& (c) & Draw a sample dataset, $y^{(j)}\sim f_{m(j)}(x\mid\theta^{(j)})$ of size $n_{\text{sim}}$ ($\leq n$).\\
& (d) & Calculate summary statistics, $s(y^{(j)})$.\\
& (e) & For each discrete $\theta_{k}^{\ast(j)}\in\theta^{(j)}$, draw $U_{k}^{(j)}\sim$ $\mathcal{U}_{(0,1)}$, set $\theta
_{k}^{(m(j))}\equiv$ $\theta_{k}^{\ast(m(j))}-U_{k}^{(j)}$.\\
\multicolumn{3}{l}{end for}\\
\multicolumn{3}{l}{Output Reference Table: $\{m(j),\theta^{(j)},s(y^{(j)}
)\}_{j=1}^{N_{M}}$;}{or $\{\theta^{(j)},s(y^{(j)})\}_{j=1}^{N}$ if $M=1$, $N_{M}\equiv N$.}\\\hline
\multicolumn{3}{l}{\textbf{(Step 2) }If $M>1$, train \texttt{drf} on $\{m(j),s(y^{(j)})\}_{j=1}^{N_{M}}$, regressing $m$ on $s$}\\
\multicolumn{3}{l}{(selected automatically by \texttt{drf}).\ Then based on trained \texttt{drf}, estimate the}\\
\multicolumn{3}{l}{posterior probabilities $\widehat{\Pi}_{x}(m)$ of models $\{m\}_{m=1}^{M}$ conditional on $s(x)$.}\\
\multicolumn{3}{l}{Select the model $\widehat{m}\in\{m\}_{m=1}^{M}$ with the highest posterior probability.}\\
\multicolumn{3}{l}{Reduce Reference Table to $\widehat{m}$: $\ \{\theta
^{(j)},s(y^{(j)})\}_{j=1}^{N}\equiv$ $\{\theta^{(j)},s(y^{(j)}):m(j)=\widehat{m}\}_{j=1}^{N}$.}\\\hline
\multicolumn{3}{l}{\textbf{(Step 3) } Estimate parameters $\{\nu,\rho,\{\pi_{k},\Pi_{k}\}_{k=1}^{d}\}$ of meta-$t$ posterior pdf:}\\
\multicolumn{3}{l}{$\pi_{\widehat{m},\nu,\rho,\pi}^{\text{ABC}}(\theta\mid
x)=c_{\nu,\rho}(u)
{\textstyle\prod\limits_{k=1}^{d}}
\pi_{x,k}(\theta_{k})$}\\
\multicolumn{3}{l}{$\ \ \ \ \ \ \ \ \ \ \ \ \ \ \ \ \ \ \ =\dfrac
{t_{d,\nu,\rho}(T_{\nu}^{-1}(u_{1}),\ldots,T_{\nu}^{-1}(u_{d}))}{
{\textstyle\prod\nolimits_{k=1}^{d}}
t_{\nu}(T_{\nu}^{-1}(u_{k}))}
{\textstyle\prod\limits_{k=1}^{d}}
\pi_{x,k}(\theta_{k})$ for $u\in(0,1)^{d}$;}\\
\multicolumn{3}{l}{$c_{\nu,\rho}(u)$ is $t$-copula pdf; $\ \ t_{\nu}$ is pdf ($T_{\nu}$ cdf) of standard univariate $t$ distribution;}\\
\multicolumn{3}{l}{$t_{d,\nu,\rho}$ is the $d$-variate $t$ pdf; with degrees of freedom $\nu$ and $d\times d$ scale matrix $\rho$;}\\
\multicolumn{3}{l}{univariate marginal posterior pdfs $\pi_{x,k}(\theta_{k})$ and cdfs $u_{k}=\Pi_{x,k}(\theta_{k})$ for $k=1,\ldots,d$.}\\
\multicolumn{3}{l}{Find estimates $\{\widehat{\nu},\widehat{\rho
},\{\widehat{\pi}_{k},\widehat{\Pi}_{k}\}_{k=1}^{d}\}$ using 2-stage semiparametric estimation:}\\
\multicolumn{3}{l}{(1) For $k=1,\ldots,d$, train \texttt{drf}\ on $\{\theta_{k}^{(j)},s(y^{(j)})\}_{j=1}^{N}$, regressing $\theta_{k}$ on $s(y)$,}\\
\multicolumn{3}{l}{with summary statistics (covariate) selection; use trained \texttt{drf} to predict estimates}\\
\multicolumn{3}{l}{of posterior cdfs\ $\widehat{u}_{k}^{(j)}=\widehat{\Pi
}_{x,k}(\theta_{k}^{(j)})$ and pdfs $\widehat{\pi}_{x,k}(\theta_{k}^{(j)})$ conditional on $s(x)$.}\\
\multicolumn{3}{l}{(2) Find: \ $(\widehat{\nu},\widehat{\rho})=$
$\underset{(\nu,\rho)\in(0,\infty)\times\{\rho\}}{\arg\max}%
{\textstyle\prod\limits_{j=1}^{N}}
c_{\nu,\rho}(\widehat{\Pi}_{x,1}(\theta_{1}^{(j)}),\ldots,\widehat{\Pi}_{x,d}(\theta_{d}^{(j)}))
{\textstyle\prod\limits_{k=1}^{d}}
\widehat{\pi}_{x,k}(\theta_{k}^{(j)})$.}\\\hline
\multicolumn{3}{l}{\textbf{(Step 4) } From estimated posterior distribution $\pi_{\widehat{m},\widehat{\nu},\widehat{\rho},\widehat{\pi}}^{\text{ABC}}(\theta\mid x)$, obtain posterior:}\\
\multicolumn{3}{l}{mean, variance, quantiles of $\theta_{k}$ from \texttt{drf}\ cdf estimates $\widehat{\Pi}_{x,k}$ for $k=1,\ldots,d$;}\\
\multicolumn{3}{l}{posterior scale matrix estimate $\widehat{\rho}$ of $\theta$ from Step 3; and the mode and MLE by:}\\
\multicolumn{3}{l}{$\widehat{\text{Mode}}_{x}(\theta)=$ $\underset{j=1,\ldots,N}{\arg\max}\text{ }\pi_{\widehat{m},\widehat{\nu},\widehat{\rho},\widehat{\pi}}^{\text{ABC}}(\theta^{(j)})$, MLE $\widehat{\theta}=$ $\underset{j=1,\ldots,N}{\arg\max}\text{ }\pi_{\widehat{m},\widehat{\nu},\widehat{\rho},\widehat{\pi}}^{\text{ABC}}(\theta^{(j)})/\pi(\theta^{(j)})$;}\\
\multicolumn{3}{l}{or by: $\widehat{\text{Mode}}_{x}(\theta
)=\underset{j=1,\ldots,N_{+}}{\arg\max}\text{ }\pi_{\widehat{m},\widehat{\nu
},\widehat{\rho},\widehat{\pi}}^{\text{ABC}}(\theta_{+}^{(j)})$,
$\widehat{\theta}=\underset{j=1,\ldots,N_{+}}{\arg\max}\text{ }\pi_{\widehat{m},\widehat{\nu},\widehat{\rho},\widehat{\pi}}^{\text{ABC}}(\theta_{+}^{(j)})/\pi(\theta_{+}^{(j)})$,}\\
\multicolumn{3}{l}{given $N_{+}\ $additional draws $\{\theta_{+}^{(j)}
\}_{j=1}^{N_{+}}$ $\overset{\text{iid}}{\sim}$ $\pi_{\widehat{m},\widehat{\nu},\widehat{\rho},\widehat{\pi}}^{\text{ABC}}$.}\\\hline\hline
&  &
\end{tabular}
\endgroup
\end{center}

\noindent possible, to define neighborhoods of relevant training data points for every covariate value $s$. Repeating this many times with randomization induces a weighting function $w_{s(x)}(s(y^{(j)}))$ quantifying the relevance of each training covariate data point $s(y^{(j)})$ (for $j=1,\ldots,N$) for a given test point $s(x)$. Specifically, the weight $w_{s(x)}(s(y^{(j)}))$ is the proportion of times out of $B$ CART subsampling randomizations that $s(y^{(j)})$ ends up in the same terminal leaf node as $s(x)$.

The \texttt{drf} estimates of the posterior conditional cdfs are given by:
\begin{subequations}
\label{cdfEst}
\begin{align}
\widehat{\Pi}_{x,k}(\theta_{k})  &  ={\displaystyle\sum\limits_{j=1}^{N}}
w_{s(x),k}\{s(y^{(j)})\}\mathbf{1}(\theta_{k}^{(j)}\leq\theta_{k})\\
&  = {\displaystyle\sum\limits_{j=1}^{N}}
\left\{  \dfrac{1}{B}%
{\displaystyle\sum\limits_{b=1}^{B}}
\frac{\mathbf{1}[s(y^{(j)})\in\mathcal{L}_{b}\{s(x)\}]}{\left\vert
\mathcal{L}_{b}\{s(x)\}\right\vert }\right\}  \mathbf{1}(\theta_{k}^{(j)}
\leq\theta_{k}),\text{  for $k=1,\ldots,d$}
\end{align}
obtained from $N$ (training) points of the Reference table $\{\theta_{k}^{(j)},s(y^{(j)})\}_{j=1}^{N}$; where the $w_{s(x),k}(s(y^{(j)}))$ (for each $k=1,\ldots,d$) are positive weights with ${\textstyle\sum\nolimits_{j=1}^{N}}w_{s(x),k}\{s(y^{(j)})\}=1$; and $B$ is the number of randomized CARTs ($T_{b}$ for $b=1,\ldots,B$) of the ensemble; $\mathcal{L}_{b}\{s(x)\}$ and $\left\vert\mathcal{L}_{b}\{s(x)\}\right\vert $ are respectively the set and number of the training data points $(\theta^{(j)},s(y^{(j)}))$ which end up in the same leaf as $s(x)$ in CART tree $T_{b}$; and $\mathbf{1}(\cdot)$ is the indicator function. The corresponding \texttt{drf} pdf estimates (for $k=1,\ldots,d$) can be computed by the following empirical histogram-type density estimator:
\end{subequations}
\begin{equation}
\widehat{\pi}_{x,k}(\theta_{k})=%
{\displaystyle\sum\limits_{l=1}^{N}}
\frac{w_{s(x),k}\{s(y_{(l)})\}}{\theta_{k,(l)}-\theta_{k,(l-1)}}\mathbf{1}%
(\theta_{k}\in(\theta_{k,(l-1)},\theta_{k,(l)}]), \label{pdfEst}
\end{equation}
where $\theta_{k,(1)},\ldots,\theta_{k,(N)}$ (with $\theta_{k,(0)}\equiv
\theta_{k,(1)}$) are the order statistics of $\theta_{k}^{(1)},\ldots
,\theta_{k}^{(N)}$, these order statistics corresponding to weights
$w_{s(x),k}\{s(y_{(1)})\},\ldots,w_{s(x),k}\{s(y_{(N)})\}$. Alternatively, the marginal posterior density estimates $\widehat{\pi}_{x,k}(\theta_{k})$, for $k=1,\ldots,d$, can, respectfully, be obtained by smoother, univariate local polynomial Gaussian-kernel density estimators (with spline interpolation to speed up computations) and automatic bandwidth selection \citep{SheatherJones91}, performed on the $\theta_{k,1},\ldots,\theta_{k,N}$, using corresponding frequency weights $N \cdot w_{s(x),k}\{s(y_{(1)})\} \ldots,N \cdot w_{s(x),k}\{s(y_{(N)})\}$, and with unbounded or bounded support \citep{Geenens14,GeenensWang18}, depending on the spaces (resp.) of the univariate parameters or the support of their priors, as appropriate, e.g., by using the \texttt{kde1d} R software package \citep{NaglerVatter24}). 

Extensive simulation studies \citep[][\S4.1]{CevidEtAl22} showed that \texttt{drf} performs well and outperforms other machine learning methods in terms of prediction accuracy, for a wide range of sample size and problem dimensionality, especially in problems where $p$ (number of covariates) is large and $d$ (dimensionality of the dependent variable) is small to moderately large, without the need for further tuning or involved numerical optimization. The \texttt{drf} training and predictions can compute fast using the \texttt{drf} R package \citep{MichelCevid21}, with minimal tuning parameters. By default, a \texttt{drf} is estimated from the given training dataset using an ensemble of $B=2,000$ CARTs, with every tree constructed (randomized) from a random subset being 50\% of the size of the training set, and with a target of 5 for the minimum number of observations in each CART tree leaf \citep[][p.33]{CevidEtAl22}. The estimated (induced) weighting function of \texttt{drf}, for each of the given model parameters $\theta_{k}^{(j)}$ (for $k=1,\ldots,d$), is used to estimate a parametric meta-$t$ multivariate copula-based distribution for these dependent variable observations, after having nonparametrically adjusted for the covariates $s(y^{(j)})$ (for $j=1,\ldots,N$) \citep[in the spirit of][]{BickelEtAl93}. All \texttt{drf} computations mentioned can be undertaken using the \texttt{drf} R package \citep{MichelCevid21}.

Step 4 of Algorithm 1, with the $d$ \texttt{drf}s estimated in Step 3, uses the weights $w_{s(x)}$ and marginal posterior cdf estimates (\ref{cdfEst}) to calculate, for each model parameter $\theta_{k}$ for $k=1,\ldots,d$, estimates of marginal posterior expectations, including estimates of the posterior mean ($\widehat{\mathbb{E}}$), variance ($\widehat{\mathbb{V}}$), and quantiles ($\widehat{Q}$):
\begin{align*}
\widehat{\mathbb{E}}_{x,k}(\theta_{k})  &  =
{\textstyle\sum\nolimits_{j=1}^{N}}
\theta_{k}^{(j)}w_{s(x)}\{s(y^{(j)})\},\\
\widehat{\mathbb{V}}_{x,k}(\theta_{k})  &  =%
{\textstyle\sum\nolimits_{j=1}^{N}}
\{\theta_{k}^{(j)}-\widehat{\mathbb{E}}_{x,k}(\theta_{k})\}^{2}w_{s(x)}%
\{s(y^{(j)})\},\\
\widehat{Q}_{x,k}(u)  &  =\widehat{\Pi}_{x,k}^{-1}(u),\text{ for }%
k=1,\ldots,d\text{ and }u\in(0,1).
\end{align*}

Step 4 also calculates estimates of the multivariate posterior mode, $\widehat{\text{Mode}}_{x}(\theta )$ and MLE $\widehat{\theta}$ of the parameters $\theta$ of the (selected) model. The MLE is the multivariate mode of the ratio of the meta-$t$ posterior pdf estimate to the prior pdf, or the posterior mode under a proper uniform prior supporting the MLE. For either the task of posterior mode or MLE estimation, a novel semiparametric mode estimator is being proposed here, which is applicable to high dimensions. Other multivariate mode estimators \citep[see][for a review]{Chacon20} seem only applicable to lower dimensions. The new multivariate mode estimator maximizes a (posterior) density estimate, and finds the MLE by maximizing a ratio of the posterior density estimate to the prior density, over a sample of points. The sample of points is the set of parameter samples $\theta^{(j)}$ from the Reference Table, or alternatively (if necessary), can be extra parameter samples $\theta _{+}^{(j)}$ drawn from the meta-t posterior density estimate. Such a sampling approach avoids a possibly costly grid search involved in multivariate mode estimation by global maximization. Similarly, other `indirect' multivariate mode estimation methods 
\citep{Devroye79,AbrahamEtAl03a,AbrahamEtAl04,HsuWu13,DasguptaKpotufe14} use the data sample points to estimate the mode from a nonparametric kernel, $k$-nearest-neighbor, or parametric density estimate, while respectively, relying on choice of bandwidth, or $k$, or on a successful normal transformation of the data. The new \texttt{copulaABCdrf} framework (Algorithm 1) estimates the MLE  (mode) while performing automatic summary statistics (variable) selection via \texttt{drf}s for the meta-$t$ marginals. Therefore, this framework advances on previous simulation-based MLE estimation methods, which do not provide this automatic selection of summaries \citep{RubioJohansen13,KajiharaEtAl18,PicchiniAnderson17,GutmannCorander16,YildirimEtAl15,DeanEtAl14,GourierouxEtAl93,McFadden89}.

In particular, the original \texttt{rejectionABC} approach to MLE and posterior mode estimation \citep{RubioJohansen13} is based on a multivariate kernel density estimate of the posterior distribution estimated from the subset of prior parameter samples from the Reference table, corresponding to simulated summary statistics being a small distance $\epsilon$ to the observed data summary statistics (and recall that $\epsilon$ can be chosen as a small \textit{q}\% (e.g., 1\%) quantile of the simulated distances). For example, for a 1- to 6-dimensional parameter, a multivariate kernel density estimate can be obtained using the \texttt{kde()} function of the \texttt{ks} R package \citep{Duong24} based on automatic selection of the bandwidth matrix. For a higher-dimensional parameter (of dimension $d$), a multivariate kernel density estimate can be obtained by a product of $d$ univariate local polynomial Gaussian-kernel density estimators \citep[][eq.6.47]{Wasserman06} based on automatic bandwidth selection \citep{SheatherJones91} performed for each of these \textit{d} densities, while the accuracy of the estimator deteriorates quickly as dimension \textit{d} increases \citep[e.g.,][p. 138]{Wasserman06}. Then, given this posterior kernel density estimate, an estimate of the posterior mode is obtained by maximizing this density, over the \textit{q}\% subsamples of the Reference Table, gives an estimate of the posterior mode. Also, the MLE is obtained by maximizing the ratio of this posterior density to the prior density over these subsample, while of course, the MLE coincides with the posterior mode under a uniform prior distribution. Later, we sometimes refer to the rejection-based ABC algorithm for estimating the posterior mode or MLE, based on kernel density estimation, as \texttt{rejectionABCkern}, \texttt{rejectionABCkern.select}, or \texttt{rejectionABCprodkern}, depending on whether the rejection-based ABC algorithm used multivariate kernel density estimation (\texttt{kern}), perhaps using pre-selection of summary statistics (kern.select), or used multivariate product kernel density estimation (\texttt{prodkern}).


\section{Results of Numerical Examples\label{NumericalExamples}}

This section (\S3) presents results of several simulation studies which evaluate and compare the \texttt{copulaABCdrf} (Algorithm 1) and methods based on the standard (small) \textit{q}\% nearest-neighbor \texttt{rejectionABC} algorithm \citep{BiauCerouGuyader15}, in terms of estimation accuracy of various features of the posterior distribution, for three models for $n$ multivariate independently and identically distributed (i.i.d.) observations in \S3.1, and for seven models for $n$-node networks. Namely, three ERGMs in \S3.3 and \S3.7, and four mechanistic network models in \S3.4-\S3.5, after \S3.2 provides a contextualizing overview of network science and modeling. To provide further illustrations of the \texttt{copulaABCdrf} and \texttt{rejectionABC} methods, these network models are applied to analyze real-life massive network datasets in \S3.6-\S3.8. Every application of \texttt{copulaABCdrf} and \texttt{rejectionABC} method is based on a Reference Table of size $N = 10,000$ samples and the same given set of summary statistics, after considering that this size is an automatic default choice in \texttt{ABC} applications, and that some summary statistics used in this study are computationally costly especially for large datasets. 

The simulation studies will in general summarize posterior estimation accuracy results of each \texttt{ABC} method by the average and standard deviation (error) of point estimates of parameters (marginal means and medians of the posterior distribution, and the posterior mode and MLE) and various fit statistics over 10 simulated data replicas from the given model generated from specified true data-generating model parameters, mentioned in the later subsections. These fit statistics, computed relative to these true data-generating model parameters, include the: Mean Absolute Error (MAE) and (the less outlier-robust) Mean Squared Error (MSE) of the above point estimates; coverage (indicator) of the marginal 95\% and 50\% posterior credible interval estimates of individual model parameters, to be compared to their nominal values; and Kolmogorov-Smirnov (KS) distance and test statistics of the null hypotheses that the estimated univariate marginal posterior distributions (respectively) match the corresponding true exact univariate marginal posterior distributions. Specifically, the weighted one-sample (two-tailed) KS distance and corresponding significance test statistic \citep[][p.358]{Monahan11}, based on the estimated \texttt{drf} weights for \texttt{copulaABCdrf}, or on equal sample weights (unweighted) for \texttt{rejectionABC}. The KS test statistic has a 95th percentile critical value of 1.358 and a 99th percentile critical value of 1.628, relative to the null hypothesis that the marginal posterior distribution of the given parameter of \texttt{copulaABCdrf} (or \texttt{rejectionABC}) equals the exact marginal posterior distribution. The KS statistics can only be computed for the three models for multivariate i.i.d. observations (in \S3.1), since they are defined by tractable (Poisson and Gaussian) likelihoods, and thus allow for computations of the exact univariate marginal posteriors, either using direct numerical computation or by using MCMC or other suitable Monte Carlo methods. In contrast, for each network model considered in this paper, the likelihood function is intractable because it is either inexplicit or not computable when the network is of a sufficiently large size $n$ (e.g., with more than a handful of nodes, $n$). This makes it impossible to compute the exact marginal posterior distributions, or at least computable in a reasonable time using Monte Carlo methods, and thus, impossible to compute the KS statistics for these models. 

Specifically, \S3.1 reports the results of simulation studies for three models for $n$ multivariate i.i.d. observations, namely: a joint Poisson and Gaussian mixture model defined by two parameters, having a unimodal joint posterior and exact univariate marginal posteriors, which can be directly calculated analytically; a bivariate Gaussian model, with five parameters (2 location and 3 covariance matrix parameters) having a multimodal joint posterior distribution; and a high-dimensional multivariate Gaussian model with 300 location parameters, with the first two parameters having a skewed posterior. Since each of the latter two Gaussian models is defined by a Gaussian likelihood pdf, the exact univariate marginal posterior distributions of the model parameters (respectively) can be calculated using standard MCMC methods. For each of these two models, this study estimates the exact univariate marginal posterior distributions using 10,000 samples generated from a componentwise MCMC Gibbs-and slice-sampling algorithm, which routinely displayed adequate mixing and convergence to the exact univariate marginal posterior distributions, according to univariate trace plots of the MCMC parameter samples. While these MCMC samples are correlated, more efficient estimation of posterior point estimates are obtained from all 10,000 stationary MCMC samples instead of from a thinned subsample \citep{MacEachernBerliner94}. 

More specifically, each sampling iteration of this MCMC algorithm: performed a Gibbs sampling update of the subset of mean location parameters by drawing a sample from its explicit multivariate Gaussian full conditional posterior distribution, given the data and the remaining model parameters which is easily derivable from the standard posterior calculus of Bayesian Gaussian models under conjugate or uniform priors \citep[e.g.,][Appendix A]{BernardoSmith94} (specifically, for the bivariate Gaussian model, the Gibbs sampling update involved simple rejection sampling from a uniform prior-truncated bivariate Gaussian for both mean parameters; and for the 300-variate Gaussian model, the sampling update remaining mean model parameters involved a direct draw from a 297-variate Gaussian with fixed diagonal covariance matrix); and for the remaining model parameters having no convenient known form for the full conditional posterior distribution(s), the MCMC algorithm employed a simple version of the slice sampler \citep{Neal03} to perform a sampling update, by repeatedly sampling from a wide uniform distribution that surely supported the entire full conditional posterior density of the parameter, until a sample was obtained that yielded a full conditional posterior density value that exceeded the corresponding value of the slice variable updated in this MCMC iteration (for the bivariate Gaussian model, a trivariate slice sampler was used for all the covariance matrix parameters; and for the 300-variate Gaussian model, a univariate slice sampling update was performed for each of the first two location parameters already known to have a skewed joint posterior, while in each update, the shrinkage procedure \citep{Neal03} was used to speed the search for the slice). For these two Gaussian models in particular, this componentwise MCMC Gibbs- and slice-sampling algorithm seems to provide the simplest posterior sampling algorithm, while making direct use of the Fundamental Theorem of Simulation \citep[][\S 2.3.1]{RobertCasella04} and using no tuning parameters. This is unlike alternative viable Monte Carlo methods, such as Metropolis-Hastings, Hamiltonian \citep{HoffmanGelman14} and affine-invariant ensemble \citep{GoodmanWeare10} sampling algorithms, which require the use of proposal covariance matrices, gradients, or other tuning parameters.

In general, for the simulation studies throughout \S3, all the evaluations and comparisons of \texttt{copulaABCdrf} and \texttt{rejectionABC} will be based on varying conditions of the number of data simulations from the likelihood, $n_{\text{sim}}$, relative to the total sample size $n$, with $n_{\text{sim}}<n$, where for the network models, $n$ is the number of nodes of a network. When constructing a Reference Table for ABC, the strategy of simulating summary statistics based on simulating datasets of size $n_{\text{sim}}<n$ can potentially be useful in situations where a large size $n$ dataset is being analyzed and/or simulating from the given model is computationally costly. In such large ($n$) network scenarios, it is prohibitively costly and practically infeasible to simulate many networks of the same size (number of nodes) as the original network dataset over iterations in an ABC or synthetic likelihood algorithm, especially considering that already it can be prohibitively costly to simulate a single network from ERGM or mechanistic network models (e.g., ERGM given parameters simulates a network using MCMC), for a network of sufficiently large size (number of nodes), let alone to compute point-estimates (e.g., MLE) of the parameters of the given network model. These issues, in the context of Algorithm 1, are addressed by a strategy that simulates network datasets of a smaller size compared to the size of the network dataset under statistical analysis, while using network summary statistics (calculated on the observed dataset and each simulated dataset) that take on values that have the same meaning and are invariant to the size of the network dataset(s) being analyzed, including Maximum Pseudo Likelihood Estimates (MPLEs) of ERGM model parameters based on network size offsets \citep{KrivitskyEtAl11}. These network size invariant summary statistics are reviewed in \S 3.2. The R packages \texttt{ergm} \citep{HandcockEtAl23}, \texttt{ergm.count} \citep{Krivitsky22count}, and \texttt{igraph} \citep{CsardiEtAl24} were used to compute network summary statistics, MPLEs, and compute Monte Carlo MLEs (MCMLEs) \citep{Snijders02} of ERGM models, with the MCMLE being a standard commonly-used MCMC approximation of the ERGM MLE. These MLEs will be compared with the MLEs of \texttt{copulaABCdrf} and \texttt{rejectionABC} by MAE and MSE. Note that some values of the ERGM parameters can concentrate probability mass on degenerate or near-degenerate networks which concentrate probability mass on a small subset of all possible network graphs with almost or exactly zero edges or almost or all possible edges among the $n$ nodes \citep{Strauss86,Handcock03}, which can lead to infinite values of MPLEs and MLEs for one or more ERGM parameters. Therefore, for the network modeling applications of \S3.3-3.8 where MPLE summary statistics are used, the results of the \texttt{copulaABCdrf} and \texttt{rejectionABC} methods will only be based on the subset of the rows of the Reference Table corresponding to finite-valued MPLE summary statistics. This provided a natural way to address this degeneracy, at least for the models and summaries used in these sections.

For each of the multivariate i.i.d. models, ERGMs, and one of the mechanistic (DMC) network model investigated in the simulation studies of \S3.1, \S3.3, \S3.5, and \S3.7, the number of summary statistics (in $s$) equals the number of model parameters (i.e., their dimensionality is equal, dim($s$) = dim($\theta$)). For each of the three other (Price, NLPA, and DMR) mechanistic network models investigated in the simulation studies of \S3.4-\S3.5, the initial number of summary statistics exceeds the number of model parameters (i.e., dim($s$) $\geq $ dim($\theta$)). The asymptotic efficiency of posterior expectations in \texttt{ABC} relies on dim($s$) = dim($\theta$) \citep{LiFearnhead18}, as mentioned in \S1. Therefore, in the simulation scenarios of \S3.4-\S3.5, where the number of summary statistics exceeds the number of model parameters, we will also consider a semi-automatic approach to \texttt{rejectionABC} based on pre-selecting the dim($\theta$) most important summary statistics (predictor variables), determined by training a \texttt{drf} regression of the prior samples of $\{\theta\}_{i=1}^{N}$ on the corresponding samples of the summary statistics $\{s\}_{i=1}^{N}$ on the Reference Table, with the importance of each summary statistic (variable) efficiently calculated by a simple weighted sum of how many times each summary statistic was split on at each depth in the forest. This semiautomatic ABC method based on preselection of summaries using \texttt{drf}, called \texttt{rejectionABCselect}, will be compared with the results obtained by \texttt{rejectionABC} using all the available summary statistics (i.e., without any pre-selection), and the results obtained by \texttt{copulaABCdrf}.

Preliminary results of simulation studies for the \textit{q}\% nearest-neighbor \texttt{rejectionABC} algorithm, for all models studied in \S3.1, \S3.3-3.5 and \S3.7, showed that the 1\% nearest-neighbor \texttt{rejectionABC} algorithm tended to perform best in terms of the MAE, MSE, and KS statistics, compared to 2\% and 3\% nearest-neighbors. Therefore, for space considerations, only the results for 1\% nearest-neighbor \texttt{rejectionABC} will be shown throughout \S3.

For the \texttt{copulaABCdrf} method, it was found through preliminary simulation studies of all eight models considered in \S 3.1 and \S 3.3-3.5, that compared to the smooth univariate kernel density estimators (mentioned in \S2), the empirical histogram density estimator (\ref{pdfEst}) had a slightly higher tendency for producing superior posterior mode and MLE estimates based on the meta-$t$ posterior, for most models considered in the simulation study, according to MAE and MSE of each individual model parameter, when the dimension dim$(\theta)$ of the given model parameters $\theta$ was less than 6. The empirical density estimator also has the advantage of not using a bandwidth parameter, while the choice of bandwidth is an important and non-trivial aspect for the accuracy of smooth kernel density estimation. However, it was found that when the given model parameter $\theta$ has a sufficiently-high dimension, the product term $\widehat{\pi}_{x,k}(\theta_{k}^{(j)})$ (for $j=1,\ldots,N$) of the meta-$t$ density estimate (see Step 4 of Algorithm 1), based on the empirical density estimator, can become zero for a large majority of the $N$ samples of  $\{\theta^{(j)}\}_{i=1}^{N}$ in the Reference Table, thereby requiring an extremely large sample $N$ (e.g., well-above, say, 10,000) to overcome this issue. Therefore, in such a high-dimensional parameter settings, the kernel density estimator has the advantage of reducing the frequencies of zero from the Reference Table, because it provides a smoother density estimator, and therefore does not require an extremely large sample for the Reference Table. Also, in a related issue, according to some preliminary simulation studies, done as a pre-cursor to the simulation studies reported later in \S 3.1 and \S 3.3-3.5, it turned out that the computation of the MLE $(\widehat{\nu},\widehat{\rho})$ of the $t$-copula density parameters (in Step 4 of Algorithm 1) based on multiplying (re-scaling) the $\widehat{u}_{k}^{(j)}$ by $N/(N+1)$, done to avoid evaluating the copula density at the edges of the unit hypercube (as advocated by \citep{GenestEtAl95,GenestNeslehova07} and others), was not obviously advantageous in terms of MAE and MSE accuracy of posterior mean, median, mode,  and MLE estimation. Therefore, \S3 throughout presents the results of the simulation studies based on Step 3 of the \texttt{copulaABCdrf} Algorithm 1, as described in \S2.

Further, recall that both of \texttt{copulaABCdrf}'s alternative approaches to posterior mode and MLE estimation are presented in Step 4 of Algorithm 1. It was found through additional preliminary simulation studies \texttt{copulaABCdrf} over all eight models considered in \S 3.1 and \S 3.3-3.5, that in terms of posterior mode and MLE estimation accuracy of \texttt{copulaABCdrf}, measured by MAE and MSE, the approach of drawing an extra set of parameter samples from the meta-$t$ posterior density estimate can be advantageous (compared to the first approach based on the original Reference Table), but only when the true posterior distribution is high-dimensional and symmetric or skewed (but not highly multimodal). This seems to be reflective of the geometry of the meta-$t$ density (distribution). Therefore, for space considerations, we primarily present the results of the posterior mode and MLE estimators of \texttt{copulaABCdrf} which is only based on the samples generated in the original Reference Table. We only focus on the extra-sample approach to mode and MLE estimation for a high-dimensional setting involving a skewed posterior distribution, and based on the smooth univariate kernel density estimators of the univariate posterior marginals.

\subsection{Simulation Study: Calculable Exact Posterior Distributions}

Here, we first study the accuracy of Algorithm 1 to estimate the directly-calculable true posterior distribution, of a joint bivariate Poisson$(\theta)$ model and a two-component scale normal mixture model with common mean parameter $\mu$, which has been studied in previous research on ABC methods. The Poisson$(\lambda)$ distribution is defined by a tractable likelihood probability mass function (p.m.f.) $f(X_{1}=x_{1}\mid\lambda)=\lambda^{x_{1}}\exp(-\lambda)/x_{1}!$, assigned a gamma prior distribution $\mathcal{G}(\lambda\mid\frac{1}{2},0.1)$ (shape $\frac{1}{2}$ and rate $0.1$), and exact gamma posterior distribution $\lambda\mid x_{1}\sim\mathcal{G}(\lambda\mid\frac{1}{2}+%
{\textstyle\sum\nolimits_{i=1}^{n}}
x_{i,1},0.1+n)$, with the sample mean $\overline{x}_{1}=\tfrac{1}{n}{\textstyle\sum\nolimits_{i=1}^{n}}
x_{i,1}$ a sufficient summary statistic for $\lambda$. The scale normal mixture model $f(X_{2}=x_{2}\mid\mu)=\frac{1}{2}\mathcal{N}(x_{2}\mid
\mu,1)+\frac{1}{2}\mathcal{N}(x_{2}\mid\mu,0.01)$, the normal distribution $\mathcal{N}(x_{2}\mid\mu,z+(1-z)0.01)$ with latent covariate $z \sim $Bernoulli$(1/2)$, is assigned a uniform prior distribution $\mu\sim\mathcal{U}_{(-10,10)}$, yielding the posterior distribution $\pi(\mu\mid x_{2})\propto\frac{1}{2}\mathcal{N}(\mu\mid\overline{x}_{2},1)+\frac{1}{2}\mathcal{N}(\mu\mid\overline{x}_{2},0.01)$ (effectively, with equality up to machine precision) with the summary statistic the sample mean $\overline{x}_{2}=\tfrac{1}{n}{\textstyle\sum\nolimits_{i=1}^{n}}x_{i,2}$. The Poisson and scale normal mixture distributions, together, define a bivariate distribution model, respectively, with parameters assigned the aforementioned independent prior distributions, and posterior distributions as marginal distributions of $\theta=(\lambda,\mu)$. This enables Algorithm 1 to estimate a bivariate posterior mode and MLE of $\theta$.

Ten replications of $n=100$ bivariate observations $\{(x_{i,1},x_{i,2})\}_{i=1}^{n=100}$ from this bivariate model were simulated from the bivariate distribution model with true data generating parameters $\theta=(\lambda=3,\mu=0)$, for each of eight conditions defined by sizes of $n_{\text{sim}}$ $=10, 25, 33, 50, 66, 75, 90$, and $100$ simulations from the exact likelihood of the model.

Table 1 presents detailed representative results for some  $n_{\text{sim}}$ $=10, 50$ and $100$ while Tables 2 and 3 present the results for all the eight conditions of $n_{\text{sim}}$. Overall, \texttt{copulaABCdrf} outperformed \texttt{rejectionABC} in terms of MAE, MAE, and KS statistics for each of the posterior distribution and mean, median, and mode estimation. The \texttt{rejectionABC} method outperformed \texttt{copulaABCdrf} only for posterior mean estimation of the mean location parameter $\mu$ of the normal mixture, especially for smaller $n_{\text{sim}}$ $= 10$ and $25$. Also,  \texttt{copulaABCdrf} and \texttt{rejectionABC} performed similarly and rather close to the nominal 95\% posterior credible intervals, while \texttt{copulaABCdrf} performed better than \texttt{rejectionABC} in producing $50\%$ posterior credible intervals, while tending to produce interval estimates that were near the nominal values.  Both \texttt{copulaABCdrf} and \texttt{rejectionABC} always produced significant KS statistics, i.e., estimates of the marginal posterior distributions of model parameters which significantly departed from the exact posterior distribution. Therefore,  the univariate \texttt{drf} regressions trained on the Reference Table of 10,000 samples were unable to very accurately estimate these marginal posterior distributions, but they estimated marginal posteriors that were more accurate than those obtained from \texttt{rejectionABC}.

Finally, for both ABC methods, the accuracy of estimation from the posterior distribution tended to improve as $n_{\text{sim}}$ is increased towards the full dataset sample size $n$, more so for the rate parameter $\lambda$. This suggests that the approach of simulating datasets from the given model of size $n_{\text{sim}}$ smaller (but not too small or zero), relative to the size $n$ of the observed dataset being analyzed, can be a reasonable approach for accurate estimation of the model's posterior distribution, using either ABC method. Intuitively, the sufficient statistic $\overline{x}_{1}$ for the Poisson distribution parameter $\lambda$ is sufficient for a dataset of size $n_{\text{sim}}$ $=100$, and only approximately sufficient for smaller simulation sample sizes $n_{\text{sim}}$. Similarly, for the statistic $\overline{x}_{2}$ summarizing the parameter $\mu$ of the scale normal mixture model. 

\begin{changemargin}{-.3in}{-.3in}
\begin{center}
\begin{tabular}{lcccccc}
\multicolumn{7}{l}{\footnotesize } \\
\multicolumn{7}{l}{\footnotesize \textbf{Table 1.} Poisson$
(\lambda )$NormalMixture$(\mu )$ model: mean (standard deviation) of estimators of posterior distribution, } \\ 
\multicolumn{7}{l}{\footnotesize and mean 95\%(50\%) credible interval coverage (95\%(50\%)c) over 10 replicas ($n$=100), under three conditions of $n_{\text{sim}}$.} \\ 
\hline\hline
& \multicolumn{1}{|c}{\footnotesize${\lambda}$} & \footnotesize${\mu}$ & 
\multicolumn{1}{|c}{\footnotesize${\lambda}$} & \footnotesize${\mu}$ & 
\multicolumn{1}{|c}{\footnotesize${\lambda}$} & \footnotesize${\mu}$ \\ 
\hline
\footnotesize${n}$ & \multicolumn{1}{|c}{\footnotesize 100} & 
{\footnotesize 100} & \multicolumn{1}{|c}{\footnotesize 100} & 
{\footnotesize 100} & \multicolumn{1}{|c}{\footnotesize 100} & 
{\footnotesize 100} \\ 
\footnotesize${n_\text{sim}}$ & \multicolumn{1}{|c}{\footnotesize 100} & 
{\footnotesize 100} & \multicolumn{1}{|c}{\footnotesize 50} & {\footnotesize %
50} & \multicolumn{1}{|c}{\footnotesize 10} & {\footnotesize 10} \\ 
{\footnotesize truth} & \multicolumn{1}{|c}{\footnotesize 3} & 
{\footnotesize 0} & \multicolumn{1}{|c}{\footnotesize 3} & 
{\footnotesize 0} & \multicolumn{1}{|c}{\footnotesize 3} & 
{\footnotesize 0} \\\hline

{\footnotesize Exact mean} & \multicolumn{1}{|c}{\footnotesize 3.02 (0.15)} & {\footnotesize 0.01 (0.10)} & \multicolumn{1}{|c}{\footnotesize 3.08 (0.17)} & {\footnotesize -0.02 (0.09)} & \multicolumn{1}{|c}{\footnotesize 2.99 (0.11)} & {\footnotesize -0.01 (0.11)} \\ 

{\footnotesize \texttt{copulaABCdrf} mean} & \multicolumn{1}{|c}{\footnotesize \textbf{3.02} (0.15)} & {\footnotesize \textbf{0.01} (0.10)} & \multicolumn{1}{|c}{\footnotesize 3.09 (0.17)} & {\footnotesize \textbf{-0.02} (0.10)} & \multicolumn{1}{|c}{\footnotesize \textbf{3.05} (0.15)} & {\footnotesize -0.03 (0.12)} \\ 

{\footnotesize \texttt{\texttt{rejectionABC}} mean} & \multicolumn{1}{|c}{\footnotesize 2.94 (0.16)} & {\footnotesize 0.02 (0.10)} & \multicolumn{1}{|c}{\footnotesize \textbf{3.03} (0.17)} & {\footnotesize -0.03 (0.10)} & 
\multicolumn{1}{|c}{\footnotesize  \textbf{2.95} (0.14)} & {\footnotesize \textbf{-0.02} (0.10)} \\\hline

{\footnotesize Exact median} & \multicolumn{1}{|c}{\footnotesize 3.02 (0.15)} & {\footnotesize 0.01 (0.10)} & \multicolumn{1}{|c}{\footnotesize 3.07 (0.17)%
} & {\footnotesize -0.02 (0.09)} & \multicolumn{1}{|c}{\footnotesize 2.99 (0.11)} & {\footnotesize -0.01 (0.11)} \\ 

{\footnotesize \texttt{copulaABCdrf} median} & \multicolumn{1}{|c}{\footnotesize \textbf{3.02} (0.15)} & {\footnotesize \textbf{0.01} (0.10)} & \multicolumn{1}{|c}{\footnotesize 3.09 (0.16)} & {\footnotesize \textbf{-0.02} (0.09)} & \multicolumn{1}{|c}{\footnotesize \textbf{3.03}
(0.15)} & {\footnotesize -0.03 (0.13)} \\ 

{\footnotesize \texttt{\texttt{rejectionABC}} median} & \multicolumn{1}{|c}{\footnotesize 2.91 (0.18)} & {\footnotesize 0.02 (0.11)} & \multicolumn{1}{|c}{\footnotesize \textbf{3.03} (0.17)} & {\footnotesize -0.05 (0.10)} & \multicolumn{1}{|c}{\footnotesize 2.91 (0.12)} & {\footnotesize \textbf{-0.02} (0.09)} \\\hline

{\footnotesize Exact mode} & \multicolumn{1}{|c}{\footnotesize 3.01 (0.15)} & {\footnotesize 0.01 (0.10)} & \multicolumn{1}{|c}{\footnotesize 3.07 (0.17)} & {\footnotesize -0.02 (0.09)} & \multicolumn{1}{|c}{\footnotesize 2.98 (0.11)} & {\footnotesize -0.01 (0.11)} \\ 

{\footnotesize \texttt{copulaABCdrf} mode} & \multicolumn{1}{|c}{\footnotesize \textbf{3.06} (0.24)} & {\footnotesize -0.14 (0.44)} & \multicolumn{1}{|c}{\footnotesize \textbf{3.07} (0.27)} & {\footnotesize -0.16 (0.27)} & \multicolumn{1}{|c}{\footnotesize 3.23 (0.60)} & {\footnotesize -0.17 (0.17)} \\ 

{\footnotesize \texttt{rejectionABCkern} mode} & \multicolumn{1}{|c}{\footnotesize 2.81 (0.34)} & {\footnotesize \textbf{0.07} (0.33)} & \multicolumn{1}{|c}{\footnotesize 2.88 (0.49)} & {\footnotesize \textbf{-0.14} (0.30)} & \multicolumn{1}{|c}{\footnotesize \textbf{2.90} (0.24)} & {\footnotesize \textbf{0.02} (0.22)} \\\hline

{\footnotesize Exact MLE} & \multicolumn{1}{|c}{\footnotesize 3.02 (0.15)} & {\footnotesize 0.01 (0.10)} & \multicolumn{1}{|c}{\footnotesize 3.08 (0.17)} & {\footnotesize -0.02 (0.09)} & \multicolumn{1}{|c}{\footnotesize 2.99 (0.11)} & {\footnotesize -0.01 (0.11)} \\ 

{\footnotesize \texttt{copulaABCdrf} MLE} & \multicolumn{1}{|c}{\footnotesize 3.06 (0.24)} & {\footnotesize -0.14 (0.44)} & \multicolumn{1}{|c}{\footnotesize \textbf{3.07} (0.27)} & {\footnotesize -0.16 (0.27)} & \multicolumn{1}{|c}{\footnotesize 3.23 (0.60)} & {\footnotesize -0.17 (0.17)} \\ 

{\footnotesize \texttt{rejectionABCkern} MLE} & \multicolumn{1}{|c}{\footnotesize \textbf{3.00} (0.29)} & {\footnotesize \textbf{0.08} (0.32)} & \multicolumn{1}{|c}{\footnotesize 3.20 (0.41)} & {\footnotesize \textbf{0.00} (0.35)} & \multicolumn{1}{|c}{\footnotesize \textbf{3.17} (0.28)} & {\footnotesize \textbf{-0.02} (0.15)} \\\hline

{\footnotesize Exact standard deviation (s.d.)} & \multicolumn{1}{|c}{\footnotesize 0.17 (0.00)} & {\footnotesize 0.71 (0.00)} & \multicolumn{1}{|c}{\footnotesize 0.18 (0.00)} & {\footnotesize 0.71 (0.00)} & \multicolumn{1}{|c}{\footnotesize 0.17 (0.00)} & {\footnotesize 0.71 (0.00)} \\ 

{\footnotesize \texttt{copulaABCdrf} s.d.} & \multicolumn{1}{|c}{\footnotesize \textbf{0.20} (0.01)} & {\footnotesize 0.22 (0.07)} & \multicolumn{1}{|c}{\footnotesize \textbf{0.26} (0.02)} & {\footnotesize 0.24 (0.08)} & \multicolumn{1}{|c}{\footnotesize \textbf{0.57} (0.04)} & {\footnotesize 0.34 (0.06)} \\ 

{\footnotesize \texttt{\texttt{rejectionABC}} s.d.} & \multicolumn{1}{|c}{\footnotesize 0.49 (0.05)} & {\footnotesize \textbf{0.46} (0.03)} & \multicolumn{1}{|c}{\footnotesize 0.54 (0.03)} & {\footnotesize \textbf{0.48} (0.03)} & \multicolumn{1}{|c}{\footnotesize 0.73 (0.06)} & {\footnotesize \textbf{0.51} (0.03)} \\\hline

{\footnotesize Exact 95\%(50\%)c} & \multicolumn{1}{|c}{\footnotesize 1.00 (0.60)} & {\footnotesize 1.00 (1.00)} & \multicolumn{1}{|c}{\footnotesize 0.90 (0.50)} & {\footnotesize 1.00 (0.90)} & \multicolumn{1}{|c}{\footnotesize 1.00 (0.60)} & {\footnotesize 1.00 (0.80)} \\

{\footnotesize \texttt{copulaABCdrf}95\%(50\%)c} & \multicolumn{1}{|c}{\footnotesize \textbf{1.00} (\textbf{0.70})} & {\footnotesize \textbf{1.00} (\textbf{0.20})} & \multicolumn{1}{|c}{\footnotesize \textbf{1.00} (\textbf{0.70})} & {\footnotesize \textbf{1.00} (\textbf{0.80})} & \multicolumn{1}{|c}{\footnotesize \textbf{1.00} (\textbf{1.00})} & {\footnotesize \textbf{1.00} (\textbf{0.90})} \\ 

{\footnotesize \texttt{\texttt{rejectionABC}} 95\%(50\%)c} & \multicolumn{1}{|c}{\footnotesize \textbf{1.00} (1.00)} & {\footnotesize \textbf{1.00} (1.00)} & \multicolumn{1}{|c}{\footnotesize \textbf{1.00} (1.00)} & {\footnotesize \textbf{1.00} (1.00)} & \multicolumn{1}{|c}{\footnotesize \textbf{1.00} (1.00)} & {\footnotesize \textbf{1.00} (1.00)} \\\hline

{\footnotesize \texttt{copulaABCdrf} KS distance} & \multicolumn{1}{|c}{\footnotesize \textbf{0.09} (0.02)} & {\footnotesize \textbf{0.20} (0.01)} & \multicolumn{1}{|c}{\footnotesize \textbf{0.14} (0.03)} & {\footnotesize \textbf{0.18} (0.01)} & \multicolumn{1}{|c}{\footnotesize \textbf{0.31} (0.03)} & {\footnotesize \textbf{0.15} (0.02)} \\ {\footnotesize \texttt{\texttt{rejectionABC}} KS distance} & \multicolumn{1}{|c}{\footnotesize 0.44 (0.03)} & {\footnotesize 0.22 (0.02)} & \multicolumn{1}{|c}{\footnotesize 0.44 (0.03)} & {\footnotesize 0.23 (0.02)} & \multicolumn{1}{|c}{\footnotesize 0.44 (0.02)} & {\footnotesize 0.22 (0.01)} \\\hline

{\footnotesize \texttt{copulaABCdrf} KS test statistic} & \multicolumn{1}{|c}{\footnotesize \textbf{9.40} (2.20)} & {\footnotesize \textbf{20.97} (1.78)} & \multicolumn{1}{|c}{\footnotesize \textbf{15.31} (3.62)} & {\footnotesize \textbf{19.10} (1.51)} & \multicolumn{1}{|c}{\footnotesize \textbf{31.83} (3.10)} & {\footnotesize \textbf{16.46} (1.84)}\\ 

{\footnotesize \texttt{\texttt{rejectionABC}}\ KS test statistic} & \multicolumn{1}{|c}{\footnotesize 132.62 (8.73)} & {\footnotesize 65.69 (4.97)} & \multicolumn{1}{|c}{\footnotesize 130.74 (9.41)} & {\footnotesize 68.65 (6.04)} & \multicolumn{1}{|c}{\footnotesize 133.49 (6.56)} & {\footnotesize 66.09 (3.93)} \\ \hline

{\footnotesize \texttt{copulaABCdrf}} & \multicolumn{1}{|c}{\footnotesize d.f.:} & {\footnotesize scale:} & \multicolumn{1}{|c}{\footnotesize d.f.:} & {\footnotesize scale:} & \multicolumn{1}{|c}{\footnotesize d.f.:} & {\footnotesize scale:} \\

{\footnotesize copula d.f. and scale} & \multicolumn{1}{|c}{\footnotesize 7.89 (7.27)} & {\footnotesize 0.19 (0.42)} & \multicolumn{1}{|c}{\footnotesize 10.74 (8.60)} & {\footnotesize 0.02 (0.40)} & \multicolumn{1}{|c}{\footnotesize 9.29 (7.98)} & {\footnotesize 0.10 (0.29)} \\ \hline\hline
\multicolumn{7}{l}{\footnotesize \textit{Note:} \textbf{Bold} indicates the more accurate ABC method for the given estimator.}\\
\end{tabular}
\end{center}
\end{changemargin}
\bigskip

\begin{center}
\begin{tabular}{cc|cc|cc}
\multicolumn{6}{l}{\footnotesize \textbf{Table 2.} MAE and MSE for the Poisson-normal mixture model over 10 replicas ($n$=100),} \\
\multicolumn{6}{l}{\footnotesize under varying conditions of $n_{\text{sim}}$.} \\ 
\hline\hline

&  & \multicolumn{2}{|c|}{\footnotesize MAE:} & 
\multicolumn{2}{|c}{\footnotesize MSE:} \\ 

&  & \multicolumn{2}{|c|}{\footnotesize \texttt{copulaABCdrf}, \texttt{\texttt{rejectionABC}} (exact)} & 
\multicolumn{2}{|c}{\footnotesize \texttt{copulaABCdrf}, \texttt{rejectionABC} (exact)} \\ \hline
{\footnotesize $n$}$_{\text{sim}}$ & {\footnotesize Posterior} & \footnotesize${\lambda}$ & \footnotesize${\mu}$ & \footnotesize${\lambda}$ & \footnotesize${\mu}$ \\ \hline
{\footnotesize 10} & \multicolumn{1}{l|}{\footnotesize Mean} & 
{\footnotesize \textbf{0.13}, \textbf{0.13} (0.09)} & {\footnotesize 0.10, \textbf{0.08} (0.10)} & 
{\footnotesize \textbf{0.02}, \textbf{0.02} (0.01)} & {\footnotesize \textbf{0.01}, \textbf{0.01} (0.01)} \\ 
& \multicolumn{1}{l|}{\footnotesize Median} & {\footnotesize \textbf{0.12}, 0.13
(0.09)} & {\footnotesize 0.11, \textbf{0.06} (0.10)} & {\footnotesize \textbf{0.02}, \textbf{0.02}
(0.01)} & {\footnotesize 0.02, \textbf{0.01} (0.01)} \\ 
& \multicolumn{1}{l|}{\footnotesize Mode} & {\footnotesize 0.47, \textbf{0.21} (0.09)}
& {\footnotesize 0.21, \textbf{0.17} (0.10)} & {\footnotesize 0.38, \textbf{0.06} (0.01)} & {\footnotesize 0.06, \textbf{0.04} (0.01)} \\ 
& \multicolumn{1}{l|}{\footnotesize MLE} & {\footnotesize 0.47, \textbf{0.27} (0.09)} & {\footnotesize 0.21, \textbf{0.12} (0.10)} & {\footnotesize 0.38, 0.10 (0.01)} & 
{\footnotesize 0.06, \textbf{0.02} (0.01)} \\ \hline
{\footnotesize 25} & \multicolumn{1}{l|}{\footnotesize Mean} & 
{\footnotesize \textbf{0.17}, \textbf{0.17} (0.15)} & {\footnotesize 0.07, \textbf{0.04} (0.07)} & 
{\footnotesize \textbf{0.04}, \textbf{0.04} (0.03)} & {\footnotesize 0.01, \textbf{0.00} (0.01)} \\ 
& \multicolumn{1}{l|}{\footnotesize Median} & {\footnotesize \textbf{0.16}, 0.17
(0.15)} & {\footnotesize 0.07, \textbf{0.05} (0.07)} & {\footnotesize \textbf{0.04}, \textbf{0.04}
(0.03)} & {\footnotesize 0.01, \textbf{0.00} (0.01)} \\ 
& \multicolumn{1}{l|}{\footnotesize Mode} & {\footnotesize 0.40, \textbf{0.27} (0.15)}
& {\footnotesize \textbf{0.19}, 0.31 (0.07)} & {\footnotesize 0.19, \textbf{0.11} (0.03)} & 
{\footnotesize \textbf{0.05}, 0.13 (0.01)} \\ 
& \multicolumn{1}{l|}{\footnotesize MLE} & {\footnotesize 0.40, \textbf{0.32} (0.15)}
& {\footnotesize \textbf{0.19}, 0.27 (0.07)} & {\footnotesize 0.19, \textbf{0.12} (0.03)} & 
{\footnotesize \textbf{0.05}, 0.11 (0.01)} \\ \hline
{\footnotesize 33} & \multicolumn{1}{l|}{\footnotesize Mean} & 
{\footnotesize \textbf{0.09}, 0.13 (0.09)} & {\footnotesize \textbf{0.06}, 0.09 (0.07)} & 
{\footnotesize \textbf{0.02}, \textbf{0.02} (0.02)} & {\footnotesize \textbf{0.00}, 0.01 (0.01)} \\ 
& \multicolumn{1}{l|}{\footnotesize Median} & {\footnotesize \textbf{0.10}, 0.16
(0.09)} & {\footnotesize \textbf{0.06}, 0.09 (0.07)} & {\footnotesize \textbf{0.02}, 0.03
(0.02)} & {\footnotesize \textbf{0.00}, 0.01 (0.01)} \\ 
& \multicolumn{1}{l|}{\footnotesize Mode} & {\footnotesize \textbf{0.21}, 0.34 (0.09)}
& {\footnotesize \textbf{0.16}, 0.34 (0.07)} & {\footnotesize \textbf{0.07}, 0.15 (0.02)} & 
{\footnotesize \textbf{0.08}, 0.13 (0.01)} \\ 
& \multicolumn{1}{l|}{\footnotesize MLE} & {\footnotesize \textbf{0.21}, 0.25 (0.09)}
& {\footnotesize \textbf{0.16}, 0.33 (0.07)} & {\footnotesize \textbf{0.07}, 0.10 (0.02)} & 
{\footnotesize \textbf{0.08}, 0.12 (0.01)} \\ \hline
{\footnotesize 50} & \multicolumn{1}{l|}{\footnotesize Mean} & 
{\footnotesize \textbf{0.14}, \textbf{0.14} (0.15)} & {\footnotesize \textbf{0.07}, 0.08 (0.06)} & 
{\footnotesize 0.04, \textbf{0.03} (0.03)} & {\footnotesize \textbf{0.01}, \textbf{0.01} (0.01)} \\ 
& \multicolumn{1}{l|}{\footnotesize Median} & {\footnotesize 0.14, \textbf{0.13}
(0.15)} & {\footnotesize \textbf{0.07}, 0.09 (0.06)} & {\footnotesize \textbf{0.03}, \textbf{0.03}
(0.03)} & {\footnotesize \textbf{0.01}, \textbf{0.01} (0.01)} \\ 
& \multicolumn{1}{l|}{\footnotesize Mode} & {\footnotesize \textbf{0.17}, 0.42 (0.15)}
& {\footnotesize \textbf{0.20}, 0.26 (0.06)} & {\footnotesize \textbf{0.07}, 0.23 (0.03)} & 
{\footnotesize \textbf{0.09}, 0.10 (0.01)} \\ 
& \multicolumn{1}{l|}{\footnotesize MLE} & {\footnotesize \textbf{0.17}, 0.37 (0.15)}
& {\footnotesize \textbf{0.20}, 0.28 (0.06)} & {\footnotesize \textbf{0.07}, 0.19 (0.03)} & 
{\footnotesize \textbf{0.09}, 0.11 (0.01)} \\ \hline
{\footnotesize 66} & \multicolumn{1}{l|}{\footnotesize Mean} & 
{\footnotesize \textbf{0.12}, \textbf{0.12} (0.12)} & {\footnotesize 0.07, \textbf{0.06} (0.06)} & 
{\footnotesize \textbf{0.02}, \textbf{0.02} (0.02)} & {\footnotesize \textbf{0.01}, \textbf{0.01} (0.00)} \\ 
& \multicolumn{1}{l|}{\footnotesize Median} & {\footnotesize \textbf{0.12}, \textbf{0.12} (0.12)} & {\footnotesize \textbf{0.06}, 0.08 (0.06)} & {\footnotesize \textbf{0.02}, \textbf{0.02}
(0.02)} & {\footnotesize \textbf{0.01}, \textbf{0.01} (0.00)} \\ 
& \multicolumn{1}{l|}{\footnotesize Mode} & {\footnotesize 0.30, \textbf{0.25} (0.12)}
& {\footnotesize \textbf{0.21}, 0.28 (0.06)} & {\footnotesize 0.17, \textbf{0.09} (0.02)} & 
{\footnotesize \textbf{0.10}, \textbf{0.10} (0.00)} \\ 
& \multicolumn{1}{l|}{\footnotesize MLE} & {\footnotesize 0.24, \textbf{0.22} (0.12)}
& {\footnotesize \textbf{0.14}, 0.32 (0.06)} & {\footnotesize 0.11, \textbf{0.07} (0.02)} & 
{\footnotesize \textbf{0.04}, 0.13 (0.00)} \\ \hline
{\footnotesize 75} & \multicolumn{1}{l|}{\footnotesize Mean} & 
{\footnotesize \textbf{0.14}, 0.17 (0.15)} & {\footnotesize \textbf{0.05}, \textbf{0.05} (0.05)} & 
{\footnotesize \textbf{0.03}, 0.04 (0.03)} & {\footnotesize \textbf{0.00}, \textbf{0.00} (0.00)} \\ 
& \multicolumn{1}{l|}{\footnotesize Median} & {\footnotesize \textbf{0.14}, 0.18
(0.15)} & {\footnotesize \textbf{0.05}, \textbf{0.05} (0.05)} & {\footnotesize \textbf{0.03}, 0.05
(0.03)} & {\footnotesize \textbf{0.00}, \textbf{0.00} (0.00)} \\ 
& \multicolumn{1}{l|}{\footnotesize Mode} & {\footnotesize \textbf{0.27}, 0.29 (0.16)}
& {\footnotesize \textbf{0.15}, 0.29 (0.05)} & {\footnotesize \textbf{0.09}, 0.12 (0.03)} & 
{\footnotesize \textbf{0.03}, 0.12 (0.00)} \\ 
& \multicolumn{1}{l|}{\footnotesize MLE} & {\footnotesize \textbf{0.28}, \textbf{0.28} (0.15)}
& {\footnotesize \textbf{0.14}, 0.28 (0.05)} & {\footnotesize \textbf{0.09}, 0.11 (0.03)} & 
{\footnotesize \textbf{0.03}, 0.12 (0.00)} \\ \hline
{\footnotesize 90} & \multicolumn{1}{l|}{\footnotesize Mean} & 
{\footnotesize \textbf{0.10}, \textbf{0.10} (0.10)} & {\footnotesize 0.06, \textbf{0.05} (0.05)} & 
{\footnotesize \textbf{0.01}, \textbf{0.01} (0.01)} & {\footnotesize \textbf{0.00}, \textbf{0.00} (0.00)} \\ 
& \multicolumn{1}{l|}{\footnotesize Median} & {\footnotesize \textbf{0.09}, 0.12
(0.10)} & {\footnotesize \textbf{0.06}, 0.08 (0.05)} & {\footnotesize \textbf{0.01}, 0.02
(0.01)} & {\footnotesize \textbf{0.00}, 0.01 (0.00)} \\ 
& \multicolumn{1}{l|}{\footnotesize Mode} & {\footnotesize \textbf{0.21}, 0.24 (0.10)} & {\footnotesize \textbf{0.12}, 0.19 (0.05)} & {\footnotesize \textbf{0.06}, 0.08 (0.01)} & {\footnotesize \textbf{0.02}, 0.05 (0.00)} \\ 
& \multicolumn{1}{l|}{\footnotesize MLE} & {\footnotesize \textbf{0.21}, 0.26 (0.10)}
& {\footnotesize \textbf{0.12}, 0.26 (0.05)} & {\footnotesize \textbf{0.06}, 0.08 (0.01)} & 
{\footnotesize \textbf{0.02}, 0.10 (0.00)} \\ \hline
{\footnotesize 100} & \multicolumn{1}{l|}{\footnotesize Mean} & 
{\footnotesize \textbf{0.10}, 0.14 (0.11)} & {\footnotesize 0.09, \textbf{0.07} (0.08)} & 
{\footnotesize \textbf{0.02}, 0.03 (0.02)} & {\footnotesize \textbf{0.01}, \textbf{0.01} (0.01)} \\ 
& \multicolumn{1}{l|}{\footnotesize Median} & {\footnotesize \textbf{0.11}, 0.15 (0.11)} & {\footnotesize \textbf{0.08}, \textbf{0.08} (0.08)} & {\footnotesize \textbf{0.02}, 0.04
(0.02)} & {\footnotesize \textbf{0.01}, \textbf{0.01} (0.01)} \\ 
& \multicolumn{1}{l|}{\footnotesize Mode} & {\footnotesize \textbf{0.20}, 0.30 (0.11)}
& {\footnotesize 0.30, \textbf{0.26} (0.08)} & {\footnotesize \textbf{0.06}, 0.14 (0.02)} & 
{\footnotesize 0.19, \textbf{0.11} (0.01)}\\ 
& \multicolumn{1}{l|}{\footnotesize MLE} & {\footnotesize \textbf{0.20}, 0.23 (0.11)}
& {\footnotesize 0.30, \textbf{0.26} (0.08)} & {\footnotesize \textbf{0.06}, 0.08 (0.02)} & {\footnotesize 0.19, \textbf{0.10} (0.01)} \\ 
\hline\hline
\multicolumn{6}{l}{\footnotesize \textit{Note:} \textbf{Bold} indicates the more accurate ABC method for the given estimator.}
\end{tabular}
\end{center}

\begin{center}
\begin{tabular}{cc|cc|cc}
\multicolumn{6}{l}{\footnotesize \textbf{Table 3.} Mean (s.d.) KS distance [test statistic] and mean interval coverage for the} \\ 
\multicolumn{6}{l}{\footnotesize Poisson-normal mixture model over 10 replicas ($n=100$), under varying ${n}_{\text{sim}}$.}\\\hline\hline
&  & \multicolumn{2}{|c|}{\footnotesize Mean KS distance [test] (s.d.)} & 
\multicolumn{2}{|c}{\footnotesize Mean 95\%(50\%) coverage} \\ \hline\footnotesize
${ n}_{\text{sim}}$ & {\footnotesize Method} & \footnotesize${\lambda }$ & \footnotesize${\mu}$ & \footnotesize${\lambda }$ & \footnotesize${\mu }$ \\ \hline
{\footnotesize 10} & \multicolumn{1}{l|}{\footnotesize \texttt{copulaABCdrf}} & 
{\footnotesize \textbf{0.31} (0.03)} & {\footnotesize \textbf{0.15} (0.02)} & {\footnotesize \textbf{1.00} (\textbf{1.00})} & {\footnotesize \textbf{1.00} (\textbf{1.00})} \\ 
& \multicolumn{1}{l|}{\footnotesize \texttt{rejectionABC}} & {\footnotesize 0.37 (0.05)} & 
\multicolumn{1}{c|}{\footnotesize 0.20 (0.04) } & {\footnotesize \textbf{1.00} (\textbf{1.00})}
& {\footnotesize \textbf{1.00} (\textbf{1.00})} \\ 
& \multicolumn{1}{l|}{\footnotesize \texttt{copulaABCdrf}} & {\footnotesize [\textbf{31.83} (3.10)]}
& {\footnotesize [\textbf{16.46} (1.84)]} & {\footnotesize exact:} & {\footnotesize %
exact:} \\ 
& \multicolumn{1}{l|}{\footnotesize \texttt{rejectionABC}} & {\footnotesize [36.79 (4.64)]} & 
\multicolumn{1}{c|}{\footnotesize [19.56 (4.04)]} & {\footnotesize 1.00
(0.60) } & {\footnotesize 1.00 (0.80)} \\ \hline
{\footnotesize 25} & \multicolumn{1}{l|}{\footnotesize \texttt{copulaABCdrf}} & 
{\footnotesize \textbf{0.22} (0.03)} & {\footnotesize \textbf{0.17} (0.01)} & {\footnotesize \textbf{1.00} (\textbf{0.80})} & {\footnotesize \textbf{1.00} (\textbf{0.70})} \\ 
& \multicolumn{1}{l|}{\footnotesize \texttt{rejectionABC}} & {\footnotesize 0.37 (0.04)} & 
{\footnotesize 0.20 (0.03)} & {\footnotesize \textbf{1.00} (1.00)} & {\footnotesize %
\textbf{1.00} (1.00)} \\ 
& \multicolumn{1}{l|}{\footnotesize \texttt{copulaABCdrf}} & {\footnotesize [\textbf{23.03} (2.45)]}
& {\footnotesize [\textbf{17.85} (1.74)]} & {\footnotesize exact:} & {\footnotesize %
exact:} \\ 
& \multicolumn{1}{l|}{\footnotesize \texttt{rejectionABC}} & {\footnotesize [36.89 (3.73)]} & 
\multicolumn{1}{c|}{\footnotesize [ 20.24 (2.97)]} & {\footnotesize 0.90
(0.40) } & {\footnotesize 1.00 (1.00) } \\ \hline
{\footnotesize 33} & \multicolumn{1}{l|}{\footnotesize \texttt{copulaABCdrf}} & 
{\footnotesize \textbf{0.20} (0.04)} & {\footnotesize \textbf{0.17} (0.01)} & {\footnotesize %
\textbf{1.00} (\textbf{0.90})} & {\footnotesize \textbf{1.00} (\textbf{0.90})} \\ 
& \multicolumn{1}{l|}{\footnotesize \texttt{rejectionABC}} & {\footnotesize 0.34 (0.05)} & 
\multicolumn{1}{c|}{\footnotesize 0.20 (0.03)} & {\footnotesize \textbf{1.00} (\textbf{0.90})}
& {\footnotesize \textbf{1.00} (1.00)} \\ 
& \multicolumn{1}{l|}{\footnotesize \texttt{copulaABCdrf}} & {\footnotesize [\textbf{21.06} (4.95)]}
& {\footnotesize [\textbf{18.70} (1.19)]} & {\footnotesize exact:} & {\footnotesize exact:} \\ 
& \multicolumn{1}{l|}{\footnotesize \texttt{rejectionABC}} & {\footnotesize [34.38 (4.83)]} & 
\multicolumn{1}{c|}{\footnotesize [19.61 (3.34)]} & {\footnotesize 1.00
(0.70) } & {\footnotesize 1.00 (1.00)} \\ \hline
{\footnotesize 50} & \multicolumn{1}{l|}{\footnotesize \texttt{copulaABCdrf}} & 
{\footnotesize \textbf{0.14} (0.03)} & {\footnotesize \textbf{0.18} (0.01)} & {\footnotesize \textbf{1.00} (\textbf{0.70})} & {\footnotesize \textbf{1.00} (\textbf{0.80})} \\ 
& \multicolumn{1}{l|}{\footnotesize \texttt{rejectionABC}} & {\footnotesize 0.33 (0.04)} & 
\multicolumn{1}{c|}{\footnotesize 0.20 (0.04)} & {\footnotesize \textbf{1.00} (1.00)}
& {\footnotesize \textbf{1.00} (1.00)} \\ 
& \multicolumn{1}{l|}{\footnotesize \texttt{copulaABCdrf}} & {\footnotesize [\textbf{15.31} (3.62)]}
& {\footnotesize [\textbf{19.10} (1.51)]} & {\footnotesize exact:} & {\footnotesize exact:} \\ 
& \multicolumn{1}{l|}{\footnotesize \texttt{rejectionABC}} & {\footnotesize [32.83 (3.5)] } & 
\multicolumn{1}{c|}{\footnotesize [20.35 (3.88)]} & {\footnotesize 0.90 (0.50)} & {\footnotesize 1.00 (0.90)} \\ \hline
{\footnotesize 66} & \multicolumn{1}{l|}{\footnotesize \texttt{copulaABCdrf}} & 
{\footnotesize \textbf{0.12} (0.03)} & {\footnotesize \textbf{0.19} (0.01)} & {\footnotesize \textbf{1.00} (\textbf{0.80})} & {\footnotesize \textbf{1.00} (\textbf{0.70})} \\ 
& \multicolumn{1}{l|}{\footnotesize \texttt{rejectionABC}} & {\footnotesize 0.34 (0.05)} & 
\multicolumn{1}{c|}{\footnotesize 0.20 (0.03)} & {\footnotesize \textbf{1.00} (1.00)}
& {\footnotesize \textbf{1.00} (1.00)} \\ 
& \multicolumn{1}{l|}{\footnotesize \texttt{copulaABCdrf}} & {\footnotesize [\textbf{13.02} (2.81)]}
& {\footnotesize [\textbf{19.98} (1.32)]} & {\footnotesize exact:} & {\footnotesize %
exact:} \\ 
& \multicolumn{1}{l|}{\footnotesize \texttt{rejectionABC}} & {\footnotesize [33.84 (5.43)]} & 
\multicolumn{1}{c|}{\footnotesize [20.35 (2.94)]} & {\footnotesize 1.00
(0.50) } & {\footnotesize 1.00 (1.00)} \\ \hline
{\footnotesize 75} & \multicolumn{1}{l|}{\footnotesize \texttt{copulaABCdrf}} & 
{\footnotesize \textbf{0.11} (0.03)} & {\footnotesize 0.19 (0.01)} & {\footnotesize %
\textbf{1.00} (\textbf{0.60})} & {\footnotesize \textbf{1.00} (\textbf{0.60})} \\ 
& \multicolumn{1}{l|}{\footnotesize \texttt{rejectionABC}} & {\footnotesize 0.35 (0.04)} & 
{\footnotesize \textbf{0.18} (0.03)} & {\footnotesize \textbf{1.00} (0.90)} & {\footnotesize \textbf{1.00} (1.00)} \\ 
& \multicolumn{1}{l|}{\footnotesize \texttt{copulaABCdrf}} & {\footnotesize [\textbf{11.61} (3.71)]}
& {\footnotesize [20.17 (1.47)]} & {\footnotesize exact:} & {\footnotesize %
exact:} \\ 
& \multicolumn{1}{l|}{\footnotesize \texttt{rejectionABC}} & {\footnotesize [34.95 (4.29)] }
& {\footnotesize [\textbf{17.79} (3.02)]} & {\footnotesize 1.00 (0.40)} & 
{\footnotesize 1.00 (1.00)} \\ \hline
{\footnotesize 90} & \multicolumn{1}{l|}{\footnotesize \texttt{copulaABCdrf}} & 
{\footnotesize \textbf{0.10} (0.03)} & {\footnotesize 0.21 (0.01)} & {\footnotesize 
\textbf{1.00} (\textbf{0.80})} & {\footnotesize \textbf{1.00} (\textbf{0.50})} \\ 
& \multicolumn{1}{l|}{\footnotesize \texttt{rejectionABC}} & {\footnotesize 0.34 (0.04)} & 
{\footnotesize \textbf{0.20} (0.05)} & {\footnotesize \textbf{1.00} (1.00)} & {\footnotesize \textbf{1.00} (1.00)} \\ 
& \multicolumn{1}{l|}{\footnotesize \texttt{copulaABCdrf}} & {\footnotesize [\textbf{10.64} (2.92)] }
& {\footnotesize [20.93 (1.56)]} & {\footnotesize exact:} & {\footnotesize %
exact:} \\ 
& \multicolumn{1}{l|}{\footnotesize \texttt{rejectionABC}} & {\footnotesize [34.12 (4.32)] }
& {\footnotesize [\textbf{20.39} (5.06)]} & {\footnotesize 1.00 (0.60) } & 
{\footnotesize 1.00 (1.00) } \\ \hline
{\footnotesize 100} & \multicolumn{1}{l|}{\footnotesize \texttt{copulaABCdrf}} & 
{\footnotesize \textbf{0.09} (0.02)} & {\footnotesize 0.20 (0.01)} & {\footnotesize %
\textbf{1.00} (\textbf{0.70})} & {\footnotesize \textbf{1.00} (\textbf{0.20})} \\ 
& \multicolumn{1}{l|}{\footnotesize \texttt{rejectionABC}} & {\footnotesize 0.36 (0.05)} & 
{\footnotesize \textbf{0.18} (0.04)} & {\footnotesize \textbf{1.00 }(1.00)} & {\footnotesize \textbf{1.00} (1.00)} \\ 
& \multicolumn{1}{l|}{\footnotesize \texttt{copulaABCdrf}} & {\footnotesize [\textbf{9.40} (2.20)]} & 
{\footnotesize [20.96 (1.78)]} & {\footnotesize exact:} & {\footnotesize %
exact:} \\ 
& \multicolumn{1}{l|}{\footnotesize \texttt{rejectionABC}} & {\footnotesize [35.63 (5.41)]} & 
{\footnotesize [\textbf{18.35} (3.88)]} & {\footnotesize 1.00 (0.60) } & 
{\footnotesize 1.00 (1.00)} \\ \hline\hline
\multicolumn{6}{l}{\footnotesize \textit{Note:} \textbf{Bold} indicates the more accurate ABC method for the given estimator.}
\end{tabular}
\end{center}
\medskip

Next, for the simulation study, we consider a simple 5-dimensional Gaussian model for $n = 4$ bivariate observations. While this model is simple, it nevertheless gives rise to a calculable multimodal posterior distribution which is challenging to estimate using traditional (e.g., rejection) ABC methods \citep{PapamakariosEtAl19, WangRockova23}. In particular, for the given set of 4 data observations $\{(x_{i1},x_{i2})\}_{i=1}^{n=4}$, the model is defined by likelihood $\prod\nolimits_{i=1}^{n=4}\mathcal{N}_2(x_{i}\mid \mu _{\theta },\Sigma _{\theta })$, with pdf $\mathcal{N}_2(x\mid\mu ,\Sigma )=(2\pi )^{-1}$det$(\Sigma)^{-1/2}\exp [-\frac{1}{2}(x-\mu )^{\intercal }\Sigma^{-1}(x-\mu )]$ based on means $\mu =(\mu _{1},\mu _{2})^{\intercal }$ and $2 \times 2$ covariance matrix $\Sigma$, and with mean parameters $\mu_{\theta }=(\theta _{1},\theta _{2})$ and covariance matrix parameters $\Sigma_{\theta }=%
\begin{pmatrix}
s_{1}^{2} & \rho s_{1}s_{2} \\ 
\rho s_{1}s_{2} & s_{2}^{2} %
\end{pmatrix}%
$, with $s_{1}=\theta _{3}^{2}$, $s_{2}=\theta _{4}^{2}$, and $\rho =$ tanh$%
(\theta _{5})$, assigned uniform prior distributions $\theta _{1}\sim 
\mathcal{U}_{(-3,3)}$, $\theta _{2}\sim \mathcal{U}_{(-4,4)}$, $\theta
_{3}\sim \mathcal{U}_{(-3,3)}$, $\theta _{4}\sim \mathcal{U}_{(-3,3)}$, and $%
\theta _{5}\sim \mathcal{U}_{(-3,3)}$ as in \citep{WangRockova23}. Approximating the posterior distribution of this model can be challenging since this posterior is complex and non-trivial, because of the signs of the parameters $\theta_3$ and $\theta_4$ are non-identifiable, and thus yield four symmetric modes (due to squaring), and because the uniform prior distributions induce cutoffs in this posterior distribution. For the simulation study for this Gaussian model, ten replicas of $n = 4$ bivariate observations $x = (x_1,x_2)^{\intercal }$  were simulated using true parameters $\theta_0 = (-0.7, -2.9, -1.0, -0.9, 0.6)^{\intercal }$ as in \citep{WangRockova23}. For analyzing data simulated from this Gaussian model using \texttt{copulaABCdrf} and \texttt{rejectionABC}, five summary statistics were used, being the MLEs of the mean and variance of the two variables, and of the covariance matrix parameter. The exact posterior distribution of this Gaussian model was estimated by 10,000 samples generated from the simple componentwise MCMC Gibbs- and slice- sampling algorithm mentioned earlier.

Table 4 presents the results of the simulation study for this Gaussian model with multimodal posterior distribution. Figure 1 shows trace plots of samples of the univariate marginals of 10,000 samples of the exact multimodal posterior distribution, generated by the MCMC Gibbs- and slice-sampler, conditionally on the first of the 10 data replications, for illustration. This trace plot shows the typical result of this MCMC sampling algorithm that was obtained for each Gaussian model studied in this subsection, and given each data replica, namely, that this MCMC sampler produced samples that quickly mixed and converged to the target posterior distribution. Table 4 shows that, on average, and according to MAE and MSE: the marginal posterior mean and median estimation of the true data-generating values of the parameters $\theta_1$, $\theta_2$, and $\theta_3$, \texttt{copulaABCdrf} produced marginal univariate means and medians that were on average near and closer to their respective (MCMC-estimated) exact values compared to 1\% nearest-neighbor rejection ABC. However, for the estimation of the marginal posterior mean and median of the true data-generating values of the parameters $\theta_4$, $\theta_5$, the marginal posterior mean and median estimates of rejection ABC were closer than those of \texttt{copulaABCdrf}, while these estimates of both methods noticeably departed from the true data-generating values. Also, the posterior mode (and MLE) estimates tended to be more accurate on average for rejection ABC based on 5-dimensional kernel density estimation, compared to \texttt{copulaABCdrf}. The marginal standard deviations tended to be closer to their (estimated) exact values for \texttt{copulaABCdrf} compared to \texttt{rejectionABC}. Both ABC methods produced mean marginal 95\% and 50\% coverage of the true data-generating parameters that were similar to those of their respective (estimated) exact values. The marginal posterior distribution estimates of \texttt{copulaABCdrf} produced smaller KS distances to the respective (estimated) exact marginal posterior distributions, compared to the distance of rejection ABC, but all values of the KS test statistics associated with these distances were significant, exceeding its 99\% critical value 1.628 for the 2-tailed (one-sample) KS test. This indicated that each of the ABC methods again produced marginal posterior distributions of the model parameter that departed with their respective exact marginal posterior distributions. Overall, \texttt{copulaABCdrf} and \texttt{rejectionABC} performed similarly with each other in terms of accuracy in estimation from the posterior distribution of this model, while their estimates were not very accurate. For  \texttt{copulaABCdrf}, this is in retrospect not a very surprising result, because it is seemingly designed to handle skewed or symmetric posterior distributions but not highly multimodal posteriors.

\begin{center}
\begin{tabular}{rccccc}
\multicolumn{6}{l}{\footnotesize\textbf{Table 4.} Simulation study results for the bivariate Gaussian model with a multimodal posterior distribution.}\\ \hline\hline
&  &  &  &  &  \\ 
\multicolumn{6}{l}{\footnotesize Mean (s.d.) of posterior estimate and 95\%(50\%) credible
interval coverage (95\%(50\%)c) over 10 replicas.} \\ \hline
& \footnotesize${ \theta }_{1}$ & \footnotesize${ \theta }_{2}$ & \footnotesize${ \theta }_{3}$ & 
\footnotesize${ \theta }_{4}$ & \footnotesize${ \theta }_{5}$ \\ \cline{2-6}
\multicolumn{1}{l}{\footnotesize truth} & {\footnotesize -0.70} & {\footnotesize -2.90} & {\footnotesize %
-1.00} & {\footnotesize -0.90} & {\footnotesize 0.60} \\\hline 

\multicolumn{1}{l}{\footnotesize Exact mean} & {\footnotesize -0.42 (0.44)} & {\footnotesize -2.55 (0.22)} & {\footnotesize -0.002 (0.02)} & {\footnotesize -0.01 (0.02)} & {\footnotesize 0.60 (0.37)} \\ 

\multicolumn{1}{l}{\footnotesize \texttt{copulaABCdrf} mean} & {\footnotesize \textbf{-0.54} (0.44)} & {\footnotesize \textbf{-2.74} (0.27)} & {\footnotesize \textbf{0.02} (0.04)} & {\footnotesize \textbf{0.03} (0.08)} & {\footnotesize 1.20 (0.29)} \\ 
\multicolumn{1}{l}{\footnotesize \texttt{rejectionABC} mean} & {\footnotesize -0.50 (0.47)} & {\footnotesize -2.73 (0.26)} & {\footnotesize 0.04 (0.10)} & {\footnotesize \textbf{0.03} (0.12)} & {\footnotesize \textbf{0.21} (0.22)} \\\hline  

\multicolumn{1}{l}{\footnotesize Exact median} & {\footnotesize -0.48 (0.48)} & {\footnotesize -2.76 (0.26)} & {\footnotesize -0.19 (0.64)} & {\footnotesize -0.29 (0.60)} & {\footnotesize 0.66 (0.42)} \\ 
\multicolumn{1}{l}{\footnotesize \texttt{copulaABCdrf} median} & {\footnotesize \textbf{-0.54} (0.47)} & {\footnotesize \textbf{-2.83} (0.29)} & {\footnotesize \textbf{0.01} (0.69)} & {\footnotesize 0.33 (0.65)} & {\footnotesize 1.24 (0.37)} \\ 
\multicolumn{1}{l}{\footnotesize \texttt{rejectionABC} median} & {\footnotesize -0.52 (0.49)} & {\footnotesize -2.79 (0.31)} & {\footnotesize 0.15 (0.35)} & {\footnotesize \textbf{0.06} (0.18)} & {\footnotesize \textbf{0.29} (0.37)}\\\hline  

\multicolumn{1}{l}{\footnotesize Exact modeMLE} & {\footnotesize -0.45 (0.50)} & {\footnotesize -2.81 (0.22)} & {\footnotesize 0.34 (0.84)} & {\footnotesize 0.05 (0.83)} & {\footnotesize 0.81 (0.68)} \\ 

\multicolumn{1}{l}{\footnotesize \texttt{copulaABCdrf} modeMLE} & {\footnotesize -0.44 (0.80)} & {\footnotesize \textbf{-3.16} (0.76)} & {\footnotesize \textbf{-0.54} (0.88)} & {\footnotesize 0.33 (1.00)} & {\footnotesize 1.25 (0.91)} \\ 

\multicolumn{1}{l}{\footnotesize \texttt{\texttt{rejectionABCkern}} modeMLE} & {\footnotesize \textbf{-0.86} (0.61)} & {\footnotesize -3.03 (0.46)} & {\footnotesize 0.29 (0.66)} & {\footnotesize \textbf{-0.12} (0.60) } & {\footnotesize \textbf{0.35} (1.29)} \\\hline

\multicolumn{1}{l}{\footnotesize Exact standard deviation (s.d.)} & {\footnotesize 0.96 (0.13)} & {\footnotesize 0.98 (0.10)} & {\footnotesize 1.45 (0.19)} & {\footnotesize 1.39 (0.14)} & {\footnotesize 1.04 (0.14)}
\\ 

\multicolumn{1}{l}{\footnotesize \texttt{copulaABCdrf} s.d.} & {\footnotesize \textbf{0.71} (0.25)} & {\footnotesize 0.60 (0.16)} & {\footnotesize \textbf{1.16} (0.30)} & {\footnotesize \textbf{1.10} (0.21)} & {\footnotesize \textbf{1.05} (0.05)}
\\ 
\multicolumn{1}{l}{\footnotesize \texttt{rejectionABC} s.d.} & {\footnotesize \textbf{0.71} (0.14)} & {\footnotesize \textbf{0.68} (0.10)} & {\footnotesize 0.96 (0.22)} & {\footnotesize 0.89 (0.14)} & {\footnotesize 1.71 (0.13)}
\\\hline

\multicolumn{1}{l}{\footnotesize Exact 95\%(50\%)c} & {\footnotesize 1.00 (0.40)} & {\footnotesize 1.00 (0.90)} & {\footnotesize 0.90 (0.10) } & {\footnotesize 1.00 (0.10)} & {\footnotesize 1.00 (1.00)} \\

\multicolumn{1}{l}{\footnotesize \texttt{copulaABCdrf} 95\%(50\%)c} & {\footnotesize \textbf{1.00} (\textbf{0.50})} & {\footnotesize \textbf{1.00} (\textbf{0.50})} & {\footnotesize \textbf{1.00} (\textbf{0.70})} & {\footnotesize \textbf{1.00} (\textbf{0.60})} & {\footnotesize \textbf{1.00} (\textbf{0.70})} \\

\multicolumn{1}{l}{\footnotesize \texttt{rejectionABC} 95\%(50\%)c} & {\footnotesize \textbf{1.00} (0.70)} & {\footnotesize \textbf{1.00} (0.80)} & {\footnotesize \textbf{1.00} (0.80) } & {\footnotesize \textbf{1.00} (0.90)} & {\footnotesize \textbf{1.00} (0.80)} \\\hline

\multicolumn{1}{l}{\footnotesize \texttt{copulaABCdrf} KS distance} & {\footnotesize 0.22 (0.27)} & {\footnotesize 0.26 (0.26)} & {\footnotesize \textbf{0.16} (0.05)} & {\footnotesize \textbf{0.23} (0.27)} & {\footnotesize \textbf{0.24} (0.06)} \\ 
\multicolumn{1}{l}{\footnotesize \texttt{rejectionABC}\ KS distance} & {\footnotesize \textbf{0.12} (0.03)} & {\footnotesize \textbf{0.13} (0.04)} & {\footnotesize 0.27 (0.04)} & {\footnotesize 0.29 (0.04)} & {\footnotesize 0.25 (0.04)} \\\hline

\multicolumn{1}{l}{\footnotesize \texttt{copulaABCdrf}\ KS test statistic} & {\footnotesize 37.31 (38.78)} & {\footnotesize 44.26 (47.38)} & {\footnotesize 44.91 (14.13)} & {\footnotesize 64.46 (72.45)} & {\footnotesize 62.17 (16.70)} \\ 
\multicolumn{1}{l}{\footnotesize \texttt{rejectionABC}\ KS test statistic} & {\footnotesize \textbf{12.42} (3.15)} & {\footnotesize \textbf{12.77} (4.29)} & {\footnotesize \textbf{27.07} (3.98)} & {\footnotesize \textbf{29.09} (4.36)} & {\footnotesize \textbf{24.68} (3.65)} \\ \hline

{\footnotesize Copula} & {\footnotesize scale matrix:} & \footnotesize${ \theta }_{2}$ & \footnotesize${\theta }_{3}$ & \footnotesize${ \theta }_{4}$ & \footnotesize${ \theta }_{5}$ \\ 
& \footnotesize${ \theta }_{1}$ & {\footnotesize 0.15 (0.12)} & {\footnotesize -0.01 (0.04)} & {\footnotesize -0.00 (0.03)} & {\footnotesize -0.09 (0.10)} \\ 
& \footnotesize${ \theta }_{2}$ &  & {\footnotesize -0.01 (0.09)} & {\footnotesize 0.00 (0.08)} & {\footnotesize -0.30 (0.13)} \\ 
& \footnotesize${ \theta }_{3}$ &  &  & {\footnotesize -0.00 (0.03)} & {\footnotesize 0.01 (0.07)}
\\ 
{\footnotesize d.f.: 8.32 (4.80)} & \footnotesize${ \theta }_{4}$ &  &  &  & {\footnotesize -0.00 (0.04)} \\ \hline
\multicolumn{6}{l}{} \\ 
\multicolumn{6}{l}{\footnotesize MAE\ (MSE) of posterior estimate over 10 replicas.}
\\ \hline
\multicolumn{1}{l}{} & \footnotesize${ \theta }_{1}$ & \footnotesize${ \theta }_{2}$ & \footnotesize$%
{ \theta }_{3}$ & \footnotesize${ \theta }_{4}$ & \footnotesize${ \theta }_{5}$ \\ 
\cline{2-6}

\multicolumn{1}{l}{\footnotesize Exact mean} & {\footnotesize 0.44 (0.25)} & {\footnotesize 0.36 (0.17)} & {\footnotesize 1.00 (1.00)} & {\footnotesize 0.89 (0.79)} & {\footnotesize 0.29 (0.12)}\\
\multicolumn{1}{l}{\footnotesize \texttt{copulaABCdrf} mean} & {\footnotesize \textbf{0.40} (\textbf{0.20})} & {\footnotesize 0.28 (\textbf{0.09})} & {\footnotesize \textbf{1.02} (\textbf{1.04})} & {\footnotesize \textbf{0.93} (0.88)} & {\footnotesize 0.60 (0.44)} \\ 
\multicolumn{1}{l}{\footnotesize \texttt{rejectionABC}\ mean} & {\footnotesize 0.44 (0.23)} & {\footnotesize \textbf{0.27} (\textbf{0.09})} & {\footnotesize 1.04 (1.09)} & {\footnotesize \textbf{0.93} (\textbf{0.87})} & {\footnotesize \textbf{0.40} (\textbf{0.20})}\\\hline

\multicolumn{1}{l}{\footnotesize Exact median} & {\footnotesize 0.45 (0.25)} & {\footnotesize 0.24 (0.08)} & {\footnotesize 0.81 (1.03)} & {\footnotesize 0.61 (0.70) } & {\footnotesize 0.33 (0.16)}\\
\multicolumn{1}{l}{\footnotesize \texttt{copulaABCdrf} median} & 
{\footnotesize \textbf{0.43} (\textbf{0.23})} & {\footnotesize \textbf{0.23} (\textbf{0.08})} & {\footnotesize \textbf{1.01} (1.45)} & {\footnotesize 1.23 (1.89)} & {\footnotesize 0.66 (0.53)}\\
\multicolumn{1}{l}{\footnotesize \texttt{rejectionABC} median} & {\footnotesize 0.47 (0.25)} & {\footnotesize 0.26 (0.10)} & {\footnotesize 1.15 (\textbf{1.43})} & {\footnotesize \textbf{0.96} (\textbf{0.95})} & {\footnotesize \textbf{0.38} (\textbf{0.22})}\\\hline

\multicolumn{1}{l}{\footnotesize Exact modeMLE} & {\footnotesize 0.45 (0.29)} & 
{\footnotesize 0.20 (0.05)} & {\footnotesize 1.34 (2.43)} & {\footnotesize 1.01 (1.53)} & {\footnotesize 0.54 (0.46)} \\

\multicolumn{1}{l}{\footnotesize \texttt{copulaABCdrf} modeMLE} & {\footnotesize 0.56 (0.65)} & 
{\footnotesize 0.64 (0.59)} & {\footnotesize \textbf{0.67} (\textbf{0.91})} & {\footnotesize 1.28 (2.40)} & {\footnotesize \textbf{0.94} (\textbf{1.17})} \\ 

\multicolumn{1}{l}{\footnotesize \texttt{rejectionABCkern} modeMLE} & {\footnotesize \textbf{0.44} (\textbf{0.36})} & {\footnotesize \textbf{0.38} (\textbf{0.21})} & {\footnotesize 1.30 (2.05)} & {\footnotesize \textbf{0.78} (\textbf{0.94})} & {\footnotesize 1.00 (1.56)} \\ 
\hline\hline
\multicolumn{6}{l}{\footnotesize \textit{Note:} \textbf{Bold} indicates the more accurate ABC method for the given estimator.}
\end{tabular}
\end{center}
\medskip

\begin{figure}[ht]
\caption{For each of the five model parameters (arranged in plots from left to right), trace plots of univariate marginal distributions, obtained from 10,000 MCMC Gibbs-and slice-samples of the exact multimodal posterior distribution, conditionally on the first data replication (for illustration). In each of the five trace plots, the x-axis refers to the sampling iteration number, and the y-axis gives the realized sample value of the given model parameter.}
\centering
\includegraphics[width=1\textwidth]{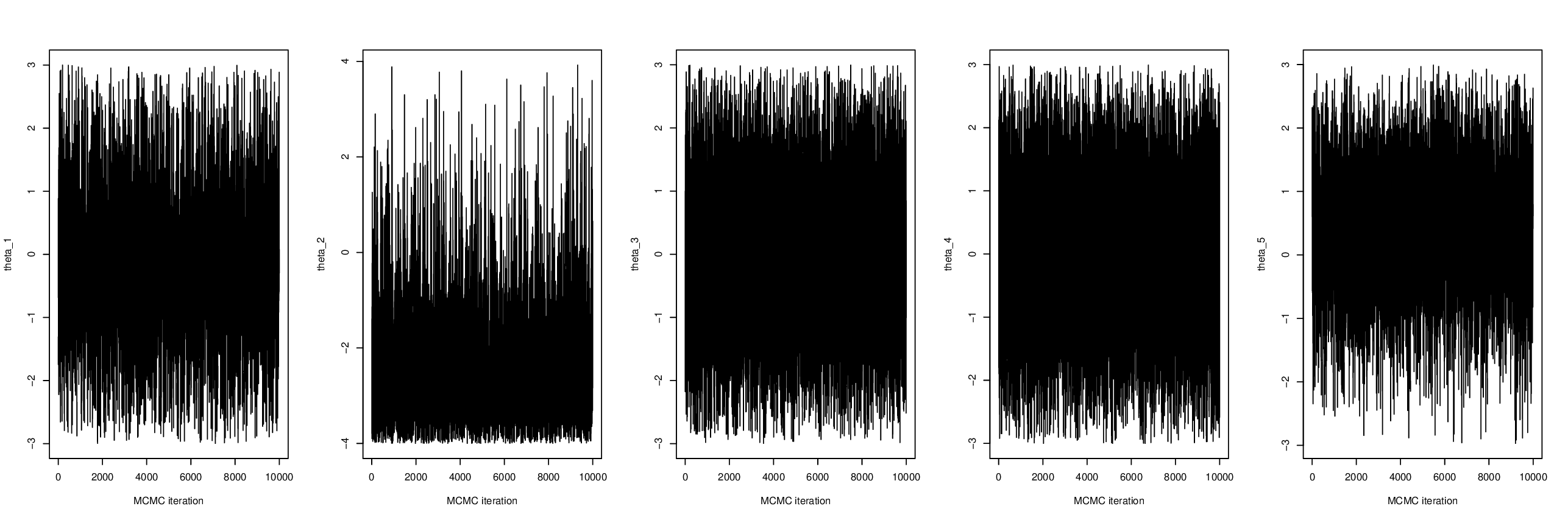}
\end{figure}

Now we consider the twisted Gaussian model \citep{HaarioEtAl99}, a high-dimensional parametric model defined by hundreds of mean location parameters, and having a calculable (and known) exact but skewed posterior distribution with a strong correlation between some of the parameters. This model has received much interest in the ABC literature since \citep{LiNottEtAl17,NottEtAl18handABC}. This is because it is challenging to estimate its posterior distribution, because the model has many (mean location) parameters, the model's likelihood function only provides location information, and information about the dependence among these parameters mainly comes from the prior distribution. In particular, here we consider one 300-dimensional observation $x=(10,0,\ldots,0)^{\intercal }$ from the twisted Gaussian model defined by likelihood $\mathcal{N}_{300}(x\mid \theta ,I)=\prod\nolimits_{k=1}^{d=300}\mathcal{N}(x_{k}\mid \theta _{k},1),$ with 300 mean (location) parameters $\theta =(\theta_{1},\theta _{2},\ldots,\theta _{300})^{\intercal }$ assigned the prior distribution (density function):
\[
\pi (\theta )\propto \exp \left\{ -\frac{\theta _{1}^{2}}{200}-\frac{(\theta_{2}-b\theta _{1}^{2}+100b)^{2}}{2}-\sum_{k=3}^{p=300}\theta
_{k}^{2}\right\} 
\]%
using $b=0.1$, as in \citep{LiNottEtAl17,NottEtAl18handABC} who considered a 250-parameter version of the model. The posterior correlation between $\theta_1$ and $\theta_2$ changes direction depending on whether the likelihood locates the posterior in the left or right tail of the prior, according to earlier graphical illustrations of the prior \citep{NottEtAl18handABC}.

For both \texttt{rejectionABC} and \texttt{copulaABCdrf}, the summary statistic is the single 300-dimensional observation, given by $x = (10,0,\ldots,0)^{\intercal }$, with corresponding pseudo-data values $(y_{1,j},y_{2,j},\ldots,y_{300,j})^{\intercal }$ (for $j=1,\ldots,N$) simulated from the model and used to construct the Reference Table for ABC, along with the prior samples of the mean locations parameters, $\theta=(\theta_1,\ldots,\theta_{300})^{\intercal }$. Preliminary runs of \texttt{copulaABCdrf} framework showed that it yielded better posterior mean estimates of model parameters when univariate \texttt{drf} regression was performed on the Reference Table in a manner such that (for $j=1,\ldots,N$) each of the $\theta_{1}^{(j)}$ and $\theta_{2}^{(j)}$ was regressed on $(y_{1}^{(j)}, y_{2}^{(j)})$, while the $\theta_{k}^{(j)}$ was regressed on the corresponding $y_{k}^{(j)}$ , for each of $k=3,\ldots,300$; instead of regressing each individual model parameter on all 300 summary statistics. Therefore, the results of the former approach will only be reported. The exact posterior distribution was calculated using 10,000 samples from the MCMC Gibbs- and slice-sampling algorithm mentioned earlier. 

Table 5 presents the results of the simulation study of the twisted Gaussian model, based on one simulated observation and true data-generating parameter, $x = \theta_0 = (10,0,\ldots,0)$ as in \citep{LiNottEtAl17,NottEtAl18handABC}. These results clearly show that \texttt{copulaABCdrf} outperformed 1\% nearest-neighbor rejection ABC in terms of estimation of the marginal posterior means median, mode, and the MLE, while producing marginal posterior distributions of the model parameters that tended to be smaller in KS distance compared to the corresponding (estimated) exact marginal posterior distributions (albeit, still with significant departures according to KS significance tests); while producing posterior means, medians, mode, and average marginal posterior interval coverages which were at least rather similar to those produced by the corresponding (estimated) exact values. In particular, for mode and MLE estimation, compared to the first mode and MLE estimation method  (described earlier in Step 4 of Algorithm 1), the accuracy of the posterior mode and MLE method was improved by basing it on an extra 100,000 samples from the meta-$t$ posterior distribution, according to the second alternative method to posterior mode and MLE estimation, as with both mode and MLE methods (described earlier in Step 4 of Algorithm 1).  

\begin{center}

\begin{tabular}{lccc}
\multicolumn{4}{l}{ \footnotesize \textbf{Table 5.} Simulation results: 300-dimensional twisted
Gaussian model.} \\ \hline\hline
& \footnotesize${ \theta }_{1}$ & \footnotesize${ \theta }_{2}$ & {\footnotesize$\theta _{3},\ldots
,\theta _{300}$}{\footnotesize \ (min, max)} \\ \hline
{\footnotesize truth} & {\footnotesize 10} & {\footnotesize 0} & {\footnotesize 0} \\ \hline

{\footnotesize Exact mean} & {\footnotesize 9.94} & {\footnotesize -0.04} & {\footnotesize (-0.02, 0.02)}
\\ 
{\footnotesize \texttt{copulaABCdrf} mean} & {\footnotesize \textbf{9.92}} & {\footnotesize \textbf{0.02}} & {\footnotesize (-0.29, \textbf{0.21})}
\\ 
{\footnotesize \texttt{rejectionABC} mean} & {\footnotesize 9.39} & {\footnotesize -0.94} & {\footnotesize (\textbf{-0.25}, 0.26)}
\\\hline

{\footnotesize Exact median} & {\footnotesize 9.95} & {\footnotesize -0.06} & {\footnotesize (-0.02, 0.02)}\\ 
{\footnotesize \texttt{copulaABCdrf} median} & {\footnotesize \textbf{9.98}} & {\footnotesize \textbf{0.03}} & {\footnotesize (-0.36, \textbf{0.31})}\\ 
{\footnotesize \texttt{rejectionABC} median} & {\footnotesize 9.61} & {\footnotesize -0.80} & {\footnotesize (\textbf{-0.27}, \textbf{0.31})}\\\hline

{\footnotesize Exact mode} & {\footnotesize 11.05} & {\footnotesize 0.89} & {\footnotesize (-1.65, 1.52)}\\

{\footnotesize \texttt{copulaABCdrf.kern} mode} & {\footnotesize \textbf{9.84}} & {\footnotesize 1.16} & {\footnotesize (-2.34, 2.96)}\\ 

{\footnotesize \texttt{copulaABCdrf.kern100k} mode} & {\footnotesize \textbf{9.80}} & {\footnotesize \textbf{-0.18}} & {\footnotesize (\textbf{-1.16}, \textbf{1.64})}\\ 

{\footnotesize \texttt{rejectionABCprodkern} mode} & {\footnotesize 10.70} & {\footnotesize 3.22} & {\footnotesize (-3.23, 2.14)}\\\hline

{\footnotesize Exact MLE} & {\footnotesize 11.05} & {\footnotesize 0.89} & {\footnotesize (-1.65, 1.52)}\\

{\footnotesize \texttt{copulaABCdrf.kern} MLE} & {\footnotesize 9.45} & {\footnotesize -1.99} & {\footnotesize (-2.82, \textbf{2.70})}\\ 

{\footnotesize \texttt{copulaABCdrf.kern100k} MLE} & {\footnotesize \textbf{10.36}} & {\footnotesize \textbf{1.07}} & {\footnotesize (\textbf{-2.57}, 3.51)}\\ 

{\footnotesize \texttt{rejectionABCprodkern} MLE} & {\footnotesize 8.60} & {\footnotesize -1.99} & {\footnotesize (-2.60, 3.18)}\\\hline

{\footnotesize Exact standard deviation (s.d.)} & {\footnotesize 0.58} & {\footnotesize 0.92} & {\footnotesize (0.69, 0.72)}\\
{\footnotesize \texttt{copulaABCdrf} s.d.} & {\footnotesize \textbf{0.63}} & {\footnotesize \textbf{1.02}} & {\footnotesize (0.56, \textbf{0.92})}\\ 
{\footnotesize \texttt{rejectionABC} s.d.} & {\footnotesize 1.58} & {\footnotesize 2.79} & {\footnotesize (\textbf{0.76}, 1.23)}\\\hline

{\footnotesize Exact 95\%} & {\footnotesize (8.79, 11.08)} & {\footnotesize (-1.78, 1.78)} & {\footnotesize 2.5\%:(-1.44, -1.32); 97.5\%:(1.33, 1.44)}\\ 
{\footnotesize \texttt{copulaABCdrf}\ 95\%} & {\footnotesize (\textbf{8.62}, \textbf{11.10})} & {\footnotesize (\textbf{-2.18}, \textbf{1.81})} & {\footnotesize 2.5\%:(\textbf{-2.44}, -0.87); 97.5\%:(0.78, \textbf{2.27})} \\ 
{\footnotesize \texttt{rejectionABC}\ 95\%} & {\footnotesize (6.02, 11.98)} & {\footnotesize (-5.20, 3.98)} & {\footnotesize 2.5\%:(-2.48, \textbf{-1.14}); 97.5\%:(\textbf{1.21}, 2.53)} \\\hline 

{\footnotesize Exact 50\%} & {\footnotesize (9.55, 10.34)} & {\footnotesize (-0.67, 0.58)} & {\footnotesize 25\%:(-0.50, -0.45); 75\%:(0.45, 0.51)} \\
{\footnotesize \texttt{copulaABCdrf}\ 50\%} & {\footnotesize (\textbf{9.49}, \textbf{10.34})} & {\footnotesize (\textbf{-0.66}, \textbf{0.72})} & {\footnotesize 25\%:(\textbf{-0.94}, -0.16); 75\%:(\textbf{0.17}, \textbf{0.82})} \\ 

{\footnotesize \texttt{rejectionABC}\ 50\%} & {\footnotesize (8.31, 10.55)} & {\footnotesize (-3.00, 1.14)} & {\footnotesize 25\%:(-1.01, \textbf{-0.29}); 75\%:(0.14, 1.00)} \\\hline

{\footnotesize \texttt{copulaABCdrf}\ KS dist} & {\footnotesize \textbf{0.07}} & {\footnotesize \textbf{0.09}} & {\footnotesize (\textbf{0.04}, \textbf{0.22})} \\ 
{\footnotesize \texttt{rejectionABC}\ KS dist} & {\footnotesize 0.32} & {\footnotesize 0.39} & {\footnotesize (0.06, 0.23)} \\\hline

{\footnotesize \texttt{copulaABCdrf}\ KS test} & {\footnotesize \textbf{7.31}} & {\footnotesize \textbf{9.71}} & {\footnotesize (\textbf{2.46}, \textbf{13.49})} \\ 
{\footnotesize \texttt{rejectionABC}\ KS test} & {\footnotesize 31.97} & {\footnotesize 39.36} & {\footnotesize 
(6.45, 22.77)} \\\hline

\multicolumn{1}{r}{\footnotesize \texttt{copulaABCdrf} copula} & {\footnotesize scale matrix} & \footnotesize${ \theta }_{2}$ & \footnotesize${\theta }_{3}{,\ldots,\theta }_{300}${\footnotesize \ (min, max)} \\ 
\multicolumn{1}{r}{\footnotesize d.f. = 117.50} & \footnotesize${ \theta }_{1}$ & {\footnotesize 0.75} & {\footnotesize (-0.12, 0.13)} \\ 

\multicolumn{1}{r}{} & \footnotesize${ \theta }_{2}$ & {\footnotesize} & {\footnotesize (-0.13, 0.13)} \\ 

\multicolumn{1}{r}{} & \footnotesize${\theta }_{3}{,\ldots,\theta }_{300}$
&  & {\footnotesize (-0.16, 0.17)} \\ \hline\hline
\multicolumn{4}{l}{\footnotesize \textit{Note:} \textbf{Bold} indicates the more accurate ABC method for the given estimator.}
\end{tabular}
\end{center}
\medskip

In summary, all the results of the simulation studies of this subsection together show that \texttt{copulaABCdrf}, as discussed in \S1, is equipped to estimate more unimodal, skewed, and high dimensional posterior distributions (as in the Poisson-normal mixture model) or skewed (as in the 300-dimensional twisted Gaussian model), especially compared to the rejection ABC method. However, both ABC methods produced estimates of marginal posterior distributions of model parameters that departed from the exact marginal posteriors according to Kolmogorov-Smirnov distances and their corresponding significant values of their test statistics, and could not easily handle the estimation of multimodal posterior distributions (as in the 5-dimensional bivariate Gaussian model). 

\subsection{An Overview of Network Data Modeling}

Network science studies systems and phenomena governing networks, which represent relational data which shape the connections of the world and are ubiquitous and focal discussion points in everyday life and the sciences. A network consists of a set of nodes (i.e., vertices, actors, individuals, or sites, etc.) which can have connections (i.e., edges, ties, lines, dyadic relationships, links, or bonds, etc.) with other nodes in the network; e.g., a social network consists of friendship ties among persons (nodes). A directed network treats each pair of nodes $(i,j)$ and $(j,i)$ as distinct, whereas an undirected network does not. Each possible edge tie connecting any given node pair in a network may be (unweighted) binary-valued, or (weighted) count-, ordinal-, categorical-, text-, ranking-, continuous valued; and/or vector valued; possibly as part of a multi-layered (or multiplex, multi-item, or multi-relational) network representing multiple relationships among the same set of nodes. Network summary statistics include network size (number of nodes, $n$), edge count and density (ratio of edge count to maximum edge count in network), node degree (a node's outdegree is the number of edges starting from it, indegree is the number of edges going into it, while in an undirected network, the outdegree and indegree coincide and equal to degree), the degree distribution over nodes and its geometric weighted form \citep{SnijdersEtAl06}, degree assortativity \citep{Newman03}, counts of 2-stars and triangles, and global \citep{WassermanFaust94} and average local \citep{WattsStrogatz98} clustering coefficients. The presence of a dyadic network tie can depend on the presence of other dyadic ties, nodal attributes, the network's evolving dynamic structure and dyadic interactions over time, or on other factors. Also, the network's structure itself may influence or predict individuals' attributes and behaviors.

Network data are collected and analyzed in various scientific fields, including the social sciences (e.g., social or friendship networks, marriage, sexual partnerships); academic paper networks (collaboration or citation networks); neuroscience (human brain networks and interactions); political science (international relations, wars between countries, insurgencies, terrorist networks, strategic alliances and friendships); education (multilevel student relation data, item response data); economics (financial markets, economic trading or resource distribution networks); epidemiology (disease spread dynamics, HIV, COVID-19); physics (Ising models, finding
parsimonious mechanisms of network formation); biology and other natural sciences (protein-protein, molecular interaction, metabolic, cellular, genetic, ecological, and food web networks); artificial intelligence and machine learning (finding missing link in a business or a terrorist network, recommender systems, Netflix, neural networks); spatial statistics (discrete Markov random fields); traffic systems and transportation networks (roads, railways, airplanes); cooperation in an organization (advice giving in an office, identity disambiguation, business organization analysis); communication patterns and networks (detecting latent terrorist cells); telecommunications (mobile phone networks); and computer science and networks (e-mail, internet, blogs, web, Facebook, Twitter(X), LinkedIn, information spread, viral marketing, gossiping, dating networks, blockchain network, virus propagation). Network science has seen many reviews due to its longtime scientific and societal impacts
\citep
[e.g.,][]{WassermanFaust94,Newman03,GoldenbergEtAl10,Snijders11,PastorSatorrasVespignani04,RavalRay13,Newman18,HorvatZweig18,BrodkaKazienko18,GhafouriKhasteh20,LoyalChen20,HammoudKramer20,KinsleyEtAl20}, while the high dimensionality of network data poses challenges to modern statistical computing methods.

Statistical and mechanistic network models are two prominent paradigms for analyzing network data. A \textit{statistical network model} specifies an explicit likelihood function for the given network dataset, available in closed-form up to a normalizing constant, making standard statistical inference tools generally available for these models, e.g., for parameter estimation and model selection. Statistical network models include the popular large family of ERGMs \citep
{FrankStrauss86,WassermanPattison96,LusherEtAl13,Harris13,SchweinbergerEtAl20,CaimoGollini22} which uses observable network configurations (e.g., $k$-stars, triangles) as sufficient statistics. An ERGM is defined by a likelihood with normalizing constant, which is intractable for a network with more than a handful of nodes $n$. The ERGM for a binary undirected or directed network is as follows:
\begin{equation}
f(X=x\mid\theta)=\dfrac{\exp\left\{  \eta(\theta)^{\intercal}h(x;z)\right\}
}{{\textstyle\sum\nolimits_{x^{\prime}\in\mathcal{X}_{n}}}\exp\left\{  \eta(\theta)^{\intercal}h(x^{\prime};z)\right\}  },
\label{ERGMbasic}
\end{equation}
with parameter vector $\theta\in\Theta\subseteq\mathbb{R}^{q}$ and its mapping to canonical parameters $\eta:\Theta\rightarrow\mathbb{R}^{d}$ (with
(\ref{ERGMbasic}) a linear ERGM if $\eta(\theta)\equiv\theta$, and is otherwise a curved ERGM); sufficient statistics $h\in
\mathbb{R}^{d}$ which are possibly dependent on any covariates $z$; and for undirected binary networks $X$, the space $\mathcal{X}_{n}$ of allowable networks on $n$ nodes is $\mathcal{X}_{n}=\{x\in%
\mathbb{R}^{n\times n},x_{i,j}=x_{j,i}\in\{0,1\},x_{i,i}=0\}$, and for directed binary networks $X$, the space $\mathcal{X}_{n}$ is the same but without the restrictions $x_{i,j}=x_{j,i}$. The ERGM likelihood (\ref{ERGMbasic}) is
intractable for a binary network with more than a handful of nodes $n$, as the normalizing constant of the model's likelihood is a sum of $2^{n(n-1)/2}$
terms (or $2^{n(n-1)}$ terms for a directed network) over the sample space of allowable networks. The basic ERGM form (\ref{ERGMbasic}) allows it to be straightforwardly extended to a general ERGM representation for multi-layered
weighted (valued) networks shown later in \S 3.8, which combines certain ERGMs \citep{Krivitsky12,KrivitskyEtAl20,CaimoGollini22}
and encompasses ERGMs: for binary networks; with size($n$) offset adjusted (invariant) parameterizations \citep{KrivitskyEtAl11,StewartEtAl19}; with nodal random effects \citep[][]{ThiemichenEtAl16}; and for valued (weighted) \citep[][]{Krivitsky12,CaimoGollini22}, multilayered \citep{KrivitskyEtAl20}, and dynamic networks \citep{HannekeEtAl10,KrivitskyHandcock14,LeeEtAl20}.

A mechanistic network model is able to incorporate domain scientific knowledge into generative mechanisms for network formation, enabling researchers to investigate complex systems using both simulation and analytical techniques. A
\textit{mechanistic network model} is an algorithmic description of network formation defined by a few domain-specific stochastic microscopic mechanistic rules by which a network grows and evolves over time, which are hypothesized and informed by understanding and related critical elements of the given scientific problem. A typical mechanistic model generates a network by starting with a small seed network, and then grows the network one node at a time according to the model's generative mechanism until some stopping condition is met, e.g., until a requisite number of nodes is reached, which can possibly number in the millions. There are easily hundreds of mechanistic network models, which originated in physics. For a long time, they were essentially the only type of network models formulated and studied using both mathematical and computer simulation methods. Mechanistic network models include the model of \citet{Price65,Price76} used to study citation patterns of scientific papers, and the \citet{BarabasiAlbert99} model, which introduce preferential attachment rules for directed and undirected networks, respectively. Also, the \citet{WattsStrogatz98} model, which produces random graphs with small-world properties, including short average path lengths and high clustering. Further, the duplication-divergence class of models, which describe the evolution of protein-protein interaction networks (where each node is a protein in an organism and two proteins are connected if they are involved in a chemical reaction together). This class includes the Duplication-Mutation-Complementation (DMC) model \citep{VazquezEtAl03,Vazquez03} and the Duplication-Mutation-Random (DMR) model \citep{SoleEtAl02,PastorSatorrasEtAl03}. Other mechanistic models include the KM model \citep{KretzschmarMorris96,MorrisKretzschmar97}, used to study HIV epidemics \citep{MorrisEtAl07,PalombiEtAl12}; along with other mechanistic models \citep{KlemmEguiluz02,KumpulaEtAl07,ProcopioEtAl23}. The Price and nonlinear preferential attachment models, and the DMC and DMR models, are further described later within the simulation studies reported in \S 3.4-\S 3.5.

We will apply Algorithm 1 to ERGM and mechanistic network models through simulation studies in \S 3.3-\S 3.5 and analyses and real-life large networks in \S 3.6-\S 3.8. Algorithm 1 can be used to analyze a network dataset $x$ using $M\geq1$ ERGM and/or mechanistic network models, based on $K$ scalar and/or vector network statistics, $g_{k}(x)$, for $k=1,\ldots,K$. As mentioned, for the analysis of a large network, it is computationally prohibitive to simulate one very large network from the ERGM or mechanistic network model. This issue can be addressed by a strategy which simulates network datasets of a smaller size compared to the size of the network dataset under statistical analysis, while using network summary statistics (calculated on the observed dataset and each simulated dataset) that take on values that are invariant to the size (number of nodes) of the network dataset being analyzed.

In terms of Algorithm 1, we can specify computationally efficient network summary statistics $s$ which can be used to directly compare between two networks that may have different node sets, size (number of nodes), and densities. In other words, summary statistics which allow for Unknown Node-Correspondence (UNC) network comparisons \citep[][p.2]{TantardiniEtAl19}, including MPLEs of size($n$) offset adjusted (invariant) parameterized ERGMs \citep{KrivitskyEtAl11,StewartEtAl19}, described below. Other computationally efficient UNC summary statistics include global and average local clustering coefficients (based on triangle counts) \citep{PrzuljEtAl04,YaverougluEtAl14,YaverougluEtAl15,FaisalEtAl17}, degree distribution including its mean and variance or geometric form, degree assortativity (propensity for similar nodes to connect), and diameter (average of shortest path lengths over all pairs of nodes in a network) \citep{YaverougluEtAl14,FaisalEtAl17}. Generally speaking, specifying the summary statistics vector $s$, by computationally efficient summary statistics which allow for UNC (size-invariant) comparisons of networks, enables Steps 1(c)-1(d) to simulate and compute these summaries of a network of a smaller size $n_{\text{sim}}$, and to directly compare these summaries with the corresponding summaries of a larger observed network dataset $x$ (of size $n$) being analyzed, for each of the $N_{M}$ sampling iterations of Algorithm 1. This lowers computational cost, compared to computing and comparing such summary based on simulating networks of the same size ($n$) as the large observed network in each sampling iteration.

As mentioned, a novel contribution is that for either the ERGM or a mechanistic network model, the UNC network summary statistics vector $s(\cdot)$ may also include MPLEs of $K$ ERGMs defined by sufficient statistics $g_{k}(x)$, respectively for $k=1,\ldots,K$, with these MPLEs adjusted by network size by specifying in each ERGM an offset term $|x|$ (edge count of network $x$) with fixed coefficient $\log\genfrac{(}{)}{}{1}{1}{n}$ \citep{KrivitskyEtAl11}. The MPLE \citep[e.g.,][]{SchmidHunter23} was introduced for lattice models \citep{Besag74}, and developed to estimate ERGM parameters \citep{StraussIkeda90,FrankStrauss86} because the exact MLE of the ERGM is intractable for a binary network with more than a handful of nodes $n$, as mentioned above. Specifically, for example, if the observed network dataset $x$ is an undirected binary network, $x=(x_{i,j}\in\{0,1\})_{n\times n}$ on $n$ nodes, described by scalar or vector valued statistics $g_{k}(x)$ for $k=1,\ldots,K$ (where each $g_{k}$ may depend on covariates $z_{k}$), then the summary statistics $s$ of $x$ (or $y$) can be specified as $s(x)=(\widehat{\beta}_{\text{MPLE},1}^{(x)},\ldots,\widehat{\beta}_{\text{MPLE},K}^{(x)})^{\intercal}$, where
$\widehat{\beta}_{\text{MPLE},k}^{(x)}$ is the MPLE of the ERGM (\ref{ERGMbasic}) using sufficient statistics $g_{k}(x)$ and offset term the network edge count $|x|$ with fixed coefficient $\log(1/n)$. Each MPLE summary statistic $\widehat{\beta}_{\text{MPLE},k}$ (for $k=1,\ldots,K$) obtained by maximizing a logistic regression likelihood, as follows:
\[
\widehat{\beta}_{\text{MPLE},k}^{(x)}=\underset{\beta_{k}\in\mathbb{R}%
^{\underline{q}_{k}}}{\arg\max}\underset{1\leq i<j\leq n}{{\textstyle\sum}{\textstyle\sum}
}\left(
\begin{array}
[c]{l}x_{i,j}\{(\eta_{k}(\beta_{k})^{\intercal},\log
\genfrac{(}{)}{}{1}{1}{n})^{\intercal}\Delta_{i,j,k}\}\\
\multicolumn{1}{c}{-\log[1+\exp\{(\eta_{k}(\beta_{k})^{\intercal},\log%
\genfrac{(}{)}{}{1}{1}{n})^{\intercal}\Delta_{i,j,k}\}]}%
\end{array}
\right),\]
\newline with $\beta_{k}\in\mathbb{R}^{\underline{q}_{k}}\text{, }\eta
_{k}:\Theta\rightarrow\mathbb{R}^{\underline{d}_{k}}\text{, }g_{k}\in\mathbb{R}^{\underline{d}_{k}}$, and network change statistics $\Delta
_{i,j,k}=((g_{k}(x_{i,j}^{+})-g_{k}(x_{i,j}^{-}))^{\intercal},|x_{i,j}^{+}|-|x_{i,j}^{-}|)^{\intercal}$, where $x_{i,j}^{+}$ and $x_{i,j}^{-}$ is network $x$ with $x_{i,j}=x_{j,i}=1$ and $x_{i,j}=x_{j,i}=0$ respectively. Likewise, MPLEs can be obtained from a directed network, or from a weighted (undirected or directed) network based on a binary representation of a polytomous network \citep[][\S4.3]{CaimoGollini22}. Such a logistic regression likelihood represents a special form of composite likelihood \citep{Lindsay88,VarinEtAl11}
which assumes (dyadic) independence of the $x_{i,j}$ observations, $\Pr_{\theta}(x_{i,j}=1\mid X_{i,j}^{c}=x_{i,j}^{c})=\Pr_{\theta}(x_{i,j}=1)$,
where $x_{i,j}^{c}$ is the network $x$ excluding $x_{i,j}$. Both the MPLE and MLE of the ERGM are consistent for a growing number of networks observed from the same set of fixed nodes \citep{ArnoldStrauss91}. The MPLE can be rapidly computed from large networks \citep[][]{SchmidDesmarais17}, using divide-and-conquer \citep{GaoEtAl22,RosenblattNadler16,Minsker19} or streaming methods \citep[][]{LuoSong19}, if necessary.

\subsection{Simulation Study: Exponential Random Graph Models (ERGMs)}

Now we consider a simulation study to evaluate and compare the ability of  \texttt{copulaABCdrf} (Algorithm 1) and \texttt{rejectionABC} to estimate the posterior distribution of ERGM parameters, based on 10 replications of $n=50$ node undirected networks from the ERGM, and of $n=300$ node undirected networks simulated from this model, for each of five conditions of $n_{\text{sim}}$,  being  at least roughly $10\%$, $33\%$, $50\%$, $75\%$ and $100\%$ of $n$ network nodes. Specifically, the conditions $n_{\text{sim}}=5, 16, 25, 37$ and $50$ for $n = 50$ nodes, and $n_{\text{sim}}= 30, 100, 150, 225$ and $300$ for $n = 300$ nodes.

Under the $n=50$ simulation conditions, each of the 10 network datasets were simulated from the ERGM (\ref{ERGMbasic}) using true data-generating parameters $\theta_0^{\intercal}=(-0.20,0.50)$ and defined by network sufficient statistics being the number of 2-stars and triangles. For each simulated network dataset $x$ analyzed by Algorithm 1 using the ERGM, this model was assigned a $g$ prior $\theta\sim\mathcal{N}_{2}(0,g(-\mathcal{H}(\widehat{\theta}_{\text{MPLE}}))^{-1})$ with $g=100$ where $\mathcal{H}(\widehat{\theta}_{\text{MPLE}})$ is the Hessian matrix of the MPLE of $\theta$, and the summary statistics were specified as $s(x)=(\widehat{\beta}_{\text{MPLE},1}^{(x)},\widehat{\beta}_{\text{MPLE},2}^{(x)})^{\intercal}$, where $\widehat{\beta}_{\text{MPLE},k}^{(x)}$ for $k=1,2$ are the MPLEs for these two network sufficient statistics respectively based on the edge count $|x|$ offset with fixed coefficient $\log(1/n)$. Likewise, for $s(y)=(\widehat{\beta}_{\text{MPLE},1}^{(y)},\widehat{\beta}_{\text{MPLE},2}^{(y)})^{\intercal}$ based on the edge count $|y|$ offset with fixed coefficient $\log(1/n_{\text{sim}})$, and on the size $n_{\text{sim}}=25$ of the network dataset $y$ simulated in each iteration of Algorithm 1.

Under the $n=300$ simulation conditions, each of 10 network datasets was simulated from the ERGM (\ref{ERGMbasic}) using true data-generating parameters $\theta_0^{\intercal}=(-0.20,0.50,0.80)$ and defined by network sufficient statistics being the number of 2-stars (\texttt{kstar(2)}), the number of triangles (\texttt{triangles}), and the geometric weighted degree distribution (\texttt{degree1.5}) with decay parameter fixed at $\alpha\equiv\log(1.5)$
\citep[][from equations 11-12, pp.112,126]{SnijdersEtAl06}. For each simulated network dataset $x$ analyzed by Algorithm 1 using the ERGM, this model was assigned a $g$ prior, given by $\theta\sim\mathcal{N}_{3}(0,g(-\mathcal{H}(\widehat{\theta}_{\text{MPLE}}))^{-1})$ with $g=100,000,$ and the summary statistics were specified as $s(x)=(\widehat{\beta}_{\text{MPLE},1}^{(x)},\widehat{\beta}_{\text{MPLE},2}^{(x)},\widehat{\beta}_{\text{MPLE},3}^{(x)})^{\intercal}$, where $\widehat{\beta}_{\text{MPLE},k}^{(x)}$ for $k=1,2,3,$ are the MPLEs for these three network sufficient statistics respectively based on the edge count $|x|$ offset with fixed coefficient $\log(1/n)$. Likewise, for $s(y)=(\widehat{\beta}_{\text{MPLE},1}^{(y)},\widehat{\beta}_{\text{MPLE},2}^{(y)},\widehat{\beta
}_{\text{MPLE},3}^{(y)})^{\intercal}$ based on the edge count $|y|$ offset with fixed coefficient $\log(1/n_{\text{sim}})$, and on the size $n_{\text{sim}}=100$ of the network dataset $y$ simulated in each iteration of Algorithm 1.

Table 6 presents some detailed representative results for $n_{\text{sim}}$ $=25$ of $n =50$ node networks, and for $n_{\text{sim}}$ $=100$ of $n =300$ node networks. Table 7 presents the results for each of all the five conditions of  $n_{\text{sim}}$ for $n =50$ nodes and for $n =300$ nodes. For $n =50$ nodes, in terms of MAE and MSE, \texttt{rejectionABC} tended to outperform \texttt{copulaABCdrf} in estimation accuracy of the posterior mean, median, and mode. Also, \texttt{copulaABCdrf} outperformed \texttt{rejectionABC} in terms of MLE accuracy. Further, \texttt{rejectionABC} performed similarly with \texttt{copulaABCdrf} in terms of accuracy in estimation of 95\% and 50\% posterior credible intervals, while being typically near the 95\% nominal values. In addition, \texttt{rejectionABC} was slightly superior in estimation of the 50\% posterior credible interval, but both methods often produced 50\% interval estimates that were far from this nominal value. Also, while $n_{\text{sim}}$ is increased towards the full network dataset sample size of $n=50 $ nodes, the MAEs and MSEs tended to decrease for each of the point estimates (posterior mean, median, mode, and MLE) for both ABC methods. For $n_{\text{sim}}=37$, \texttt{copulaABCdrf} produced MLEs that were competitive with MCMLEs. 

For $n =300$ nodes, \texttt{rejectionABC} tended to outperform \texttt{copulaABCdrf} in terms of estimation accuracy of the posterior mean, median, and mode, especially for parameters $\theta_1$ and $\theta_3$. Also,  \texttt{copulaABCdrf} outperformed \texttt{rejectionABC} in terms of MLE accuracy for all model parameters, and in terms of accuracy in estimation of 95\% posterior credible intervals, while being typically near the nominal values. Further, \texttt{rejectionABC} tended to outperform \texttt{copulaABCdrf} in terms of accuracy in estimation of 50\% posterior credible intervals. Again, both ABC methods often produced 50\% interval estimates far away from this nominal value. Finally, as $n_{\text{sim}}$ is increased towards the full network dataset sample size of $n=300$ nodes, for each of the two ABC methods, there was no clear pattern of the MAEs and MSEs decreasing for each of the point estimates, and for the posterior credible intervals to approach their nominal rates.

\begin{center}
\begin{tabular}{lcc|ccc}
\multicolumn{6}{l}{\footnotesize \textbf{Table 6.} Results of simulation study for the ERGM.  Mean (standard deviation) of posterior estimates}\\ 
\multicolumn{6}{l}{\footnotesize over 10 replicas, under some conditions of $n$ and ${n}_{\text{sim}}$.} \\ \hline\hline
& {\footnotesize kstar(2)} & {\footnotesize triangles} & {\footnotesize kstar(2)} & {\footnotesize %
triangles} & {\footnotesize degree1.5} \\ \hline
{\footnotesize \textit{n}} & {\footnotesize 50} & {\footnotesize 50} & {\footnotesize 300} & {\footnotesize 300} & 
{\footnotesize 300} \\ 
{\footnotesize \textit{n}}$_{\text{sim}}$ & {\footnotesize 25} & {\footnotesize 25} & {\footnotesize 100} & 
{\footnotesize 100} & {\footnotesize 100} \\ 
{\footnotesize true $\theta $} & {\footnotesize -0.20} & {\footnotesize 0.50} & {\footnotesize -0.20} & 
{\footnotesize 0.50} & {\footnotesize 0.80} \\\hline

{\footnotesize \texttt{copulaABCdrf} mean} & {\footnotesize -0.22 (0.03)} & {\footnotesize 0.95 (0.35)} & {\footnotesize -0.03 (0.02)} & {\footnotesize \textbf{0.49} (0.10)} & {\footnotesize -0.15 (0.12)} \\ 
{\footnotesize \texttt{rejectionABC} mean} & {\footnotesize \textbf{-0.21} (0.04)} & {\footnotesize \textbf{0.79} (0.39)} & {\footnotesize \textbf{-0.05} (0.01) } & {\footnotesize -0.2 (0.07)} & {\footnotesize \textbf{0.33} (0.06)} \\\hline

{\footnotesize \texttt{copulaABCdrf} median} & {\footnotesize -0.22 (0.03)} & {\footnotesize 0.86 (0.33)} & {\footnotesize 0.01 (0.03)} & {\footnotesize \textbf{0.28} (0.07)} & {\footnotesize -0.85 (0.28)} \\ 
{\footnotesize \texttt{rejectionABC} median} & {\footnotesize  \textbf{-0.20} (0.04)} & {\footnotesize \textbf{0.67} (0.32)} & {\footnotesize \textbf{-0.04} (0.01)} & {\footnotesize -0.30 (0.03)} & {\footnotesize \textbf{0.30} (0.07)} \\\hline

{\footnotesize \texttt{copulaABCdrf} mode} & {\footnotesize -0.18 (0.06)} & {\footnotesize \textbf{0.57} (0.51)} & {\footnotesize 0.06 (0.07)} & {\footnotesize \textbf{1.18} (0.64)} & {\footnotesize -1.14 (0.07)} \\ 

{\footnotesize \texttt{rejectionABCkern} mode} & {\footnotesize \textbf{-0.19} (0.04)} & {\footnotesize 0.67 (0.37)} & {\footnotesize \textbf{-0.03} (0.05)} & {\footnotesize -0.27 (0.11)} & {\footnotesize \textbf{0.29} (0.22)} \\\hline

{\footnotesize \texttt{copulaABCdrf} MLE} & {\footnotesize \textbf{-0.22} (0.06)} & {\footnotesize \textbf{0.79} (0.48)} & {\footnotesize 0.01 (0.07)} & {\footnotesize 1.61 (0.58)} & {\footnotesize -1.09 (0.09)}\\ 

{\footnotesize \texttt{\texttt{rejectionABCkern}} MLE} & {\footnotesize -0.33 (0.06)} & {\footnotesize 1.98 (1.69)} & {\footnotesize \textbf{-0.31} (0.21)} & {\footnotesize \textbf{0.89} (1.47)} & {\footnotesize \textbf{0.82} (0.89)} \\

{\footnotesize MCMLE} & {\footnotesize -0.20 (0.02)} & {\footnotesize 0.46 (0.16)} & {\footnotesize %
-0.03 (0.01)} & {\footnotesize 0.09 (0.01)} & {\footnotesize 0.08 (0.03)} \\ 
{\footnotesize MPLE} & {\footnotesize 0.15 (0.00)} & {\footnotesize 1.18 (0.08)} & {\footnotesize 0.03
(0.00)} & {\footnotesize 0.26 (0.00)} & {\footnotesize 0.18 (0.00)} \\\hline 
{\footnotesize \texttt{copulaABCdrf} s.d.} & {\footnotesize 0.07 (0.01)} & {\footnotesize 0.66 (0.12)} & {\footnotesize 0.27 (0.02)} & {\footnotesize 0.79 (0.10)} & {\footnotesize 1.51 (0.21)} \\ 
{\footnotesize \texttt{rejectionABC} s.d.} & {\footnotesize 0.05 (0.01)} & {\footnotesize 0.62 (0.19)} & {\footnotesize 0.10 (0.01)} & {\footnotesize 0.46 (0.14)} & {\footnotesize 0.38 (0.04)} \\
{\footnotesize SE(MCMLE)} & {\footnotesize 0.02 (0.00)} & {\footnotesize 0.19 (0.01)} & {\footnotesize 0.00 (0.00)} & {\footnotesize 0.01 (0.00)} & {\footnotesize 0.02 (0.00)} \\ \hline
{\footnotesize \texttt{copulaABCdrf}} & {\footnotesize d.f.} & {\footnotesize scale} & {\footnotesize d.f.:} & 
\multicolumn{2}{c}{\footnotesize scale matrix:} \\ 
{\footnotesize Copula d.f. and scale} & {\footnotesize 14.55\ (13.33)} & {\footnotesize -0.48 (0.12)} & {\footnotesize 20.48\ (24.78)} & {\footnotesize triangles} & {\footnotesize degree1.5} \\ 
&  &  & {\footnotesize kstar(2)} & {\footnotesize -0.07 (0.08)} & {\footnotesize -0.97 (0.01)} \\ 
&  &  & {\footnotesize triangles} &  & {\footnotesize -0.09 (0.09)} \\ \hline\hline
\multicolumn{6}{l}{\footnotesize \textit{Note:} \textbf{Bold} indicates the more accurate ABC method for the given estimator.}
\end{tabular}
\end{center}

\begin{changemargin}{-.7in}{-.7in}
\begin{center}
\begin{tabular}{c|ccc|c|ccc}
\multicolumn{8}{l}{\footnotesize\textbf{Table 7.} MAE (MSE), mean percent interval coverage (95\%(50\%)c) for ERGM over 10 replicas, varying $n$ and ${n}_{\text{sim}}$.} \\ \hline\hline
&  & \multicolumn{2}{c|}{{\footnotesize \textit{n} = 50}} &  &  & 
{\footnotesize \textit{n} = 300} &  \\ 
{\footnotesize ${n}_{\text{sim}}$} & {\footnotesize Posterior} & 
{\footnotesize kstar(2)} & {\footnotesize triangles} & {\footnotesize ${n}_{%
\text{sim}}$} & {\footnotesize kstar(2)} & {\footnotesize triangles} & 
{\footnotesize degree1.5} \\ \hline
{\footnotesize 5} & \multicolumn{1}{|l}{\footnotesize \texttt{copulaABCdrf}~mean} & 
{\footnotesize \textbf{0.22} (\textbf{0.05})} & {\footnotesize \textbf{0.80} (\textbf{0.64})} & {\footnotesize 30} & {\footnotesize 0.29 (0.08)} & {\footnotesize \textbf{1.43} (\textbf{2.07})} & 
{\footnotesize 1.27 (1.60)} \\ 
{\footnotesize} & \multicolumn{1}{|l}{\footnotesize \texttt{rejectionABC} mean} & 
{\footnotesize \textbf{0.22} (\textbf{0.05})} & {\footnotesize 0.84 (0.71)} & {\footnotesize } & {\footnotesize \textbf{0.08} (\textbf{0.01})} & {\footnotesize 1.58 (2.50)} & 
{\footnotesize \textbf{0.56} (\textbf{0.36})} \\ 
& \multicolumn{1}{|l}{\footnotesize \texttt{copulaABCdrf}~median} & {\footnotesize \textbf{0.22}
(\textbf{0.05})} & {\footnotesize \textbf{0.73} (\textbf{0.54})} &  & {\footnotesize 0.30 (0.09)} & 
{\footnotesize \textbf{1.47} (\textbf{2.20})} & {\footnotesize 1.45 (2.09)} \\ 
& \multicolumn{1}{|l}{\footnotesize \texttt{rejectionABC} median} & {\footnotesize \textbf{0.22} (\textbf{0.05})} & {\footnotesize 0.77 (0.60)} &  & {\footnotesize \textbf{0.04} (\textbf{0.00})} & 
{\footnotesize 1.58 (2.51)} & {\footnotesize \textbf{0.83} (\textbf{0.70})} \\ 

& \multicolumn{1}{|l}{\footnotesize \texttt{copulaABCdrf}~mode} & {\footnotesize \textbf{0.17} (\textbf{0.04})}
& {\footnotesize \textbf{0.58} (0.54)} &  & {\footnotesize 0.19 (0.04)} & 
{\footnotesize \textbf{0.61} (\textbf{0.51})} & {\footnotesize 1.23 (1.63)} \\ 

& \multicolumn{1}{|l}{\footnotesize \texttt{\texttt{rejectionABCkern}} mode} & {\footnotesize 0.20 (\textbf{0.04})}
& {\footnotesize \textbf{0.58} (\textbf{0.37})} &  & {\footnotesize \textbf{0.04} (\textbf{0.00})} & 
{\footnotesize 1.54 (2.40)} & {\footnotesize \textbf{0.76} (\textbf{0.62})} \\ 

& \multicolumn{1}{|l}{\footnotesize \texttt{copulaABCdrf} MLE} & {\footnotesize \textbf{0.16} (\textbf{0.04})}
& {\footnotesize \textbf{0.62} (\textbf{0.55})} &  & {\footnotesize \textbf{0.15} (\textbf{0.03})} & 
{\footnotesize \textbf{0.50} (\textbf{0.39})} & {\footnotesize \textbf{1.23} (\textbf{1.60})} \\ 

& \multicolumn{1}{|l}{\footnotesize \texttt{rejectionABCkern} MLE} & {\footnotesize 0.36 (0.17)}
& {\footnotesize 2.47 (7.31)} &  & {\footnotesize 0.18 (0.04)} & 
{\footnotesize 2.48 (6.76)} & {\footnotesize 2.20 (5.67)} \\ 
& \multicolumn{1}{|l}{\footnotesize MCMLE} & {\footnotesize 0.01 (0.00)} & 
{\footnotesize 0.17 (0.04)} &  & {\footnotesize 0.16 (0.03)} & 
{\footnotesize 0.40 (0.16)} & {\footnotesize 0.72 (0.52)} \\ 
& \multicolumn{1}{|l}{\footnotesize MPLE} & {\footnotesize 0.35 (0.12)} & 
{\footnotesize 0.67 (0.45)} &  & {\footnotesize 0.23 (0.05)} & 
{\footnotesize 0.24 (0.06)} & {\footnotesize 0.62 (0.38)} \\ 
& \multicolumn{1}{|l}{\footnotesize \texttt{copulaABCdrf} 95\%(50\%)c} & {\footnotesize 0.00 (\textbf{0.00})} & {\footnotesize \textbf{1.00} (\textbf{0.00})} &  & {\footnotesize \textbf{0.90} (0.00)} & 
{\footnotesize \textbf{0.90} (\textbf{0.00})} & {\footnotesize \textbf{1.00} (0.00)} \\ 
& \multicolumn{1}{|l}{\footnotesize \texttt{rejectionABC} 95\%(50\%)c} & {\footnotesize 0.10 (\textbf{0.00})} & {\footnotesize \textbf{1.00} (\textbf{0.00})} &  & {\footnotesize \textbf{1.00} (\textbf{1.00})} & 
{\footnotesize 0.50 (\textbf{0.00})} & {\footnotesize \textbf{1.00} (\textbf{0.40})} \\ 
& \multicolumn{1}{|l}{\footnotesize MCMLE 95\%c} & {\footnotesize 1.00} & 
{\footnotesize 1.00} &  & {\footnotesize 0.00} & {\footnotesize 0.00} & {\footnotesize 0.00} \\ \hline
{\footnotesize 16} & \multicolumn{1}{|l}{\footnotesize \texttt{copulaABCdrf}~mean} & 
{\footnotesize 0.07 (0.01)} & {\footnotesize 0.49 (0.25)} & {\footnotesize %
100} & {\footnotesize 0.17 (0.03)} & {\footnotesize \textbf{0.07} (\textbf{0.01})} & 
{\footnotesize 0.95 (0.92)} \\ 
{\footnotesize } & \multicolumn{1}{|l}{\footnotesize \texttt{rejectionABC}~mean} & 
{\footnotesize \textbf{0.06} (\textbf{0.00})} & {\footnotesize \textbf{0.37} (\textbf{0.14})} & {\footnotesize } & {\footnotesize \textbf{0.15} (\textbf{0.02})} & {\footnotesize 0.70 (0.49)} & 
{\footnotesize \textbf{0.47} (\textbf{0.22})} \\ 
& \multicolumn{1}{|l}{\footnotesize \texttt{copulaABCdrf}~median} & {\footnotesize 0.07
(0.01)} & {\footnotesize 0.51 (0.26)} &  & {\footnotesize 0.21 (0.04)} & 
{\footnotesize \textbf{0.22} (\textbf{0.05})} & {\footnotesize 1.65 (2.80)} \\ 
& \multicolumn{1}{|l}{\footnotesize \texttt{rejectionABC}~median} & {\footnotesize \textbf{0.06} (\textbf{0.00})} & {\footnotesize \textbf{0.33} (\textbf{0.11})} &  & {\footnotesize \textbf{0.16} (\textbf{0.02})} & 
{\footnotesize 0.80 (0.64)} & {\footnotesize \textbf{0.50} (\textbf{0.25})} \\ 

& \multicolumn{1}{|l}{\footnotesize \texttt{copulaABCdrf}~mode} & {\footnotesize 0.09 (0.01)}
& {\footnotesize 0.46 (0.29)} &  & {\footnotesize 0.26 (0.07)} & 
{\footnotesize \textbf{0.72} (0.83)} & {\footnotesize 1.94 (3.77)} \\ 

& \multicolumn{1}{|l}{\footnotesize \texttt{rejectionABCkern} mode} & {\footnotesize \textbf{0.05} (\textbf{0.00})}
& {\footnotesize \textbf{0.19} (\textbf{0.05})} &  & {\footnotesize \textbf{0.17} (\textbf{0.03})} & 
{\footnotesize 0.77 (\textbf{0.60})} & {\footnotesize \textbf{0.51} (\textbf{0.30})} \\ 

& \multicolumn{1}{|l}{\footnotesize \texttt{copulaABCdrf}~MLE} & {\footnotesize \textbf{0.07} (\textbf{0.01})}
& {\footnotesize 0.53 (0.43)} &  & {\footnotesize 0.21 (\textbf{0.05})} & 
{\footnotesize \textbf{1.11} (\textbf{1.53})} & {\footnotesize 1.89 (3.60)} \\ 

& \multicolumn{1}{|l}{\footnotesize \texttt{rejectionABCkern} MLE} & {\footnotesize 0.09 (0.01)}
& {\footnotesize \textbf{0.35} (\textbf{0.16})} &  & {\footnotesize \textbf{0.20} (\textbf{0.05})} & 
{\footnotesize 1.26 (2.10)} & {\footnotesize \textbf{0.72} (\textbf{0.71})} \\ 

& \multicolumn{1}{|l}{\footnotesize MCMLE} & {\footnotesize 0.01 (0.00)} & 
{\footnotesize 0.14 (0.03)} &  & {\footnotesize \textbf{0.17} (\textbf{0.03})} & 
{\footnotesize 0.41 (0.16)} & {\footnotesize 0.72 (0.52)} \\ 
& \multicolumn{1}{|l}{\footnotesize MPLE} & {\footnotesize 0.35 (0.12)} & 
{\footnotesize 0.66 (0.44)} &  & {\footnotesize 0.23 (0.05)} & 
{\footnotesize 0.24 (0.06)} & {\footnotesize 0.62 (0.38)} \\ 
& \multicolumn{1}{|l}{\footnotesize \texttt{copulaABCdrf} 95\%(50\%)c} & {\footnotesize \textbf{1.00}
(\textbf{0.20})} & {\footnotesize \textbf{1.00} (0.00)} &  & {\footnotesize \textbf{1.00} (\textbf{0.00})} & 
{\footnotesize \textbf{1.00} (\textbf{0.90})} & {\footnotesize \textbf{1.00} (\textbf{0.20})} \\ 
& \multicolumn{1}{|l}{\footnotesize \texttt{rejectionABC} 95\%(50\%)c} & {\footnotesize \textbf{1.00}
(0.10)} & {\footnotesize \textbf{1.00} (\textbf{0.10})} &  & {\footnotesize \textbf{0.90} (\textbf{0.00})} & 
{\footnotesize 0.60 (0.00)} & {\footnotesize \textbf{1.00} (0.00)} \\ 
& \multicolumn{1}{|l}{\footnotesize MCMLE 95\%c} & {\footnotesize 0.90} & 
{\footnotesize 0.90} &  & {\footnotesize 0.00} & {\footnotesize 0.00} & 
{\footnotesize 0.00} \\ \hline
{\footnotesize 25} & \multicolumn{1}{|l}{\footnotesize \texttt{copulaABCdrf}~mean} & 
{\footnotesize \textbf{0.03} (\textbf{0.00})} & {\footnotesize 0.47 (0.31)} & {\footnotesize %
150} & {\footnotesize \textbf{0.11} (\textbf{0.01})} & {\footnotesize \textbf{0.04} (\textbf{0.00})} & 
{\footnotesize 0.66 (0.46)} \\ 
{\footnotesize } & \multicolumn{1}{|l}{\footnotesize \texttt{rejectionABC}~mean} & 
{\footnotesize \textbf{0.03} (\textbf{0.00})} & {\footnotesize \textbf{0.41} (\textbf{0.22})} & {\footnotesize } & {\footnotesize \textbf{0.11} (\textbf{0.01})} & {\footnotesize 0.68 (0.47)} & 
{\footnotesize \textbf{0.29} (\textbf{0.09})} \\ 
& \multicolumn{1}{|l}{\footnotesize \texttt{copulaABCdrf}~median} & {\footnotesize \textbf{0.03}
(\textbf{0.00})} & {\footnotesize 0.42 (0.23)} &  & {\footnotesize 0.14 (0.02)} & 
{\footnotesize \textbf{0.03} (\textbf{0.00})} & {\footnotesize 0.67 (0.57)} \\ 
& \multicolumn{1}{|l}{\footnotesize \texttt{rejectionABC}~median} & {\footnotesize \textbf{0.03} (\textbf{0.00})} & {\footnotesize \textbf{0.31} (\textbf{0.12})} &  & {\footnotesize \textbf{0.11} (\textbf{0.01})} & 
{\footnotesize 0.82 (0.69)} & {\footnotesize \textbf{0.32} (\textbf{0.11})} \\ 

& \multicolumn{1}{|l}{\footnotesize \texttt{copulaABCdrf}~mode} & {\footnotesize 0.05 (\textbf{0.00})}
& {\footnotesize 0.41 (0.24)} &  & {\footnotesize 0.29 (0.09)} & 
{\footnotesize \textbf{0.60} (\textbf{0.51})} & {\footnotesize 1.98 (3.94)} \\ 

& \multicolumn{1}{|l}{\footnotesize \texttt{rejectionABCkern} mode} & {\footnotesize \textbf{0.03} (\textbf{0.00})}
& {\footnotesize \textbf{0.35} (\textbf{0.15})} &  & {\footnotesize \textbf{0.19} (\textbf{0.04})} & 
{\footnotesize 0.94 (0.90)} & {\footnotesize \textbf{0.54} (\textbf{0.38})} \\ 

& \multicolumn{1}{|l}{\footnotesize \texttt{copulaABCdrf}~MLE} & {\footnotesize \textbf{0.05} (\textbf{0.00})}
& {\footnotesize \textbf{0.43} (\textbf{0.29})} &  & {\footnotesize 0.25 (\textbf{0.06})} & 
{\footnotesize \textbf{0.93} (\textbf{1.04})} & {\footnotesize 2.01 (4.06)} \\ 

& \multicolumn{1}{|l}{\footnotesize \texttt{rejectionABCkern} MLE} & {\footnotesize 0.13 (0.02)}
& {\footnotesize 1.66 (4.77)} &  & {\footnotesize \textbf{0.24} (0.07)} & 
{\footnotesize 0.94 (1.33)} & {\footnotesize \textbf{0.86} (\textbf{1.28})} \\ 
& \multicolumn{1}{|l}{\footnotesize MCMLE} & {\footnotesize 0.02 (0.00)} & 
{\footnotesize 0.12 (0.02)} &  & {\footnotesize 0.17 (0.03)} & 
{\footnotesize 0.40 (0.16)} & {\footnotesize 0.72 (0.52)} \\ 
& \multicolumn{1}{|l}{\footnotesize MPLE} & {\footnotesize 0.35 (0.12)} & 
{\footnotesize 0.68 (0.47)} &  & {\footnotesize 0.23 (0.05)} & 
{\footnotesize 0.24 (0.06)} & {\footnotesize 0.62 (0.38)} \\ 
& \multicolumn{1}{|l}{\footnotesize \texttt{copulaABCdrf} 95\%(50\%)c} & {\footnotesize \textbf{1.00}
(\textbf{0.60})} & {\footnotesize \textbf{1.00} (\textbf{0.50})} &  & {\footnotesize \textbf{1.00} (1.00)} & 
{\footnotesize \textbf{1.00} (\textbf{1.00})} & {\footnotesize \textbf{1.00} (0.90)} \\ 
& \multicolumn{1}{|l}{\footnotesize \texttt{rejectionABC} 95\%(50\%)c} & {\footnotesize \textbf{1.00}
(\textbf{0.60})} & {\footnotesize \textbf{1.00} (\textbf{0.50})} &  & {\footnotesize \textbf{1.00} (\textbf{0.20})} & 
{\footnotesize \textbf{1.00} (\textbf{0.00})} & {\footnotesize \textbf{1.00} (\textbf{0.60})} \\ 
& \multicolumn{1}{|l}{\footnotesize MCMLE 95\%c} & {\footnotesize 0.90} & 
{\footnotesize 1.00} &  & {\footnotesize 0.00} & {\footnotesize 0.00} & 
{\footnotesize 0.00}  \\\hline\hline
\multicolumn{8}{l}{\footnotesize \textit{Note:} \textbf{Bold} indicates the more accurate ABC method for the given estimator. The table is continued on the next page.}\\
\end{tabular}
\end{center}
\end{changemargin}

\begin{changemargin}{-.7in}{-.7in}
\begin{center}
\begin{tabular}{c|ccc|c|ccc}
\multicolumn{8}{l}{\footnotesize\textbf{Table 7 (continued).} MAE (MSE), mean percent interval coverage for ERGM over 10 replicas, varying $n$ and ${n}_{\text{sim}}$.}\\\hline\hline
&  & \multicolumn{2}{c|}{{\footnotesize \textit{n} = 50}} &  & 
\multicolumn{3}{|c}{{\footnotesize \textit{n} = 300}} \\ 
{\footnotesize ${n}_{\text{sim}}$} & {\footnotesize Posterior} & 
{\footnotesize kstar(2)} & {\footnotesize triangles} & {\footnotesize ${n}_{%
\text{sim}}$} & {\footnotesize kstar(2)} & {\footnotesize triangles} & 
{\footnotesize degree1.5} \\ \hline

{\footnotesize 37} & \multicolumn{1}{|l}{\footnotesize \texttt{copulaABCdrf}~mean} & 
{\footnotesize \textbf{0.01} (\textbf{0.00})} & {\footnotesize 0.16 (0.03)} & {\footnotesize %
225} & {\footnotesize 0.44 (0.20)} & {\footnotesize 0.29 (0.09)} & 
{\footnotesize 2.12 (4.59)} \\ 
{\footnotesize } & \multicolumn{1}{|l}{\footnotesize \texttt{rejectionABC}~mean} & 
{\footnotesize \textbf{0.01} (\textbf{0.00})} & {\footnotesize \textbf{0.08} (\textbf{0.01})} & {\footnotesize } & {\footnotesize \textbf{0.20} (\textbf{0.04})} & {\footnotesize \textbf{0.16} (\textbf{0.04})} & 
{\footnotesize \textbf{0.97} (\textbf{0.96})} \\ 
& \multicolumn{1}{|l}{\footnotesize \texttt{copulaABCdrf}~median} & {\footnotesize \textbf{0.01}
(\textbf{0.00})} & {\footnotesize 0.16 (0.04)} &  & {\footnotesize 0.45 (0.22)} & 
{\footnotesize \textbf{0.25} (\textbf{0.07})} & {\footnotesize 2.12 (4.64)} \\ 
& \multicolumn{1}{|l}{\footnotesize \texttt{rejectionABC}~median} & {\footnotesize \textbf{0.01} (\textbf{0.00})} & {\footnotesize \textbf{0.09} (\textbf{0.01})} &  & {\footnotesize \textbf{0.18} (\textbf{0.04})} & 
{\footnotesize 0.32 (0.11)} & {\footnotesize \textbf{0.79} (\textbf{0.64})} \\ 
& \multicolumn{1}{|l}{\footnotesize \texttt{copulaABCdrf}~mode} & {\footnotesize 0.04 (\textbf{0.00})}
& {\footnotesize 0.15 (0.03)} &  & {\footnotesize 0.43 (0.20)} & 
{\footnotesize \textbf{0.34} (\textbf{0.15})} & {\footnotesize 2.44 (6.26)} \\ 
& \multicolumn{1}{|l}{\footnotesize \texttt{rejectionABCkern} mode} & {\footnotesize \textbf{0.02} (\textbf{0.00})}
& {\footnotesize \textbf{0.07} (\textbf{0.01})} &  & {\footnotesize \textbf{0.11} (\textbf{0.02})} & 
{\footnotesize 0.71 (0.54)} & {\footnotesize \textbf{0.55} (\textbf{0.51})} \\ 
& \multicolumn{1}{|l}{\footnotesize \texttt{copulaABCdrf}~MLE} & {\footnotesize \textbf{0.02} (\textbf{0.00})}
& {\footnotesize \textbf{0.17} (\textbf{0.05})} &  & {\footnotesize \textbf{0.59} (\textbf{0.47})} & 
{\footnotesize \textbf{0.42} (\textbf{0.28})} & {\footnotesize \textbf{3.21} (\textbf{12.45})} \\ 
& \multicolumn{1}{|l}{\footnotesize \texttt{rejectionABCkern} MLE} & {\footnotesize 0.10 (0.01)}
& {\footnotesize 0.32 (0.22)} &  & {\footnotesize 0.75 (0.60)} & 
{\footnotesize 0.87 (1.17)} & {\footnotesize 3.89 (15.68)} \\ 
& \multicolumn{1}{|l}{\footnotesize MCMLE} & {\footnotesize 0.01 (0.00)} & 
{\footnotesize 0.11 (0.02)} &  & {\footnotesize 0.17 (0.03)} & 
{\footnotesize 0.40 (0.16)} & {\footnotesize 0.73 (0.53)} \\ 
& \multicolumn{1}{|l}{\footnotesize MPLE} & {\footnotesize 0.35 (0.12)} & 
{\footnotesize 0.73 (0.53)} &  & {\footnotesize 0.23 (0.05)} & 
{\footnotesize 0.24 (0.06)} & {\footnotesize 0.62 (0.38)} \\ 
& \multicolumn{1}{|l}{\footnotesize \texttt{copulaABCdrf} 95\%(50\%)c} & {\footnotesize \textbf{1.00}
(\textbf{1.00})} & {\footnotesize \textbf{1.00} (\textbf{0.70})} &  & {\footnotesize \textbf{1.00} (1.00)} & 
{\footnotesize \textbf{1.00} (1.00)} & {\footnotesize \textbf{1.00} (1.00)} \\ 
& \multicolumn{1}{|l}{\footnotesize \texttt{rejectionABC} 95\%(50\%)c} & {\footnotesize \textbf{1.00}
(\textbf{1.00})} & {\footnotesize \textbf{1.00} (0.90)} &  & {\footnotesize \textbf{1.00} (\textbf{0.40})} & 
{\footnotesize \textbf{1.00} (\textbf{0.90})} & {\footnotesize \textbf{1.00} (\textbf{0.10})} \\ 
& \multicolumn{1}{|l}{\footnotesize MCMLE 95\%c} & {\footnotesize 1.00} & 
{\footnotesize 1.00} &  & {\footnotesize 0.00} & {\footnotesize 0.00} & 
{\footnotesize 0.00} \\\hline

{\footnotesize 50} & \multicolumn{1}{|l}{\footnotesize \texttt{copulaABCdrf}~mean} & 
{\footnotesize \textbf{0.01} (\textbf{0.00})} & {\footnotesize 0.09 (0.02)} & {\footnotesize 300} & {\footnotesize \textbf{0.52} (\textbf{0.28})} & {\footnotesize \textbf{0.34} (\textbf{0.15})} & 
{\footnotesize \textbf{2.91} (\textbf{8.78})} \\ 
{\footnotesize } & \multicolumn{1}{|l}{\footnotesize \texttt{rejectionABC}~mean} & {\footnotesize \textbf{0.01} (\textbf{0.00})} & {\footnotesize \textbf{0.07} (\textbf{0.01})} & {\footnotesize } & {\footnotesize 0.75 (0.57)} & {\footnotesize 0.44 (0.20)} & {\footnotesize 4.37 (19.42)} \\ 
& \multicolumn{1}{|l}{\footnotesize \texttt{copulaABCdrf}~median} & {\footnotesize \textbf{0.01}
(\textbf{0.00})} & {\footnotesize 0.10 (\textbf{0.01})} &  & {\footnotesize \textbf{0.43} (\textbf{0.19})} & 
{\footnotesize \textbf{0.44} (\textbf{0.23})} & {\footnotesize \textbf{2.33} (\textbf{5.76})} \\ 
& \multicolumn{1}{|l}{\footnotesize \texttt{rejectionABC}~median} & {\footnotesize 0.02 (\textbf{0.00})} & {\footnotesize \textbf{0.07} (\textbf{0.01})} &  & {\footnotesize 0.70 (0.51)} & 
{\footnotesize 0.49 (0.26)} & {\footnotesize 4.09 (17.20)} \\ 
& \multicolumn{1}{|l}{\footnotesize \texttt{copulaABCdrf}~mode} & {\footnotesize 0.04 (\textbf{0.00})}
& {\footnotesize 0.32 (0.14)} &  & {\footnotesize \textbf{0.17} (\textbf{0.06})} & 
{\footnotesize \textbf{0.33} (\textbf{0.15})} & {\footnotesize \textbf{0.92} (\textbf{1.92})} \\ 
& \multicolumn{1}{|l}{\footnotesize \texttt{rejectionABCkern} mode} & {\footnotesize \textbf{0.02} (\textbf{0.00})}
& {\footnotesize \textbf{0.12} (\textbf{0.02})} &  & {\footnotesize 0.60 (0.42)} & 
{\footnotesize 0.58 (0.43)} & {\footnotesize 3.60 (14.50)} \\ 
& \multicolumn{1}{|l}{\footnotesize \texttt{copulaABCdrf}~MLE} & {\footnotesize \textbf{0.04} (\textbf{0.00})}
& {\footnotesize \textbf{0.33} (\textbf{0.17})} &  & {\footnotesize \textbf{0.19} (\textbf{0.07})} & 
{\footnotesize \textbf{0.39} (\textbf{0.23})} & {\footnotesize \textbf{0.96} (\textbf{2.04})} \\ 
& \multicolumn{1}{|l}{\footnotesize \texttt{rejectionABCkern} MLE} & {\footnotesize 0.11 (0.01)}
& {\footnotesize 0.44 (0.45)} &  & {\footnotesize 2.05 (4.46)} & 
{\footnotesize 1.47 (2.61)} & {\footnotesize 11.67 (141.65)} \\ 
& \multicolumn{1}{|l}{\footnotesize MCMLE} & {\footnotesize 0.01 (0.00)} & 
{\footnotesize 0.14 (0.03)} &  & {\footnotesize 0.17 (0.03)} & 
{\footnotesize 0.40 (0.16)} & {\footnotesize 0.74 (0.55)} \\ 
& \multicolumn{1}{|l}{\footnotesize MPLE} & {\footnotesize 0.35 (0.12)} & 
{\footnotesize 0.68 (0.46)} &  & {\footnotesize 0.23 (0.05)} & 
{\footnotesize 0.24 (0.06)} & {\footnotesize 0.62 (0.38)} \\ 
& \multicolumn{1}{|l}{\footnotesize \texttt{copulaABCdrf} 95\%(50\%)c} & {\footnotesize \textbf{1.00}
(0.80)} & {\footnotesize \textbf{1.00} (\textbf{0.90})} &  & {\footnotesize \textbf{1.00} (\textbf{0.10})} & 
{\footnotesize \textbf{1.00} (\textbf{0.80})} & {\footnotesize \textbf{1.00} (\textbf{0.00})} \\ 
& \multicolumn{1}{|l}{\footnotesize \texttt{rejectionABC} 95\%(50\%)c} & {\footnotesize \textbf{1.00}
(\textbf{0.70})} & {\footnotesize \textbf{1.00} (1.00)} &  & {\footnotesize 0.70 (0.00)} & 
{\footnotesize \textbf{1.00} (1.00)} & {\footnotesize 0.30 (\textbf{0.00})}\\ 
& \multicolumn{1}{|l}{\footnotesize MCMLE 95\%c} & {\footnotesize 1.00} & 
{\footnotesize 1.00} &  & {\footnotesize 0.00} & {\footnotesize 0.00} & {\footnotesize 0.00} \\ \hline\hline
\multicolumn{8}{l}{\footnotesize \textit{Note:} \textbf{Bold} indicates the more accurate ABC method for the given estimator.}\\
\end{tabular}
\end{center}
\end{changemargin}

\subsection{Simulation Study: Preferential Attachment (PA) Models}

Mechanistic network models defined by preferential attachment rules, including the Price model and the Nonlinear Preferential Attachment (NLPA) model, grows a directed or undirected network by introducing a new node at each stage and linking the new node to any given existing node $i$ with probability $\Pr($New node attaches to old node $i$) $\propto k_{0}+k_{i}^{\alpha}$, with some constant parameter $k_{0}\in\mathbb{R}$, and $k_{i}$ is the degree of node $i$ (indegree for a directed network); and power parameter $\alpha\in(0,\infty)$. In particular, the Price model grows a directed binary citation network one article (node) at a time (each edge being a paper citing another paper), with $\Pr($new article cites existing article $i)\propto k_{0}+k_{i}$, with $\alpha\equiv1$ and constant parameter $k_{0}$ centered around 1, and $k_{i}$ is the in-degree $k_{i}$ of existing article $i$, such that each new article cites $m$ existing articles on average, and the number of articles a new article cites follows a Binomial$(n_{0},p)$ distribution, with success probability parameter $p$, and with $n_{0}$ set as the maximum outdegree of $x$ \citep{RaynalEtAl22}. The unknown Price model parameters to be estimated from the observed directed network dataset $x$ are $(k_{0},p)$. For an undirected (binary) network $x$, the \citet{BarabasiAlbert99} (BA) model has the same form, but without the assumptions $k_{0}=0$ and $\alpha\equiv1$, and based on $k_{i}$ defined as the degree of existing node $i$.

The NLPA model \citep{KrapivskyEtAl00} generalizes the BA model, using $\Pr($new node attaches to existing node $i)=\dfrac{k_{i}^{\alpha}}{{\textstyle\sum\nolimits_{\text{Existing nodes }j}}
k_{j}^{\alpha}},$ \ $\alpha>0$, with $k_{i}$ the degree of existing node $i$. Since possibly $\alpha\neq1$, the number of nodes a new node attaches was assumed to follow a truncated Binomial$(n_{0},p)$ distribution which supports only positive counts. For the NLPA model, the unknown parameters to be estimated from the given undirected network dataset are $(\alpha,p)$. Thus from an observed network dataset $x$, the unknown model parameters to be estimated are $(\alpha,p)$, and $n_{0}$ is chosen as the maximum degree of $x$.

Now we consider simulation studies that evaluate and compare the ability of \texttt{copulaABCdrf} (Algorithm 1) and \texttt{rejectionABC} to estimate the posterior distribution of the parameters of the Price model, and the parameters of the posterior distribution of the NLPA model. The simulation studies are based on 10 replications of $n=50$ node directed networks from the Price model (undirected networks from the NLPA model, respectively); and of $n=300$ node undirected networks simulated from the Price model (undirected networks from the NLPA model, respectively). These simulations were done for each of five conditions of $n_{\text{sim}}$, being  at least roughly $10\%$, $33\%$, $50\%$, $75\%$ and $100\%$ of $n$ network nodes. Specifically, the conditions $n_{\text{sim}}=5, 16, 25, 37$ and $50$ for $n = 50$ nodes, and $n_{\text{sim}}= 30, 100, 150, 225$ and $300$ for $n = 300$ nodes. The true data-generating parameters of the Price model were specified as $\theta_0^{\intercal}=(k_{0},p)=(1,0.02)$, while the true data-generating parameters of the NLPA model were  $\theta_0^{\intercal}=(\alpha,p)=(1.2,0.02)$. 

For each network dataset $x$ simulated from the Price model and analyzed by Algorithm 1 using this model, this model was assigned a uniform prior $(k_{0},p)\sim$ $\mathcal{U}_{(0.9,1.1)}\mathcal{U}_{(0,b)}$ (with $b=0.10$ under $n=50$ and $b=0.20$ for $n=300$) and used vector of summaries $s(\cdot)$ specified by network size invariant offset MPLEs of geometrically weighted degree, its decay estimate, and triangle count. For each network dataset $x$ simulated from the NLPA model, and analyzed by Algorithm 1 using this model, the NLPA model was assigned a uniform prior $(\alpha,p)\sim\mathcal{U}_{(0,3)}\mathcal{U}_{(0,0.20)}$ and using a vector of summaries $s(\cdot)$ specified by the network size invariant offset MPLEs of density, average clustering coefficient, and diameter. The MPLE summaries $s(x)$ of each network $x$ used offsets based on $n$ nodes ($n=50$ or $300$). The MPLE summaries $s(y)$ of each network $y$ of smaller size $n_{\text{sim}}<n$ (being $n_{\text{sim}}=25$ for $n=50$, and $n_{\text{sim}}=100$ for $n=300$) were simulated in each iteration of the ABC algorithm used offsets based on $n_{\text{sim}}$ nodes, and based on a Binomial$(n_{\text{sim}}-1,p)$ distribution (or the truncated binomial for NLPA).  

Table 8 presents some representative detailed results of the simulation study of \\\texttt{copulaABCdrf} and \texttt{rejectionABC}, for the Price and NLPA models, for pairs of values of $n_{\text{sim}}$ and $n$ with $n_{\text{sim}}<n$, namely, $n_{\text{sim}}=25$ for $n=50$ , and $n_{\text{sim}}=100$ for $n=300$. Tables 9 and 10 present the results of MAE, MSE, accuracy of 95\% and 50\% posterior interval credible interval estimates of \texttt{copulaABCdrf} and \texttt{rejectionABC} for the Price and NLPA models, for each of the five $n_{\text{sim}}$ conditions within each of  $n=50$ and $n=300$. 

In particular, since for each of the Price and NLPA models, there were 3 summary statistics relative to the two models parameters, we not only considered for each model implementations of \texttt{rejectionABC} using all 3 summary statistics. We  also considered, for comparison purposes, each implementation of \texttt{rejectionABC} based on pre-selecting two of the most important of the 3 summary statistics, according to the results obtained from a variable importance analysis based on training a \texttt{drf} multivariate regression on a Reference Table of simulated model parameters on all 3 summary statistics (covariate variables), as described earlier in \S3. For the Price model, \texttt{drf} variable importance analyses found that over all ten replicas of network datasets under the $n_{\text{sim}}=5$ and $n=50$ condition, that the network size invariant offset MPLEs of geometrically weighted degree, and its decay estimate, were always the most important summary statistics of the three total summaries. Also, for each of the all other simulation conditions of $n_{\text{sim}}$ and $n$, the \texttt{drf} variable importance analyses found that over all ten replicas of network datasets, the most important summaries were always the network size invariant offset MPLEs of geometrically weighted degree and triangle counts. For the NLPA model, the \texttt{drf} variable importance analyses always found that network density and average clustering coefficient were the two most important network summaries of the three total summaries, compared to network diameter summary, for each of all the 10 network replicas, and within each of all conditions $n_{\text{sim}}=5, 16, 25, 37$ and $50$ for $n = 50$ nodes, and all conditions of $n_{\text{sim}}= 30, 100, 150, 225$ and $300$ for $n = 300$ nodes.

From the results of Table 9, on the Price model, it can be concluded overall that \texttt{copulaABCdrf} and \texttt{rejectionABC} performed similarly. Also,  \texttt{rejectionABC} based on using \texttt{drf} to preselect two of the three most important summaries, produced results that sometimes slightly improved but were usually similar to the results of \texttt{rejectionABC} using all three summaries. More specifically, for the parameter $k_0$ and $n=50$ nodes, \texttt{copulaABCdrf} slightly outperformed \texttt{rejectionABC} in MSE of posterior mode estimation and MLE, and \texttt{rejectionABC} outperformed \texttt{copulaABCdrf} in MAE of posterior mode and MLE. For the parameter $p$ and $n=50$ nodes, \texttt{copulaABCdrf} slightly outperformed \texttt{rejectionABC} in estimation accuracy of the 95\% credible interval, while being close to the nominal level, while \texttt{rejectionABC} slightly outperformed \texttt{copulaABCdrf} in terms of MSE of estimation of the posterior mode and MLE. For the parameter $k_0$ and $n=300$ nodes, \texttt{copulaABCdrf} slightly outperformed \texttt{rejectionABC} in MSE of estimation of the posterior mode and MLE, \texttt{rejectionABC} outperformed \texttt{copulaABCdrf} in MAE of estimation of the posterior mode and MLE. For the parameter $p$ and $n=300$ nodes,  \texttt{copulaABCdrf} outperformed \texttt{rejectionABC} in accuracy in estimation of the 95\% posterior credible interval, and slightly outperformed \texttt{rejectionABC} in terms of accuracy in estimation of the 50\% posterior credible interval, while not being near the nominal rate a few times.

From the results of Table 10 on the NLPA model, it can be concluded that, overall, \texttt{copulaABCdrf} and \texttt{rejectionABC} performed similarly, while \texttt{rejectionABC} based on using \texttt{drf} to preselect two of the three most important summaries producing results that often improved on results of \texttt{rejectionABC} using all three summaries. More specifically, for the parameter $\alpha$ and $n=50$ nodes, \texttt{copulaABCdrf} slightly outperformed \texttt{rejectionABC} in MAE and MSE of posterior mean estimation, and MAE of posterior median estimation. Also, \texttt{rejectionABC} slightly outperformed \texttt{copulaABCdrf} in MSE of posterior median estimation and in MAE and MSE in estimation of the posterior mode and MLE. For the parameter $p$ and $n=50$ nodes, \texttt{copulaABCdrf} slightly outperformed \texttt{rejectionABC} in accuracy of the 50\% credible interval estimation, while not often being close to the nominal rate. In addition, \texttt{rejectionABC} outperformed \texttt{copulaABCdrf} in MAE of posterior mean and median estimation, and in MAE and MSE of estimation of the posterior mode and the MLE, and in estimation of the 95\% posterior credible interval, while typically being close to the nominal rate; while for the parameter $\alpha$ and $n=300$ nodes, \texttt{copulaABCdrf} slightly outperformed \texttt{rejectionABC} in accuracy of the 50\% credible interval estimation, while \texttt{rejectionABC} outperformed \texttt{copulaABCdrf} in terms of MAE and MSE in estimation of the posterior mead, median, mode, and the MLE, and in accuracy of estimation of the 50\% posterior credible interval, while often being far from the nominal level. For the parameter $p$ and $n=300$ nodes, \texttt{copulaABCdrf} slightly outperformed \texttt{rejectionABC} in MAE in posterior mean estimation, MSE in estimation of the posterior median, the posterior mode and the MLE, and in accuracy in estimation of the 50\% posterior credible interval, while not always close to the nominal level. Also, \texttt{rejectionABC} outperformed \texttt{copulaABCdrf} in MAE of estimation of the posterior mean, median, and 95\% posterior credible interval coverage, while usually being close to the nominal level.

Tables 9 and 10 showed that, for each of the \texttt{copulaABCdrf} and \texttt{rejectionABC} methods, as $n_{\text{sim}}$ gets smaller relative to $n$, that roughly, and on the most part, MAE and MSE of estimates tend to get smaller, and the 95\% and 50\% posterior credible intervals more closely approach their respective nominal rates. Intuitively, the summary statistics $s$ become less sufficient as $n_{\text{sim}}$ gets smaller relative to $n$.

\begin{center}
\begin{tabular}{lcc|cc}
\multicolumn{5}{l}{\footnotesize \textbf{Table 8.} Simulation study results for Price and NLPA models. Mean (standard deviation) of} \\ 
\multicolumn{5}{l}{\footnotesize posterior estimates over 10 replicas, under some conditions $n$ and ${n}_{\text{sim}}$.} \\ 
\hline\hline
{\footnotesize Price model} & \footnotesize${ k}_{0}$ & \footnotesize${ p}$ & \footnotesize${ k}_{0}$ & \footnotesize ${p}$ \\ \hline \hline
{\footnotesize \textit{n}} & {\footnotesize 50} & {\footnotesize 50} & {\footnotesize 300} & {\footnotesize 300} \\ 
{\footnotesize \textit{n}}$_{\text{sim}}$ & {\footnotesize 25} & {\footnotesize 25} & {\footnotesize 100} & 
{\footnotesize 100} \\ 
{\footnotesize true $\theta $} & {\footnotesize 1.00} & {\footnotesize 0.02} & {\footnotesize 1.00} & 
{\footnotesize 0.02} \\ \hline

{\footnotesize \texttt{copulaABCdrf} mean} 
& {\footnotesize \textbf{1.00} (0.00)} 
& {\footnotesize \textbf{0.05} (0.01)} 
& {\footnotesize 0.99 (0.01)} 
& {\footnotesize \textbf{0.05} (0.00)} \\

{\footnotesize \texttt{rejectionABC} mean} 
& {\footnotesize \textbf{1.00} (0.01)} 
& {\footnotesize \textbf{0.05} (0.01)} 
& {\footnotesize \textbf{1.00} (0.01)} 
& {\footnotesize \textbf{0.05} (0.00)} \\

{\footnotesize \texttt{\texttt{rejectionABCselect}} mean} 
& {\footnotesize \textbf{1.00} (0.01)} 
& {\footnotesize \textbf{0.05} (0.01)} 
& {\footnotesize \textbf{1.00} (0.01)} 
& {\footnotesize \textbf{0.05} (0.00)} \\\hline

{\footnotesize \texttt{copulaABCdrf} median} 
& {\footnotesize \textbf{1.00} (0.01)} 
& {\footnotesize 0.05 (0.01)} 
& {\footnotesize 0.99 (0.01)} 
& {\footnotesize \textbf{0.05} (0.00)} \\ 

{\footnotesize \texttt{rejectionABC} median} 
& {\footnotesize 0.99 (0.01)} 
& {\footnotesize 0.05 (0.01)} 
& {\footnotesize \textbf{1.00} (0.01)} 
& {\footnotesize \textbf{0.05} (0.00)} \\

{\footnotesize \texttt{\texttt{rejectionABCselect}} median} 
& {\footnotesize 0.99 (0.01)} 
& {\footnotesize \textbf{0.04} (0.02)} 
& {\footnotesize \textbf{1.00} (0.01)} 
& {\footnotesize \textbf{0.05} (0.00)} \\\hline

{\footnotesize \texttt{copulaABCdrf} modeMLE}
& {\footnotesize 0.97 (0.05)} 
& {\footnotesize \textbf{0.05} (0.02)} 
& {\footnotesize 1.02 (0.06)} 
& {\footnotesize \textbf{0.05} (0.00)} \\ 

{\footnotesize \texttt{rejectionABCkern} modeMLE}
& {\footnotesize \textbf{1.00} (0.04)} 
& {\footnotesize \textbf{0.05} (0.01)} 
& {\footnotesize \textbf{0.99} (0.05)} 
& {\footnotesize \textbf{0.05} (0.00)} \\ 

{\footnotesize \texttt{rejectionABCkern.select} modeMLE}
& {\footnotesize 0.99 (0.04)} 
& {\footnotesize 0.06 (0.01)} 
& {\footnotesize \textbf{0.99} (0.05)} 
& {\footnotesize \textbf{0.05} (0.00)}\\\hline

{\footnotesize \texttt{copulaABCdrf} s.d.} 
& {\footnotesize 0.06 (0.00)} 
& {\footnotesize 0.02 (0.00)} 
& {\footnotesize 0.06 (0.00)} 
& {\footnotesize 0.01 (0.00)} \\ 

{\footnotesize \texttt{rejectionABC} s.d.} 
& {\footnotesize 0.06 (0.00)} 
& {\footnotesize 0.02 (0.00)} 
& {\footnotesize 0.06 (0.00)} 
& {\footnotesize 0.01 (0.00)} \\

{\footnotesize \texttt{\texttt{rejectionABCselect}} s.d.} 
& {\footnotesize 0.06 (0.00)} 
& {\footnotesize 0.02 (0.00)} 
& {\footnotesize 0.06 (0.00)} 
& {\footnotesize 0.01 (0.00)}\\\hline

{\footnotesize \texttt{copulaABCdrf}} 
& {\footnotesize d.f.} 
& {\footnotesize scale} 
& {\footnotesize d.f.} & 
{\footnotesize scale} \\ 

{\footnotesize Copula d.f. and scale} 
& {\footnotesize 633.32 (474.75)} 
& {\footnotesize 0.06 (0.05)} 
& {\footnotesize 568.93 (456.18)} 
& {\footnotesize 0.04\ (0.07)} \\\hline

\multicolumn{5}{l}{} \\ \hline \hline

{\footnotesize NLPA model} & \footnotesize${\alpha }$ & \footnotesize${ p}$ & \footnotesize${ \alpha }$
& \footnotesize${ p}$ \\ \hline \hline
{\footnotesize \textit{n}} & {\footnotesize 50} & {\footnotesize 50} & {\footnotesize 300} & {\footnotesize 300} \\ 
{\footnotesize \textit{n}}$_{\text{sim}}$ & {\footnotesize 25} & {\footnotesize 25} & {\footnotesize 100} & 
{\footnotesize 100} \\ 
{\footnotesize true $\theta$} & {\footnotesize 1.20} & {\footnotesize 0.02} & {\footnotesize 1.20} & 
{\footnotesize 0.02} \\\hline

{\footnotesize \texttt{copulaABCdrf} mean} & {\footnotesize \textbf{1.24} (0.30)} & {\footnotesize 0.01 (0.00)} & {\footnotesize \textbf{1.12} (0.09)} & {\footnotesize \textbf{0.02} (0.00)} \\ 

{\footnotesize \texttt{rejectionABC} mean} 
& {\footnotesize 0.94 (0.30)} 
& {\footnotesize 0.03 (0.00)} 
& {\footnotesize 0.89 (0.17)} 
& {\footnotesize 0.04 (0.01)} \\ 

{\footnotesize \texttt{\texttt{rejectionABCselect}} mean} 
& {\footnotesize 1.37 (0.21)} 
& {\footnotesize \textbf{0.02} (0.00)} 
& {\footnotesize \textbf{1.12} (0.07)} 
& {\footnotesize \textbf{0.02} (0.00)} \\\hline

{\footnotesize \texttt{copulaABCdrf} median} 
& {\footnotesize \textbf{1.23} (0.30)} 
& {\footnotesize 0.01 (0.00)} 
& {\footnotesize \textbf{1.12} (0.08)} 
& {\footnotesize \textbf{0.02} (0.00)} \\ 

{\footnotesize \texttt{rejectionABC} median} 
& {\footnotesize 0.90 (0.33)} 
& {\footnotesize 0.03 (0.00)} 
& {\footnotesize 0.87 (0.19)} 
& {\footnotesize 0.04 (0.01)} \\ 

{\footnotesize \texttt{\texttt{rejectionABCselect}} median} 
& {\footnotesize 1.36 (0.22)} 
& {\footnotesize \textbf{0.02} (0.00)} 
& {\footnotesize 1.11 (0.07)} 
& {\footnotesize \textbf{0.02} (0.00)} \\\hline

{\footnotesize \texttt{copulaABCdrf} modeMLE} 
& {\footnotesize \textbf{1.35} (0.50)} 
& {\footnotesize \textbf{0.01} (0.00)} 
& {\footnotesize 1.11 (0.22)} 
& {\footnotesize \textbf{0.02} (0.00)} \\ 

{\footnotesize \texttt{rejectionABCkern} modeMLE} 
& {\footnotesize 0.86 (0.43)} 
& {\footnotesize \textbf{0.03} (0.01)} 
& {\footnotesize 0.82 (0.19)} 
& {\footnotesize 0.04 (0.01)} \\ 

{\footnotesize \texttt{rejectionABCkern.select} modeMLE} 
& {\footnotesize 1.42 (0.33)} 
& {\footnotesize \textbf{0.01} (0.01)} 
& {\footnotesize \textbf{1.12} (0.09)} 
& {\footnotesize \textbf{0.02} (0.00)} \\\hline

{\footnotesize \texttt{copulaABCdrf} standard deviation (s.d.)} 
& {\footnotesize 0.51 (0.02)} 
& {\footnotesize 0.01 (0.00)} 
& {\footnotesize 0.27 (0.04)} 
& {\footnotesize 0.00 (0.00)} \\ 

{\footnotesize \texttt{rejectionABC} s.d.} 
& {\footnotesize 0.48 (0.05)} 
& {\footnotesize 0.01 (0.00)} 
& {\footnotesize 0.25 (0.03)} 
& {\footnotesize 0.01 (0.00)} \\ 

{\footnotesize \texttt{\texttt{rejectionABCselect}} s.d.} 
& {\footnotesize 0.59 (0.06)} 
& {\footnotesize 0.01 (0.00)} 
& {\footnotesize 0.25 (0.04)} 
& {\footnotesize 0.01 (0.00)} \\\hline

{\footnotesize \texttt{copulaABCdrf}} & {\footnotesize d.f.} & {\footnotesize scale} & {\footnotesize d.f.} & 
{\footnotesize scale} \\ 
{\footnotesize Copula d.f. and scale} & {\footnotesize 13.52 (9.78)} & {\footnotesize 0.21 (0.25)} & 
{\footnotesize 7.39 (9.64)} & {\footnotesize 0.10 (0.10)} \\ \hline\hline
\multicolumn{5}{l}{\footnotesize \textit{Note:} \textbf{Bold} indicates the more accurate ABC method for the given estimator.}
\end{tabular}
\end{center}

\begin{changemargin}{-.7in}{-.7in}
\begin{center}
\begin{tabular}{c|ccc|c|cc}
\multicolumn{7}{l}{\footnotesize \textbf{Table 9.} MAE (MSE), mean percent interval coverage (95\%(50\%)c) for Price model over 10 replicas,  under varying $n$ and ${n}_{\text{sim}}$.} \\ \hline\hline
&  & \multicolumn{2}{c|}{{\footnotesize \textit{n} = 50}} &  & 
\multicolumn{2}{|c}{{\footnotesize \textit{n} = 300}} \\ 
{\footnotesize ${n}_{\text{sim}}$} & {\footnotesize Posterior} & 
{\footnotesize ${k}_{0}$} & {\footnotesize ${p}$} & {\footnotesize ${n}_{%
\text{sim}}$} & {\footnotesize ${k}_{0}$} & {\footnotesize ${p}$} \\ \hline
{\footnotesize 5 \ } & \multicolumn{1}{|l}{\footnotesize \texttt{copulaABCdrf} mean} & 
{\footnotesize \textbf{0.00} (\textbf{0.00})} & {\footnotesize \textbf{0.06} (\textbf{0.00})} & {\footnotesize 30} & {\footnotesize \textbf{0.01} (\textbf{0.00})} & {\footnotesize \textbf{0.10} (\textbf{0.01})} \\ 
{\footnotesize } & \multicolumn{1}{|l}{\footnotesize \texttt{\texttt{rejectionABCselect}} mean} & 
{\footnotesize \textbf{0.01}\textsubscript{-.02}(\textbf{0.00})} & {\footnotesize \textbf{0.06}\textsubscript{-.01}(\textbf{0.00})} & {\footnotesize }{\footnotesize} & {\footnotesize \textbf{0.01} (\textbf{0.00})} & {\footnotesize \textbf{0.11} (\textbf{0.01})} \\ 
& \multicolumn{1}{|l}{\footnotesize \texttt{copulaABCdrf} median} & {\footnotesize \textbf{0.01} (\textbf{0.00})} & {\footnotesize \textbf{0.06} (\textbf{0.00})} &  & {\footnotesize \textbf{0.01} (\textbf{0.00})} & 
{\footnotesize \textbf{0.11} (\textbf{0.01})} \\ 
& \multicolumn{1}{|l}{\footnotesize \texttt{rejectionABCselect} median} & {\footnotesize \textbf{0.01}\textsubscript{-.02}(\textbf{0.00})} & {\footnotesize \textbf{0.06}\textsubscript{-.01}(\textbf{0.00})} &  & {\footnotesize \textbf{0.01} (\textbf{0.00})} & 
{\footnotesize \textbf{0.11} (\textbf{0.01})} \\ 
& \multicolumn{1}{|l}{\footnotesize \texttt{copulaABCdrf} modeMLE} & {\footnotesize 0.06
(\textbf{0.00})} & {\footnotesize \textbf{0.07} (\textbf{0.00})} &  & {\footnotesize 0.05 (\textbf{0.00})} & 
{\footnotesize \textbf{0.11} (\textbf{0.01})} \\ 
& \multicolumn{1}{|l}{\footnotesize \texttt{rejectionABCkern.select} modeMLE} & {\footnotesize \textbf{0.04}\textsubscript{-.03}(\textbf{0.00})}
& {\footnotesize \textbf{0.07}(\textbf{0.00}\textsubscript{-.01})} &  & {\footnotesize \textbf{0.04} (\textbf{0.00})} & {\footnotesize \textbf{0.11}\textsubscript{-.01}(\textbf{0.01})} \\ 
& \multicolumn{1}{|l}{\footnotesize \texttt{copulaABCdrf} 95\%(50\%)c} & {\footnotesize \textbf{1.00}
(1.00)} & {\footnotesize \textbf{0.00} (\textbf{0.00})} &  & {\footnotesize \textbf{1.00} (\textbf{1.00})} & 
{\footnotesize \textbf{0.10} (\textbf{0.00})} \\ 
& \multicolumn{1}{|l}{\footnotesize \texttt{rejectionABCselect} 95\%(50\%)c} & {\footnotesize \textbf{0.90}\textsubscript{+.10}(\textbf{0.90}\textsubscript{+.60})} & {\footnotesize \textbf{0.00} (\textbf{0.00})} &  & {\footnotesize \textbf{1.00} (\textbf{1.00})} & {\footnotesize 0.00 (\textbf{0.00})} \\ \hline
{\footnotesize 16} & \multicolumn{1}{|l}{\footnotesize \texttt{copulaABCdrf} mean} & 
{\footnotesize \textbf{0.00} (\textbf{0.00})} & {\footnotesize \textbf{0.05} (\textbf{0.00})} & {\footnotesize 100} & {\footnotesize \textbf{0.01} (\textbf{0.00})} & {\footnotesize \textbf{0.03} (\textbf{0.00})} \\ 
{\footnotesize } & \multicolumn{1}{|l}{\footnotesize \texttt{rejectionABCselect} mean} & 
{\footnotesize \textbf{0.01} (\textbf{0.00})} & {\footnotesize \textbf{0.05} (\textbf{0.00})} & {\footnotesize } & {\footnotesize \textbf{0.01} (\textbf{0.00)}} & {\footnotesize \textbf{0.03} (\textbf{0.00})} \\ 
& \multicolumn{1}{|l}{\footnotesize \texttt{copulaABCdrf} median} & {\footnotesize \textbf{0.01}
(\textbf{0.00})} & {\footnotesize \textbf{0.05} (\textbf{0.00})} &  & {\footnotesize \textbf{0.01} (\textbf{0.00})} & 
{\footnotesize \textbf{0.03} (\textbf{0.00})} \\ 
& \multicolumn{1}{|l}{\footnotesize \texttt{rejectionABCselect} median} & {\footnotesize \textbf{0.01} (\textbf{0.00})} & {\footnotesize \textbf{0.05} (\textbf{0.00})} &  & {\footnotesize \textbf{0.01} (\textbf{0.00})} & {\footnotesize \textbf{0.03} (\textbf{0.00})} \\ 
& \multicolumn{1}{|l}{\footnotesize \texttt{copulaABCdrf} modeMLE} & {\footnotesize 0.06
(\textbf{0.00})} & {\footnotesize \textbf{0.05} (\textbf{0.00})} &  & {\footnotesize \textbf{0.05} (\textbf{0.00})} & 
{\footnotesize \textbf{0.03} (\textbf{0.00})} \\ 
& \multicolumn{1}{|l}{\footnotesize \texttt{rejectionABCkern.select} modeMLE} & {\footnotesize \textbf{0.04}\textsubscript{+.01}(\textbf{0.00})} & {\footnotesize \textbf{0.05} (\textbf{0.00})} &  & {\footnotesize \textbf{0.05} (\textbf{0.00})} & {\footnotesize \textbf{0.03} (\textbf{0.00})} \\ 
& \multicolumn{1}{|l}{\footnotesize \texttt{copulaABCdrf} 95\%(50\%)c} & {\footnotesize \textbf{1.00}
(\textbf{1.00})} & {\footnotesize \textbf{0.00} (\textbf{0.00})} &  & {\footnotesize \textbf{1.00} (\textbf{1.00})} & 
{\footnotesize \textbf{0.20} (\textbf{0.00})} \\ 
& \multicolumn{1}{|l}{\footnotesize \texttt{rejectionABCselect} 95\%(50\%)c} & {\footnotesize \textbf{1.00} (\textbf{1.00})} & {\footnotesize \textbf{0.00} (\textbf{0.00})} &  & {\footnotesize \textbf{1.00} (\textbf{1.00})} & 
{\footnotesize 0.00 (\textbf{0.00})} \\ \hline
{\footnotesize 25} & \multicolumn{1}{|l}{\footnotesize \texttt{copulaABCdrf} mean} & 
{\footnotesize \textbf{0.00} (\textbf{0.00})} & {\footnotesize \textbf{0.03} (\textbf{0.00})} & {\footnotesize 150} & {\footnotesize \textbf{0.01} (\textbf{0.00})} & {\footnotesize \textbf{0.01} (\textbf{0.00})} \\ 
{\footnotesize }{\footnotesize} & \multicolumn{1}{|l}{\footnotesize \texttt{rejectionABCselect} mean} & {\footnotesize \textbf{0.00}\textsubscript{-.01}(\textbf{0.00})} & {\footnotesize \textbf{0.03} (\textbf{0.00})} & {\footnotesize } & {\footnotesize \textbf{0.01} (\textbf{0.00})} & {\footnotesize 0.02 (\textbf{0.00})} \\ 
& \multicolumn{1}{|l}{\footnotesize \texttt{copulaABCdrf} median} & {\footnotesize \textbf{0.00}
(\textbf{0.00})} & {\footnotesize \textbf{0.03} (\textbf{0.00})} &  & {\footnotesize \textbf{0.01} (\textbf{0.00})} & 
{\footnotesize \textbf{0.01} (\textbf{0.00})} \\ 
& \multicolumn{1}{|l}{\footnotesize \texttt{rejectionABCselect} median} & {\footnotesize \textbf{0.01} (\textbf{0.00})} & {\footnotesize \textbf{0.03} (\textbf{0.00})} &  & {\footnotesize \textbf{0.01} (\textbf{0.00})} & 
{\footnotesize 0.02 (\textbf{0.00})} \\ 
& \multicolumn{1}{|l}{\footnotesize \texttt{copulaABCdrf} modeMLE} & {\footnotesize 0.05
(\textbf{0.00})} & {\footnotesize \textbf{0.03} (\textbf{0.00})} &  & {\footnotesize 0.05 (\textbf{0.00})} & 
{\footnotesize \textbf{0.01} (\textbf{0.00})} \\ 
& \multicolumn{1}{|l}{\footnotesize \texttt{rejectionABCkern.select} modeMLE} & {\footnotesize \textbf{0.04} (\textbf{0.00})} & {\footnotesize 0.04\textsubscript{+.01}(\textbf{0.00})} &  & {\footnotesize \textbf{0.03}\textsubscript{-.01}(\textbf{0.00})} & {\footnotesize \textbf{0.01} (\textbf{0.00})} \\ 
& \multicolumn{1}{|l}{\footnotesize \texttt{copulaABCdrf} 95\%(50\%)c} & {\footnotesize \textbf{1.00}
(\textbf{1.00})} & {\footnotesize \textbf{0.40} (\textbf{0.00})} &  & {\footnotesize \textbf{1.00} (\textbf{1.00})} & 
{\footnotesize 0.00 (\textbf{0.00})} \\ 
& \multicolumn{1}{|l}{\footnotesize \texttt{rejectionABCselect} 95\%(50\%)c} & {\footnotesize \textbf{1.00} (\textbf{1.00})} & {\footnotesize 0.20 (\textbf{0.00})} &  & {\footnotesize \textbf{1.00} (\textbf{1.00})} & {\footnotesize \textbf{0.30}\textsubscript{+.20}(\textbf{0.00})} \\ \hline
{\footnotesize 37} & \multicolumn{1}{|l}{\footnotesize \texttt{copulaABCdrf} mean} & 
{\footnotesize \textbf{0.00} (\textbf{0.00})} & {\footnotesize \textbf{0.02} (\textbf{0.00})} & {\footnotesize 225} & {\footnotesize \textbf{0.01} (\textbf{0.00})} & {\footnotesize \textbf{0.01} (\textbf{0.00})} \\ 
{\footnotesize }{\footnotesize} & \multicolumn{1}{|l}{\footnotesize \texttt{rejectionABC} mean} & {\footnotesize \textbf{0.00} (\textbf{0.00})} & {\footnotesize \textbf{0.02} (\textbf{0.00})} & 
{\footnotesize }{\footnotesize} & {\footnotesize \textbf{0.01} (\textbf{0.00})} & 
{\footnotesize \textbf{0.01} (\textbf{0.00})} \\ 
& \multicolumn{1}{|l}{\footnotesize \texttt{copulaABCdrf} median} & {\footnotesize \textbf{0.00}
(\textbf{0.00})} & {\footnotesize \textbf{0.02} (\textbf{0.00})} &  & {\footnotesize \textbf{0.01} (\textbf{0.00})} & 
{\footnotesize \textbf{0.01} (\textbf{0.00})} \\ 
& \multicolumn{1}{|l}{\footnotesize \texttt{rejectionABC} median} & {\footnotesize \textbf{0.01} (\textbf{0.00})} & {\footnotesize \textbf{0.02} (\textbf{0.00})} &  & {\footnotesize \textbf{0.01} (\textbf{0.00})} & {\footnotesize \textbf{0.01} (\textbf{0.00})} \\ 
& \multicolumn{1}{|l}{\footnotesize \texttt{copulaABCdrf} modeMLE} & {\footnotesize 0.05 (\textbf{0.00})} & {\footnotesize \textbf{0.02} (\textbf{0.00})} &  & {\footnotesize \textbf{0.05} (\textbf{0.00})} & 
{\footnotesize \textbf{0.01} (\textbf{0.00})} \\ 
& \multicolumn{1}{|l}{\footnotesize \texttt{rejectionABCkern.select} modeMLE} & {\footnotesize \textbf{0.04} (\textbf{0.00})} & {\footnotesize \textbf{0.02} (\textbf{0.00})} &  & {\footnotesize \textbf{0.05}\textsubscript{+.02}(0.00)} & {\footnotesize \textbf{0.01} (\textbf{0.00})} \\ 
& \multicolumn{1}{|l}{\footnotesize \texttt{copulaABCdrf} 95\%(50\%)c} & {\footnotesize \textbf{1.00}
(\textbf{1.00})} & {\footnotesize \textbf{0.80} (\textbf{0.10})} &  & {\footnotesize \textbf{1.00} (\textbf{1.00})} & 
{\footnotesize \textbf{0.60} (\textbf{0.00})} \\ 
& \multicolumn{1}{|l}{\footnotesize \texttt{rejectionABCselect} 95\%(50\%)c} & {\footnotesize \textbf{1.00}
(\textbf{1.00})} & {\footnotesize \textbf{0.80}\textsubscript{+.10}(\textbf{0.10})} &  & {\footnotesize \textbf{1.00} (\textbf{1.00})} & 
{\footnotesize 0.40\textsubscript{-.20}(\textbf{0.00})} \\ \hline
{\footnotesize 50} & \multicolumn{1}{|l}{\footnotesize \texttt{copulaABCdrf} mean} & 
{\footnotesize \textbf{0.01} (\textbf{0.00})} & {\footnotesize \textbf{0.02} (\textbf{0.00})} & {\footnotesize 300} & {\footnotesize \textbf{0.01} (\textbf{0.00})} & {\footnotesize \textbf{0.00} (\textbf{0.00})} \\ 
{\footnotesize } & \multicolumn{1}{|l}{\footnotesize \texttt{rejectionABCselect} mean} & 
{\footnotesize \textbf{0.01} (\textbf{0.00})} & {\footnotesize \textbf{0.02} (\textbf{0.00})} & {\footnotesize } & {\footnotesize \textbf{0.01} (\textbf{0.00})} & {\footnotesize \textbf{0.00} (\textbf{0.00})} \\ 
& \multicolumn{1}{|l}{\footnotesize \texttt{copulaABCdrf} median} & {\footnotesize \textbf{0.01}
(\textbf{0.00})} & {\footnotesize \textbf{0.02} (\textbf{0.00})} &  & {\footnotesize \textbf{0.01} (\textbf{0.00})} & 
{\footnotesize \textbf{0.00} (\textbf{0.00})} \\ 
& \multicolumn{1}{|l}{\footnotesize \texttt{rejectionABCselect} median} & {\footnotesize \textbf{0.01} (\textbf{0.00})} & {\footnotesize \textbf{0.02} (\textbf{0.00})} &  & {\footnotesize \textbf{0.01} (\textbf{0.00})} & 
{\footnotesize \textbf{0.00} (\textbf{0.00})} \\ 
& \multicolumn{1}{|l}{\footnotesize \texttt{copulaABCdrf} modeMLE} & {\footnotesize 0.05
(\textbf{0.00})} & {\footnotesize 0.12 (0.10)} &  & {\footnotesize 0.05 (\textbf{0.00})} & 
{\footnotesize \textbf{0.00} (\textbf{0.00})} \\ 
& \multicolumn{1}{|l}{\footnotesize \texttt{rejectionABCkern.select} modeMLE} & {\footnotesize \textbf{0.04}\textsubscript{+.01}(0.00)} & {\footnotesize \textbf{0.02} (\textbf{0.00})} &  & {\footnotesize \textbf{0.03} (\textbf{0.00})} & {\footnotesize \textbf{0.00} (\textbf{0.00})} \\ 
& \multicolumn{1}{|l}{\footnotesize \texttt{copulaABCdrf} 95\%(50\%)c} & {\footnotesize \textbf{1.00}
(\textbf{1.00})} & {\footnotesize \textbf{1.00} (\textbf{0.50})} &  & {\footnotesize \textbf{1.00} (\textbf{1.00})} & 
{\footnotesize \textbf{1.00} (\textbf{0.80})} \\ 
& \multicolumn{1}{|l}{\footnotesize \texttt{rejectionABCselect} 95\%(50\%)c} & {\footnotesize \textbf{1.00}
(\textbf{1.00})} & {\footnotesize 0.80\textsubscript{-.10}(\textbf{0.50}\textsubscript{+.10})} &  & {\footnotesize \textbf{1.00} (\textbf{1.00})} & {\footnotesize \textbf{1.00} (0.90)} \\\hline\hline
\multicolumn{7}{l}{\footnotesize \textit{Note:}  Subscript is change in MAE (MSE) based on \texttt{drf} selecting the 2 best of the 3 summaries.}\\
\multicolumn{7}{l}{\footnotesize \textbf{Bold} indicates the more accurate ABC method for the given estimator.}
\end{tabular}
\end{center}
\end{changemargin}

\begin{changemargin}{-.7in}{-.7in}
\begin{center}
\begin{tabular}{c|ccc|c|cc}
\multicolumn{7}{l}{\footnotesize \textbf{Table 10.} MAE (MSE), mean credible interval coverage (95\%(50\%)c) for NLPA model over 10 replicas, varying $n$ and ${n}_{\text{sim}}$.}\\\hline\hline
&  & \multicolumn{2}{c|}{{\footnotesize \textit{n} = 50}} &  & \multicolumn{2}{|c}{%
{\footnotesize \textit{n} = 300}} \\ 
{\footnotesize ${n}_{\text{sim}}$} & {\footnotesize Posterior} & {\footnotesize ${\alpha }$} & 
{\footnotesize ${p}$} & {\footnotesize ${n}_{\text{sim}}$} & {\footnotesize ${\alpha }$} & 
{\footnotesize ${p}$} \\ \hline
{\footnotesize 5} & \multicolumn{1}{|l}{\footnotesize \texttt{copulaABCdrf} mean} & {\footnotesize \textbf{0.20} (\textbf{0.04})} & 
{\footnotesize \textbf{0.07} (\textbf{0.00})} & {\footnotesize 30} & {\footnotesize 0.67 (0.48)} & {\footnotesize 0.02 (\textbf{0.00})} \\ 
{\footnotesize }{\footnotesize } & \multicolumn{1}{|l}{\footnotesize \texttt{rejectionABCselect} mean} & {\footnotesize 0.29\textsubscript{+.01}(0.09\textsubscript{+.01})} & {\footnotesize \textbf{0.07}\textsubscript{+.01}(\textbf{0.00})} & {\footnotesize }{\footnotesize } & {\footnotesize \textbf{0.07}\textsubscript{-.03}(\textbf{0.01})} & {\footnotesize \textbf{0.01} (\textbf{0.00})} \\ 
& \multicolumn{1}{|l}{\footnotesize \texttt{copulaABCdrf} median} & {\footnotesize 0.37 (0.14)} & {\footnotesize %
\textbf{0.06} (\textbf{0.00})} &  & {\footnotesize 0.65 (0.46)} & {\footnotesize 0.02 (\textbf{0.00})} \\ 
& \multicolumn{1}{|l}{\footnotesize \texttt{rejectionABCselect} median} & {\footnotesize \textbf{0.29}\textsubscript{-.16}(\textbf{0.08}\textsubscript{-.12})} & {\footnotesize \textbf{0.06}\textsubscript{+.01}(\textbf{0.00})} &  & {\footnotesize \textbf{0.07}\textsubscript{-.05}(\textbf{0.01}\textsubscript{-.01})} & {\footnotesize \textbf{0.01}\textsubscript{-.01}(\textbf{0.00})} \\ 
& \multicolumn{1}{|l}{\footnotesize \texttt{copulaABCdrf} modeMLE} & {\footnotesize 0.76 (0.68)} & 
{\footnotesize 0.10 (0.01)} &  & {\footnotesize 0.55 (0.37)} & {\footnotesize 0.02 (\textbf{0.00})} \\ 
& \multicolumn{1}{|l}{\footnotesize \texttt{rejectionABCkern.select} modeMLE} & {\footnotesize \textbf{0.61}\textsubscript{-.27}(\textbf{0.52}\textsubscript{-.25})} & 
{\footnotesize \textbf{0.02}\textsubscript{+.01}(\textbf{0.00})} &  & {\footnotesize \textbf{0.21}\textsubscript{-.15}(\textbf{0.06}\textsubscript{-.11})} & {\footnotesize \textbf{0.01}\textsubscript{-.01}(\textbf{0.00})} \\ 
& \multicolumn{1}{|l}{\footnotesize \texttt{copulaABCdrf}95\%(50\%)c} & {\footnotesize \textbf{1.00} (\textbf{1.00})} & 
{\footnotesize \textbf{1.00} (\textbf{0.00})} &  & {\footnotesize \textbf{1.00} (\textbf{0.00})} & {\footnotesize 0.00 (0.00)} \\ 
& \multicolumn{1}{|l}{\footnotesize \texttt{rejectionABCselect} 95\%(50\%)c} & {\footnotesize \textbf{1.00} (\textbf{1.00})} & 
{\footnotesize \textbf{0.90}\textsubscript{-.10}(\textbf{0.00})} &  & {\footnotesize \textbf{1.00} (\textbf{1.00})} & {\footnotesize \textbf{1.00} (\textbf{0.30}\textsubscript{-.70})} \\
\hline
{\footnotesize 16} & \multicolumn{1}{|l}{\footnotesize \texttt{copulaABCdrf} mean} & {\footnotesize 0.35 (0.16)}
& {\footnotesize 0.01 (\textbf{0.00})} & {\footnotesize 100} & {\footnotesize 0.09 (\textbf{0.01})} & {\footnotesize \textbf{0.00}
(\textbf{0.00})} \\ 
{\footnotesize }{\footnotesize } & \multicolumn{1}{|l}{\footnotesize \texttt{rejectionABCselect} mean} & {\footnotesize \textbf{0.25}\textsubscript{-.14}(\textbf{0.07}\textsubscript{-.14})} & {\footnotesize \textbf{0.00}\textsubscript{-.01}(\textbf{0.00})} & {\footnotesize }{\footnotesize } & {\footnotesize \textbf{0.08}\textsubscript{-.23}(\textbf{0.01}\textsubscript{-.11})} & {\footnotesize \textbf{0.00}\textsubscript{-.02}(\textbf{0.00})} \\ 
& \multicolumn{1}{|l}{\footnotesize \texttt{copulaABCdrf} median} & {\footnotesize 0.40 (0.21)} & {\footnotesize 0.01 (\textbf{0.00})} &  & {\footnotesize \textbf{0.08} (\textbf{0.01})} & {\footnotesize \textbf{0.00} (\textbf{0.00})} \\ 
& \multicolumn{1}{|l}{\footnotesize \texttt{rejectionABCselect} median} & {\footnotesize \textbf{0.26}\textsubscript{-.18}(\textbf{0.08}\textsubscript{-.19})} & {\footnotesize \textbf{0.00}\textsubscript{-.01}(\textbf{0.00})} &  & {\footnotesize \textbf{0.09}\textsubscript{-.24}(\textbf{0.01}\textsubscript{-.13})} & {\footnotesize \textbf{0.00}\textsubscript{-.02}(\textbf{0.00})} \\ 
& \multicolumn{1}{|l}{\footnotesize \texttt{copulaABCdrf} modeMLE} & {\footnotesize 0.61 (0.53)} & 
{\footnotesize 0.02 (\textbf{0.00})} &  & {\footnotesize 0.16 (0.05)} & {\footnotesize \textbf{0.00} (\textbf{0.00})} \\ 
& \multicolumn{1}{|l}{\footnotesize \texttt{rejectionABCkern.select} modeMLE} & {\footnotesize \textbf{0.34}\textsubscript{-.24}(\textbf{0.14}\textsubscript{-.27})} & {\footnotesize \textbf{0.01}\textsubscript{+.01}(\textbf{0.00})} &  & {\footnotesize \textbf{0.09}\textsubscript{-.29}(\textbf{0.01}\textsubscript{-.17})} & {\footnotesize \textbf{0.00}\textsubscript{-.02}(\textbf{0.00})} \\ 
& \multicolumn{1}{|l}{\footnotesize \texttt{copulaABCdrf}95\%(50\%)c} & {\footnotesize \textbf{1.00} (\textbf{0.30})} & 
{\footnotesize \textbf{1.00} (\textbf{0.30})} &  & {\footnotesize \textbf{1.00} (\textbf{0.80})} & {\footnotesize \textbf{1.00} (\textbf{0.00})} \\ 
& \multicolumn{1}{|l}{\footnotesize \texttt{rejectionABCselect} 95\%(50\%)c} & {\footnotesize \textbf{1.00} (1.00\textsubscript{+.70})} & 
{\footnotesize \textbf{1.00}(1.00\textsubscript{+.10})} &  & \multicolumn{1}{|c}{\footnotesize \textbf{1.00}\textsubscript{+.10}(\textbf{0.80}\textsubscript{+.80})} & {\footnotesize \textbf{1.00}\textsubscript{+.20}(\textbf{1.00}\textsubscript{+1.00})} \\ \hline
{\footnotesize 25 } & \multicolumn{1}{|l}{\footnotesize \texttt{copulaABCdrf} mean} & {\footnotesize 0.23 (0.08)}
& {\footnotesize \textbf{0.01} (\textbf{0.00})} & {\footnotesize 150} & {\footnotesize 0.12 (0.02)} & {\footnotesize \textbf{0.00} (\textbf{0.00})} \\ 
{\footnotesize }{\footnotesize } & \multicolumn{1}{|l}{\footnotesize \texttt{rejectionABCselect} mean} & {\footnotesize \textbf{0.22}\textsubscript{-.13}(\textbf{0.07}\textsubscript{-.08})} & {\footnotesize \textbf{0.01}\textsubscript{-.01}(\textbf{0.00})} & {\footnotesize }{\footnotesize } & {\footnotesize \textbf{0.10}\textsubscript{-.14}(\textbf{0.01}\textsubscript{-.06})} & {\footnotesize \textbf{0.00}\textsubscript{-.01}(\textbf{0.00})} \\ 
& \multicolumn{1}{|l}{\footnotesize \texttt{copulaABCdrf} median} & {\footnotesize \textbf{0.23} (0.08)} & {\footnotesize 0.02 (\textbf{0.00})} &  & {\footnotesize 0.11 (0.02)} & {\footnotesize \textbf{0.00} (\textbf{0.00})} \\ 
& \multicolumn{1}{|l}{\footnotesize \texttt{rejectionABCselect} median} & {\footnotesize 0.24\textsubscript{-.14}(\textbf{0.07}\textsubscript{-.12})} & {\footnotesize \textbf{0.00}\textsubscript{-.01}(\textbf{0.00})} &  & {\footnotesize \textbf{0.10}\textsubscript{-.13}(\textbf{0.01}\textsubscript{-.06})} & {\footnotesize \textbf{0.00}\textsubscript{-.01}(\textbf{0.00})} \\ 
& \multicolumn{1}{|l}{\footnotesize \texttt{copulaABCdrf} modeMLE} & {\footnotesize 0.41 (0.25)} & {\footnotesize \textbf{0.01} (\textbf{0.00})} &  & {\footnotesize \textbf{0.12} (\textbf{0.02})} & {\footnotesize \textbf{0.00} (\textbf{0.00})} \\ 
& \multicolumn{1}{|l}{\footnotesize \texttt{rejectionABCkern.select} modeMLE} & {\footnotesize \textbf{0.34}\textsubscript{-.10}(0.15\textsubscript{-.13})} & 
{\footnotesize \textbf{0.01} (\textbf{0.00})} &  & {\footnotesize \textbf{0.12}\textsubscript{-.10}(\textbf{0.02}\textsubscript{-.05})} & {\footnotesize 0.01 (\textbf{0.00})} \\ 
& \multicolumn{1}{|l}{\footnotesize \texttt{copulaABCdrf}95\%(50\%)c} & {\footnotesize \textbf{1.00} (\textbf{0.70})} & 
{\footnotesize 0.60 (0.00)} &  & {\footnotesize \textbf{1.00} (\textbf{0.50})} & {\footnotesize \textbf{1.00} (\textbf{0.40})} \\ 
& \multicolumn{1}{|l}{\footnotesize \texttt{rejectionABCselect} 95\%(50\%)c} & {\footnotesize \textbf{1.00} (1.00\textsubscript{+.70})} & 
{\footnotesize \textbf{1.00} (\textbf{0.90}\textsubscript{+.40})} &  & {\footnotesize \textbf{1.00}\textsubscript{+.30}(0.60\textsubscript{+.30})} & {\footnotesize \textbf{1.00}\textsubscript{+.30}(1.00\textsubscript{+.90})} \\ 
\hline
{\footnotesize 37} & \multicolumn{1}{|l}{\footnotesize \texttt{copulaABCdrf} mean} & {\footnotesize \textbf{0.21} (\textbf{0.05})}
& {\footnotesize 0.01 (\textbf{0.00})} & {\footnotesize 225} & {\footnotesize \textbf{0.05} (\textbf{0.00})} & {\footnotesize \textbf{0.00} (\textbf{0.00})} \\ 
{\footnotesize }{\footnotesize } & \multicolumn{1}{|l}{\footnotesize \texttt{rejectionABCselect} mean} & {\footnotesize 0.24\textsubscript{-.04}(0.08\textsubscript{-.04})} & {\footnotesize \textbf{0.00}\textsubscript{-.01}(\textbf{0.00})} & {\footnotesize }{\footnotesize} & {\footnotesize 0.06\textsubscript{-.09}(\textbf{0.00}\textsubscript{-.03})} & {\footnotesize \textbf{0.00} (\textbf{0.00})} \\ 
& \multicolumn{1}{|l}{\footnotesize \texttt{copulaABCdrf} median} & {\footnotesize \textbf{0.21} (\textbf{0.05})} & {\footnotesize 0.01 (\textbf{0.00})} &  & {\footnotesize \textbf{0.05} (\textbf{0.00})} & {\footnotesize \textbf{0.00} (\textbf{0.00})} \\ 
& \multicolumn{1}{|l}{\footnotesize \texttt{rejectionABCselect} median} & {\footnotesize 0.25\textsubscript{-.07}(0.09\textsubscript{-.07})} & {\footnotesize \textbf{0.00}\textsubscript{-.01}(\textbf{0.00})} &  & {\footnotesize \textbf{0.06}\textsubscript{-.08}(\textbf{0.00}\textsubscript{-.02})} & {\footnotesize \textbf{0.00} (\textbf{0.00})} \\ 
& \multicolumn{1}{|l}{\footnotesize \texttt{copulaABCdrf} modeMLE} & {\footnotesize 0.28 (0.11)} & 
{\footnotesize \textbf{0.01} (\textbf{0.00})} &  & {\footnotesize 0.11 (0.04)} & {\footnotesize \textbf{0.00} (\textbf{0.00})} \\ 
& \multicolumn{1}{|l}{\footnotesize \texttt{rejectionABCkern.select} modeMLE} & {\footnotesize \textbf{0.24}\textsubscript{-.24}(\textbf{0.09}\textsubscript{-.21})} & 
{\footnotesize \textbf{0.01} (\textbf{0.00})} &  & {\footnotesize \textbf{0.06}\textsubscript{-.10}(\textbf{0.01}\textsubscript{-.03})} & {\footnotesize \textbf{0.00}\textsubscript{-.01}(\textbf{0.00})} \\ 
& \multicolumn{1}{|l}{\footnotesize \texttt{copulaABCdrf}95\%(50\%)c} & {\footnotesize \textbf{1.00} (0.70)} & 
{\footnotesize 0.50 (\textbf{0.10})} &  & {\footnotesize \textbf{1.00} (\textbf{0.80})} & {\footnotesize \textbf{1.00} (\textbf{0.90})} \\ 
& \multicolumn{1}{|l}{\footnotesize \texttt{rejectionABCselect} 95\%(50\%)c} & {\footnotesize \textbf{1.00} (\textbf{0.60})} & 
{\footnotesize \textbf{1.00} (0.90\textsubscript{+.10})} &  & {\footnotesize \textbf{1.00} (\textbf{0.80}\textsubscript{+.30})} & {\footnotesize \textbf{1.00}\textsubscript{+.20}(1.00\textsubscript{+.20})} \\ 
\hline
{\footnotesize 50} & \multicolumn{1}{|l}{\footnotesize \texttt{copulaABCdrf} mean} & {\footnotesize \textbf{0.11} (\textbf{0.02})}
& {\footnotesize \textbf{0.00} (\textbf{0.00})} & {\footnotesize 300} & {\footnotesize \textbf{0.04} (\textbf{0.00})} & {\footnotesize \textbf{0.00} (\textbf{0.00})} \\ 
{\footnotesize}{\footnotesize} & \multicolumn{1}{|l}{\footnotesize \texttt{rejectionABCselect} mean} & {\footnotesize 0.17\textsubscript{-.05}(0.03\textsubscript{-.03})} & {\footnotesize \textbf{0.00}\textsubscript{-.01}(\textbf{0.00})} & {\footnotesize}{\footnotesize} & {\footnotesize 0.05\textsubscript{-.10}(\textbf{0.00}\textsubscript{-.02})} & {\footnotesize \textbf{0.00}\textsubscript{-.01}(\textbf{0.00})} \\ 
& \multicolumn{1}{|l}{\footnotesize \texttt{copulaABCdrf} median} & {\footnotesize \textbf{0.10} (\textbf{0.02})} & {\footnotesize \textbf{0.00} (\textbf{0.00})} &  & {\footnotesize \textbf{0.04} (\textbf{0.00})} & {\footnotesize \textbf{0.00} (\textbf{0.00})} \\ 
& \multicolumn{1}{|l}{\footnotesize \texttt{rejectionABCselect} median} & {\footnotesize 0.16\textsubscript{-.06}(0.03\textsubscript{-.04})} & {\footnotesize \textbf{0.00}\textsubscript{-.01}(\textbf{0.00})} &  & {\footnotesize 0.05\textsubscript{-.09}(\textbf{0.00}\textsubscript{-.02}}) & {\footnotesize \textbf{0.00}\textsubscript{-.01}(0.00)} \\ 
& \multicolumn{1}{|l}{\footnotesize \texttt{copulaABCdrf} modeMLE} & {\footnotesize 0.26 (0.09)} & 
{\footnotesize 0.12 (0.14)} &  & {\footnotesize 0.07 (\textbf{0.01})} & {\footnotesize \textbf{0.00} (\textbf{0.00})} \\ 
& \multicolumn{1}{|l}{\footnotesize \texttt{rejectionABCkern.select} modeMLE} & {\footnotesize \textbf{0.18}\textsubscript{-.05}(\textbf{0.05}\textsubscript{-.03})} & 
{\footnotesize \textbf{0.01} (\textbf{0.00})} &  & {\footnotesize \textbf{0.05}\textsubscript{-.12}(\textbf{0.01}\textsubscript{-.03})} & {\footnotesize \textbf{0.00}\textsubscript{-.01}(0.00)} \\ 
& \multicolumn{1}{|l}{\footnotesize \texttt{copulaABCdrf}95\%(50\%)c} & {\footnotesize \textbf{1.00} (0.80)} & 
{\footnotesize \textbf{0.90} (\textbf{0.60})} &  & {\footnotesize \textbf{1.00} (\textbf{0.80})} & {\footnotesize \textbf{1.00} (\textbf{0.60})} \\ 
& \multicolumn{1}{|l}{\footnotesize \texttt{rejectionABCselect} 95\%(50\%)c} & {\footnotesize \textbf{1.00} (\textbf{0.70}\textsubscript{+.10})} & 
{\footnotesize \textbf{1.00} (0.90\textsubscript{+.40})} &  & {\footnotesize \textbf{1.00} (\textbf{0.80}\textsubscript{+.20})} 
& {\footnotesize \textbf{1.00}\textsubscript{+.10}(1.00\textsubscript{+.40})} \\ 
\hline\hline
\multicolumn{7}{l}{\footnotesize \textit{Note:}  Subscript is change in MAE (MSE) based on \texttt{drf} selecting the 2 best of the 3 summaries.}\\
\multicolumn{7}{l}{\footnotesize \textbf{Bold} indicates the more accurate ABC method for the given estimator.}\\
\multicolumn{7}{l}{\footnotesize }\\
\end{tabular}
\end{center}
\end{changemargin}

\subsection{Simulation Study: Duplication-Divergence Models}

The DMC model and the DMR model each grows an undirected network of $n$ nodes by starting with a seed network of $n^{\prime}\geq1$ nodes, and then repeats the following steps until requisite number of nodes ($n=n^{\prime}$) is reached: Add new node ($n^{\prime}\equiv n^{\prime}+1$), then add edge between the new node and each neighbor of a randomly chosen existing node, $k\sim\mathcal{U}_{\{1,n^{\prime}-1\}}$. Then, for the DMC model, the remaining steps are as follows: for each neighbor of the chosen node, randomly select either the edge between the chosen node and the neighbor, or the edge between the new node and the neighbor, and remove that edge with probability $q_{\operatorname{mod}}$; and then add an edge between the chosen node and the new node with probability $q_{\text{con}}$. Alternatively, for the DMR model, the remaining steps are as follows: each edge connected to the new node is removed independently with probability $q_{\text{del}}$, and an edge between any existing node and the new node is added with probability $q_{\text{new}}/(n^{\prime}-1)$.

Now we consider simulation studies which evaluate and compare the ability of  \texttt{copulaABCdrf} (Algorithm 1) and \texttt{rejectionABC} to estimate the posterior distribution of the parameters of the DMC model, and the parameters of the posterior distribution of the DMR model. As before, the simulation studies are based on 10 replications of $n=50$ node undirected networks from the DMC model (undirected networks from the DMR model, respectively); and of $n=300$ node undirected networks simulated from the DMC model (undirected networks from the DMR model, respectively). These simulations are done for each of five conditions of $n_{\text{sim}}$, being at least roughly $10\%$, $33\%$, $50\%$, $75\%$ and $100\%$ of $n$ network nodes. Specifically, the conditions $n_{\text{sim}}=5, 16, 25, 37$ and $50$ for $n = 50$ nodes, and $n_{\text{sim}}= 30, 100, 150, 225$ and $300$ for $n = 300$ nodes. The true data-generating parameters for the DMC model were set as $\theta_0^{\intercal}=(q_{\operatorname{mod}},q_{\text{con}})=(0.20,0.10)$, while the true data-generating parameters of the DMR model were $\theta_0^{\intercal}=(q_{\text{del}},q_{\text{new}})=(0.20,0.10)$.

For each network dataset $x$ simulated from the DMC model (DMR model, respectively) and analyzed by \texttt{copulaABCdrf} and \texttt{rejectionABCdrf}, the DMC model (DMR model, resp.) was assigned a uniform prior $(q_{\operatorname{mod}},q_{\text{con}})\sim\mathcal{U}_{(0.15,0.35)}\mathcal{U}_{(0,1)}$ (uniform prior $(q_{\text{del}},q_{\text{new}})\sim\mathcal{U}_{(0.15,0.35)}\mathcal{U}_{(0,1)}$, resp.) and using vector of summaries $s(\cdot)$ specified by network size invariant offset MPLEs of undirected network summary statistics of mean degree and triangles (of local and global average clustering coefficients, and degree assortativity, resp.). The MPLE summaries $s(x)$ of each network $x$ used offsets based on $n$ nodes ($n=50$ or $300$), while the MPLE summaries $s(y)$ of each network $y$ of smaller size $n_{\text{sim}}<n$ (being $n_{\text{sim}}=25$ for $n=50$, and $n_{\text{sim}}=100$ for $n=300$) were simulated in each iteration of the ABC algorithm using offsets based on $n_{\text{sim}}$ nodes.

Table 11 presents some representative detailed results of the simulation study of \texttt{copulaABCdrf} and \texttt{rejectionABC}, for the DMC and DMR models, for pairs of values of $n_{\text{sim}}$ and $n$. Specifically, for the DMC model, $n_{\text{sim}}=50$ for $n=50$, and $n_{\text{sim}}=300$ for $n=300$; and for the DMR model, $n_{\text{sim}}=50$ for $n=150$, and $n_{\text{sim}}=300$ for $n=300$. Tables 12 and 13 present the results of MAE, MSE, accuracy of 95\% and 50\% posterior interval credible interval estimates of \texttt{copulaABCdrf} and \texttt{rejectionABC} for the DMC and DMR models, for each of the five $n_{\text{sim}}$ conditions within each of $n=50$ and $n=300$.

In particular, since for the DMR models, there were 3 summary statistics relative to the two models parameters, we not only considered for each model implementations of \texttt{rejectionABC} using all 3 summary statistics. We also considered, for comparison purposes, each implementation of \texttt{rejectionABC} based on pre-selecting two of the most important of the 3 summary statistics, according to the results obtained from a variable importance analysis, based on training a \texttt{drf} multivariate regression on a Reference Table of simulated model parameters on all 3 summary statistics (covariate variables). For this network model, \texttt{drf} variable importance analyses found that over all ten replicas of network datasets under the $n_{\text{sim}}=16 $ and $n=50$ condition, that the network size invariant offset MPLEs of global average clustering coefficients (importance measure of 1) and degree assortativity (importance measure of 0.60) were the two most important summary statistics. Meanwhile, for each of the all other simulation conditions of $n_{\text{sim}}$ and $n$, \texttt{drf} variable importance analyses found that over all ten replicas of network datasets, the network size invariant offset MPLEs of undirected network summary statistics, of local and global average clustering coefficients, were always the most important two summaries of the 3 total summary statistics. 

From the results of Table 12, on the DMC model, it can be concluded that overall \texttt{copulaABCdrf} and \texttt{rejectionABC} performed fairly similarly. Specifically, for the parameter $q_\text{mod}$ and $n=50$ nodes, \texttt{copulaABCdrf} slightly outperformed \texttt{rejectionABC} in MAE of posterior mean and median estimation, and in accuracy in estimation of the 50\% credible interval, while usually being around the nominal rate. Also, \texttt{rejectionABC} slightly outperformed \texttt{copulaABCdrf} in MSE of posterior mean estimation, and in MAE and MSE in the estimation of the posterior mode and MLE. For the parameter $q_\text{con}$ and $n=50$ nodes, \texttt{copulaABCdrf} outperformed \texttt{rejectionABC} in MAE and MSE of estimation of the posterior mean, median, mode and MLE, as well as accuracy in estimation of the 95\% credible interval and the 50\% credible interval, while usually being around the nominal rate in each case, especially for $n_\text{sim} = 37$ and $50$. For the parameter $q_\text{mod}$ and $n=300$ nodes, \texttt{copulaABCdrf} slightly outperformed \texttt{rejectionABC} in accuracy in estimation of the 50\% credible interval, while often not being around the nominal rate. Further, \texttt{rejectionABC} slightly outperformed \texttt{copulaABCdrf} in MSE of the estimation of the posterior mode and MLE. For the parameter $q_\text{con}$ and $n=300$ nodes, \texttt{copulaABCdrf} outperformed \texttt{rejectionABC} in MAE and MSE of the estimation of the posterior median, median, mode, and the MLE.

From the results of Table 13, on the DMR model, it can be concluded that, overall, \texttt{copulaABCdrf} and \texttt{rejectionABC} performed rather similarly. Meanwhile, \texttt{rejectionABC} based on using \texttt{drf} to preselect two of the three most important summaries, produced results that often noticeably improved on results of \texttt{rejectionABC} using all three summaries. More specifically, for the parameter $q_\text{del}$ and $n=50$ nodes, \texttt{copulaABCdrf} slightly outperformed \texttt{rejectionABC} in MSE of posterior mean estimation, MAE and MSE of posterior mode estimation, and in accuracy of the 50\% credible interval estimation sometimes being close to the nominal rate. For the parameter $q_\text{new}$ and $n=50$ nodes, \texttt{rejectionABC} outperformed \texttt{copulaABCdrf} in MAE and MSE in the estimation of the posterior mean, median, and mode, MSE in posterior median estimation, and slightly outperformed in accuracy of the 95\% credible interval estimation while usually being close to the nominal rate. Also, \texttt{copulaABCdrf} outperformed \texttt{rejectionABC} in accuracy of the 50\% credible interval estimation, while usually being not close to the nominal rate. For the parameter $q_\text{del}$ and $n=300$ nodes, \texttt{copulaABCdrf} slightly outperformed \texttt{rejectionABC} in MSE of posterior mean estimation, MAE in the estimation of the posterior median, mode, and the MLE, and in accuracy of the 50\% credible interval estimation, while sometimes being rather near the nominal rate. Meanwhile, \texttt{rejectionABC} slightly outperformed \texttt{copulaABCdrf} in terms of MSE in estimation of the posterior mode and MLE. Finally, for the parameter $q_\text{new}$ and $n=300$ nodes, \texttt{rejectionABC} slightly outperformed \texttt{copulaABCdrf} in MAE and MSE in estimation of the posterior mode and the MLE, and in accuracy of  95\% credible interval estimation, while usually being close to the nominal rate.  

Tables 12 and 13 showed that, for each of the \texttt{copulaABCdrf} and \texttt{rejectionABC} methods, as $n_{\text{sim}}$ gets smaller relative to $n$, that at least roughly and on the most part, MAE and MSE of estimates tend to get smaller, and the 95\% and 50\% posterior credible intervals more closely approach their respective nominal rates. Intuitively, the summary statistics $s$ become less sufficient as $n_{\text{sim}}$ gets smaller relative to $n$.

\begin{center}
\begin{tabular}{lcccc}
\multicolumn{5}{l}{\footnotesize \textbf{Table 11.} Simulation study results of \texttt{\texttt{copulaABCdrf}} for the DMC and DMR models.} \\ 
\multicolumn{5}{l}{\footnotesize Mean (standard deviation) of posterior estimates over 10 replicas, under some varying $n$ and  ${n}_{\text{sim}}$.} \\\hline\hline
{\footnotesize DMC\ model} & \footnotesize${q}_{\text{mod}}$ & \footnotesize${q}_{\text{con}}$
& \multicolumn{1}{|c}{\footnotesize${q}_{\text{mod}}$} & \footnotesize${q}_{\text{con}}$
\\ \hline \hline
{\footnotesize \textit{n}} & {\footnotesize 50} & {\footnotesize 50} & \multicolumn{1}{|c}{\footnotesize 300} & 
{\footnotesize 300} \\ 
{\footnotesize \textit{n}}$_{\text{sim}}$ & {\footnotesize 50} & {\footnotesize 50} & \multicolumn{1}{|c}%
{\footnotesize 300} & {\footnotesize 300} \\ 
{\footnotesize true $\theta$} & {\footnotesize 0.20} & {\footnotesize 0.10} & \multicolumn{1}{|c}%
{\footnotesize 0.20} & {\footnotesize 0.10} \\ \hline

{\footnotesize \texttt{copulaABCdrf} mean} 
& {\footnotesize \textbf{0.24} (0.02)} 
& {\footnotesize \textbf{0.16} (0.05)} 
& \multicolumn{1}{|c}{\footnotesize \textbf{0.22} (0.03)} 
& {\footnotesize \textbf{0.16} (0.05)} \\ 

{\footnotesize \texttt{rejectionABC} mean} 
& {\footnotesize 0.26 (0.01)} 
& {\footnotesize 0.32 (0.17)} 
& \multicolumn{1}{|c}{\footnotesize 0.26 (0.01)} 
& {\footnotesize 0.34 (0.21)} \\ \hline

{\footnotesize \texttt{copulaABCdrf} median} 
& {\footnotesize \textbf{0.24} (0.03)} 
& {\footnotesize \textbf{0.15} (0.05)} 
& 
\multicolumn{1}{|c}{\footnotesize \textbf{0.22} (0.04)} 
& {\footnotesize \textbf{0.14} (0.04)} \\ 

{\footnotesize \texttt{rejectionABC} median} 
& {\footnotesize 0.26 (0.01)} 
& {\footnotesize 0.32 (0.18)} 
& \multicolumn{1}{|c}{\footnotesize 0.27 (0.01)} 
& {\footnotesize 0.33 (0.22)} \\ \hline

{\footnotesize \texttt{copulaABCdrf} modeMLE} 
& {\footnotesize \textbf{0.25} (0.06)} 
& {\footnotesize \textbf{0.17} (0.09)} 
& \multicolumn{1}{|c}{\footnotesize \textbf{0.26} (0.04)} 
& {\footnotesize \textbf{0.13} (0.09)} \\ 

{\footnotesize \texttt{rejectionABCkern} modeMLE} 
& {\footnotesize \textbf{0.25} (0.05)} 
& {\footnotesize 0.32 (0.23)} 
& \multicolumn{1}{|c}{\footnotesize 0.28 (0.04)} 
& {\footnotesize 0.38 (0.31)} \\ \hline

{\footnotesize \texttt{copulaABCdrf} s.d.} 
& {\footnotesize 0.05 (0.01)} 
& {\footnotesize 0.07 (0.02)} 
& \multicolumn{1}{|c}{\footnotesize 0.04 (0.01)} 
& {\footnotesize 0.07 (0.04)} \\ 

{\footnotesize \texttt{rejectionABC} s.d.} 
& {\footnotesize 0.06 (0.00)} 
& {\footnotesize 0.14 (0.08)} 
& \multicolumn{1}{|c}{\footnotesize 0.06 (0.00)} 
& {\footnotesize 0.13 (0.07)} \\ \hline

{\footnotesize \texttt{copulaABCdrf}} & {\footnotesize d.f.} & {\footnotesize scale} & \multicolumn{1}{|c}%
{\footnotesize d.f.} & {\footnotesize scale} \\ 
{\footnotesize Copula d.f. and scale} & {\footnotesize 339.94}$\ ${\footnotesize (456.81)} & {\footnotesize 0.09
(0.24)} & \multicolumn{1}{|c}{\footnotesize 131.75 (289.23)} & {\footnotesize 0.21 (0.30)}
\\ \hline
\multicolumn{5}{l}{} \\ \hline \hline
{\footnotesize DMR\ model} & \footnotesize${q}_{\text{del}}$ & \footnotesize${q}_{\text{new}}$
& \multicolumn{1}{|c}{\footnotesize${q}_{\text{del}}$} & \footnotesize${q}_{\text{new}}$
\\ \hline \hline
{\footnotesize \textit{n}} & {\footnotesize 50} & {\footnotesize 50} & \multicolumn{1}{|c}{\footnotesize 300} & 
{\footnotesize 300} \\ 
{\footnotesize \textit{n}}$_{\text{sim}}$ & {\footnotesize 50} & {\footnotesize 50} & \multicolumn{1}{|c}%
{\footnotesize 150} & {\footnotesize 150} \\ 
{\footnotesize true $\theta$} & {\footnotesize \ 0.20} & {\footnotesize 0.10} & 
\multicolumn{1}{|c}{\footnotesize \ 0.20} & {\footnotesize 0.10} \\ \hline

{\footnotesize \texttt{copulaABCdrf} mean} 
& {\footnotesize \textbf{0.24} (0.03)} 
& {\footnotesize 0.24 (0.19)}
& \multicolumn{1}{|c}{\footnotesize 0.26 (0.01)}
& {\footnotesize 0.15 (0.11)} \\ 

{\footnotesize \texttt{rejectionABC} mean} 
& {\footnotesize \textbf{0.24} (0.02)} 
& {\footnotesize \textbf{0.23} (0.16)}
& \multicolumn{1}{|c}{\footnotesize \textbf{0.25} (0.02)}
& {\footnotesize \textbf{0.13} (0.04)} \\

{\footnotesize \texttt{rejectionABCselect} mean} 
& {\footnotesize 0.25 (0.02)} 
& {\footnotesize 0.27 (0.23)}
& \multicolumn{1}{|c}{\footnotesize \textbf{0.25} (\textbf{0.00})} & {\footnotesize 0.14 (0.11)}\\\hline

{\footnotesize \texttt{copulaABCdrf} median} 
& {\footnotesize \textbf{0.24} (0.03)} 
& {\footnotesize 0.23 (0.21)} 
& \multicolumn{1}{|c}{\footnotesize 0.26 (0.01)} 
& {\footnotesize 0.14 (0.11)} \\ 

{\footnotesize \texttt{rejectionABC} median} 
& {\footnotesize \textbf{0.24} (0.03)} 
& {\footnotesize \textbf{0.21} (0.18)} 
& \multicolumn{1}{|c}{\footnotesize \textbf{0.25} (0.02)} 
& {\footnotesize \textbf{0.11} (0.03)} \\ 

{\footnotesize \texttt{rejectionABCselect} median} 
& {\footnotesize 0.25 (0.03)} 
& {\footnotesize 0.26 (0.26)} 
& \multicolumn{1}{|c}{\footnotesize 0.26 (0.01)} & {\footnotesize 0.20 (0.10)}\\\hline

{\footnotesize \texttt{copulaABCdrf} modeMLE} 
& {\footnotesize \textbf{0.23} (0.06)} 
& {\footnotesize 0.28 (0.20)} 
& \multicolumn{1}{|c}{\footnotesize \textbf{0.24} (0.05)} & {\footnotesize 0.19 (0.14)} \\ 

{\footnotesize \texttt{rejectionABCkern} modeMLE} 
& {\footnotesize 0.25 (0.05)} 
& {\footnotesize \textbf{0.22} (0.24)} 
& \multicolumn{1}{|c}{\footnotesize \textbf{0.24} (0.05)} 
& {\footnotesize 0.09 (0.03)} \\ 

{\footnotesize \texttt{rejectionABCkern.select} modeMLE} 
& {\footnotesize 0.27 (0.05)} 
& {\footnotesize \textbf{0.22} (0.29)} 
& \multicolumn{1}{|c}{\footnotesize \textbf{0.26} (0.03)} & {\footnotesize \textbf{0.10} (0.13)} \\\hline

{\footnotesize \texttt{copulaABCdrf} s.d.} 
& {\footnotesize 0.05 (0.01)} 
& {\footnotesize 0.12 (0.04)} 
& \multicolumn{1}{|c}{\footnotesize 0.06 (0.00)} 
& {\footnotesize 0.08 (0.03)} \\ 

{\footnotesize \texttt{rejectionABC} s.d.} 
& {\footnotesize 0.05 (0.01)} 
& {\footnotesize 0.12 (0.05)} 
& \multicolumn{1}{|c}{\footnotesize 0.06 (0.01)} 
& {\footnotesize 0.09 (0.02)} \\ 

{\footnotesize \texttt{rejectionABCselect} s.d.} 
& {\footnotesize 0.06 (0.01)} 
& {\footnotesize 0.13 (0.03)} 
& \multicolumn{1}{|c}{\footnotesize 0.06 (0.00)}& {\footnotesize 0.08 (0.03)}\\\hline

{\footnotesize \texttt{copulaABCdrf}} & {\footnotesize d.f.} & {\footnotesize scale} & \multicolumn{1}{|c}%
{\footnotesize d.f.} & {\footnotesize scale} \\ 
{\footnotesize Copula d.f. and scale} & {\footnotesize 426.42\ (425.06)} & {\footnotesize -0.05\ (0.19)}
& \multicolumn{1}{|c}{\footnotesize 440.06 (399.70)} & {\footnotesize -0.07\ (0.06)} \\ 
\hline\hline
\multicolumn{5}{l}{\footnotesize \textit{Note:} \textbf{Bold} indicates the more accurate ABC method for the given estimator.}
\end{tabular}
\end{center}

\begin{center}
\begin{tabular}{c|ccc|c|cc}
\multicolumn{7}{l}{\footnotesize \textbf{Table 12.} MAE (MSE), mean credible interval coverage (95\%(50\%)c) for the DMC model} \\ 
\multicolumn{7}{l}{\footnotesize over 10 replicas, under varying conditions of $n$ and ${n}_{\text{sim}}$.}\\ 
\hline\hline
&  & \multicolumn{2}{c|}{\footnotesize\textit{n}{\footnotesize \ = 50}} &  & 
\multicolumn{2}{|c}{\footnotesize\textit{n}{\footnotesize \ = 300}} \\ \footnotesize${ n}_{\text{sim}}$ & {\footnotesize Posterior} & \footnotesize$%
{ q}_{\text{mod}}$ & \footnotesize${ q}_{\text{con}}$ & \footnotesize$%
{ n}_{\text{sim}}$ & \footnotesize${ q}_{\text{mod}}$ & \footnotesize${ q}_{\text{con}}$ \\ \hline
{\footnotesize 5} & \multicolumn{1}{|l}{\footnotesize \texttt{copulaABCdrf} mean} & 
{\footnotesize 0.05 (\textbf{0.00})} & {\footnotesize \textbf{0.57} (\textbf{0.33})} & {\footnotesize 30} & {\footnotesize 0.04 (\textbf{0.00})} & {\footnotesize \textbf{0.69} (\textbf{0.52})} \\ 
{\footnotesize} & \multicolumn{1}{|l}{\footnotesize \texttt{rejectionABC} mean} & 
{\footnotesize \textbf{0.04} (\textbf{0.00}) } & {\footnotesize 0.61 (0.39)} & 
{\footnotesize} & {\footnotesize \textbf{0.03} (\textbf{0.00})} & {\footnotesize 0.71
(0.56)} \\ 
& \multicolumn{1}{|l}{\footnotesize \texttt{copulaABCdrf} median} & {\footnotesize 0.05
(\textbf{0.00})} & {\footnotesize \textbf{0.60} (\textbf{0.37})} &  & {\footnotesize \textbf{0.04} (\textbf{0.00})} & 
{\footnotesize \textbf{0.73} (\textbf{0.59})} \\ 
& \multicolumn{1}{|l}{\footnotesize \texttt{rejectionABC} median} & {\footnotesize \textbf{0.04}
(\textbf{0.00})} & {\footnotesize 0.65 (0.43)} &  & {\footnotesize \textbf{0.04} (\textbf{0.00})} & 
{\footnotesize 0.74 (0.60)} \\ 
& \multicolumn{1}{|l}{\footnotesize \texttt{copulaABCdrf} modeMLE} & {\footnotesize 0.05
(0.00)} & {\footnotesize \textbf{0.61} (\textbf{0.42})} &  & {\footnotesize 0.06 (0.01)} & 
{\footnotesize \textbf{0.71} (\textbf{0.57})} \\ 
& \multicolumn{1}{|l}{\footnotesize \texttt{rejectionABCkern} modeMLE} & {\footnotesize \textbf{0.03} (\textbf{0.00})} & {\footnotesize 0.75 (0.58)} &  & {\footnotesize \textbf{0.04} (\textbf{0.00})} & 
{\footnotesize 0.76 (0.64)} \\ 
& \multicolumn{1}{|l}{\footnotesize \texttt{copulaABCdrf} 95\%(50\%)c} & {\footnotesize \textbf{1.00}
(\textbf{0.60})} & {\footnotesize \textbf{0.00} (\textbf{0.00})} &  & {\footnotesize \textbf{1.00} (\textbf{0.30})} & 
{\footnotesize \textbf{0.20} (\textbf{0.10})} \\ 
& \multicolumn{1}{|l}{\footnotesize \texttt{rejectionABC} 95\%(50\%)c} & {\footnotesize \textbf{1.00} (0.90)} 
& {\footnotesize \textbf{0.00} (\textbf{0.00})} &  & {\footnotesize \textbf{1.00} (0.00)} 
& {\footnotesize 0.10 (\textbf{0.10})} \\ \hline
{\footnotesize 16} & \multicolumn{1}{|l}{\footnotesize \texttt{copulaABCdrf} mean} & 
{\footnotesize \textbf{0.05} (\textbf{0.00})} & {\footnotesize \textbf{0.24} (\textbf{0.09})} & {\footnotesize %
100} & {\footnotesize 0.05 (\textbf{0.00})} & {\footnotesize 0.57 (0.42)} \\ 
{\footnotesize } & \multicolumn{1}{|l}{\footnotesize \texttt{rejectionABC}\ mean} & 
{\footnotesize \textbf{0.05} (\textbf{0.00})} & {\footnotesize 0.31 (0.14)} & {\footnotesize} 
& {\footnotesize \textbf{0.04} (\textbf{0.00})} & {\footnotesize \textbf{0.53} (\textbf{0.37})} \\ 
& \multicolumn{1}{|l}{\footnotesize \texttt{copulaABCdrf} median} & {\footnotesize \textbf{0.05} (\textbf{0.00})} & {\footnotesize \textbf{0.23} (\textbf{0.09})} &  & {\footnotesize \textbf{0.05} (\textbf{0.00})} & 
{\footnotesize 0.59 (0.44)} \\ 
& \multicolumn{1}{|l}{\footnotesize \texttt{rejectionABC} median} & {\footnotesize \textbf{0.05}
(\textbf{0.00})} & {\footnotesize 0.31 (0.14)} &  & {\footnotesize \textbf{0.05} (\textbf{0.00})} & 
{\footnotesize \textbf{0.55} (\textbf{0.40})} \\ 
& \multicolumn{1}{|l}{\footnotesize \texttt{copulaABCdrf} modeMLE} & {\footnotesize \textbf{0.06}
(\textbf{0.01})} & {\footnotesize 0.30 (\textbf{0.14})} &  & {\footnotesize 0.08 (0.01)} & 
{\footnotesize 0.61 (0.46)} \\ 
& \multicolumn{1}{|l}{\footnotesize \texttt{rejectionABCkern} modeMLE} & {\footnotesize 0.07
(\textbf{0.01})} & {\footnotesize \textbf{0.29} (\textbf{0.14})} &  & {\footnotesize \textbf{0.06} (\textbf{0.00})} & 
{\footnotesize \textbf{0.59} (\textbf{0.45})} \\ 
& \multicolumn{1}{|l}{\footnotesize \texttt{copulaABCdrf} 95\%(50\%)c} & {\footnotesize \textbf{1.00}
(0.30)} & {\footnotesize \textbf{0.80} (\textbf{0.00})} &  & {\footnotesize \textbf{0.80} (\textbf{0.30})} & 
{\footnotesize \textbf{0.20} (\textbf{0.20})} \\ 
& \multicolumn{1}{|l}{\footnotesize \texttt{rejectionABC} 95\%(50\%)c} & {\footnotesize \textbf{1.00}
(\textbf{0.40})} & {\footnotesize 0.60 (\textbf{0.00})} &  & {\footnotesize 0.50 (0.10)} & 
{\footnotesize \textbf{0.20} (0.10)} \\ \hline
{\footnotesize 25} & \multicolumn{1}{|l}{\footnotesize \texttt{copulaABCdrf} mean} & 
{\footnotesize \textbf{0.05} (\textbf{0.00})} & {\footnotesize \textbf{0.10} (\textbf{0.01})} & {\footnotesize %
150} & {\footnotesize \textbf{0.05} (\textbf{0.00})} & {\footnotesize \textbf{0.30} (\textbf{0.18})} \\ 
{\footnotesize} & \multicolumn{1}{|l}{\footnotesize \texttt{rejectionABC}\ mean} & 
{\footnotesize \textbf{0.05} (\textbf{0.00})} & {\footnotesize 0.18 (0.04)} & {\footnotesize } & {\footnotesize \textbf{0.05} (\textbf{0.00})} & {\footnotesize 0.31 (0.19)} \\ 
& \multicolumn{1}{|l}{\footnotesize \texttt{copulaABCdrf} median} & {\footnotesize \textbf{0.05}
(\textbf{0.00})} & {\footnotesize \textbf{0.09} (\textbf{0.01})} &  & {\footnotesize \textbf{0.05} (0.00)} & 
{\footnotesize \textbf{0.30} (\textbf{0.19})} \\ 
& \multicolumn{1}{|l}{\footnotesize \texttt{rejectionABC}\ median} & {\footnotesize 0.06
(\textbf{0.00})} & {\footnotesize 0.16 (0.04)} &  & {\footnotesize 0.06 (\textbf{0.00})} & 
{\footnotesize 0.32 (0.20)} \\ 
& \multicolumn{1}{|l}{\footnotesize \texttt{copulaABCdrf} modeMLE} & {\footnotesize 0.10 (\textbf{0.01})} 
& {\footnotesize \textbf{0.13} (\textbf{0.02})} &  & {\footnotesize \textbf{0.06} (\textbf{0.01})} & 
{\footnotesize \textbf{0.32} (\textbf{0.22})} \\ 
& \multicolumn{1}{|l}{\footnotesize \texttt{rejectionABCkern} modeMLE} & {\footnotesize \textbf{0.07}
(\textbf{0.01})} & {\footnotesize 0.18 (0.08)} &  & {\footnotesize 0.07 (\textbf{0.01})} & 
{\footnotesize 0.34 (0.23)} \\ 
& \multicolumn{1}{|l}{\footnotesize \texttt{copulaABCdrf} 95\%(50\%)c} & {\footnotesize \textbf{1.00}
(\textbf{0.50})} & {\footnotesize \textbf{1.00} (\textbf{0.30})} &  & {\footnotesize \textbf{0.80} (\textbf{0.30})} & 
{\footnotesize 0.40 (0.10)} \\ 
& \multicolumn{1}{|l}{\footnotesize \texttt{rejectionABC} 95\%(50\%)c} & {\footnotesize \textbf{1.00}
(0.20)} & {\footnotesize \textbf{0.90} (0.10)} &  & {\footnotesize \textbf{0.80} (0.00)} & 
{\footnotesize \textbf{0.60} (\textbf{0.20})} \\ \hline
{\footnotesize 37 \ } & \multicolumn{1}{|l}{\footnotesize \texttt{copulaABCdrf} mean} & 
{\footnotesize \textbf{0.04} (\textbf{0.00})} & {\footnotesize \textbf{0.11} (\textbf{0.03})} & {\footnotesize %
225 } & {\footnotesize \textbf{0.06} (\textbf{0.00})} & {\footnotesize \textbf{0.05} (\textbf{0.00})} \\ 
{\footnotesize} & \multicolumn{1}{|l}{\footnotesize \texttt{rejectionABC}\ mean} & 
{\footnotesize 0.05 (\textbf{0.00})} & {\footnotesize 0.36 (0.16)} & {\footnotesize } & {\footnotesize 0.07 (\textbf{0.00})} & {\footnotesize 0.10 (0.03)} \\ 
& \multicolumn{1}{|l}{\footnotesize \texttt{copulaABCdrf} median} & {\footnotesize \textbf{0.04}
(\textbf{0.00})} & {\footnotesize \textbf{0.10} (\textbf{0.03})} &  & {\footnotesize \textbf{0.07} (\textbf{0.00})} & 
{\footnotesize \textbf{0.05} (\textbf{0.00})} \\ 
& \multicolumn{1}{|l}{\footnotesize \texttt{rejectionABC}\ median} & {\footnotesize 0.05
(\textbf{0.00})} & {\footnotesize 0.36 (0.16)} &  & {\footnotesize \textbf{0.07} (0.01)} & 
{\footnotesize 0.10 (0.03)} \\ 
& \multicolumn{1}{|l}{\footnotesize \texttt{copulaABCdrf} modeMLE} & {\footnotesize 0.07
(0.01)} & {\footnotesize \textbf{0.12} (\textbf{0.02})} &  & {\footnotesize \textbf{0.08} (\textbf{0.01})} & 
{\footnotesize \textbf{0.04} (\textbf{0.00})} \\ 
& \multicolumn{1}{|l}{\footnotesize \texttt{rejectionABCkern} modeMLE} & {\footnotesize \textbf{0.05}
(\textbf{0.00})} & {\footnotesize 0.38 (0.20)} &  & {\footnotesize \textbf{0.08} (\textbf{0.01})} & 
{\footnotesize 0.10 (0.03)} \\ 
& \multicolumn{1}{|l}{\footnotesize \texttt{copulaABCdrf} 95\%(50\%)c} & {\footnotesize \textbf{1.00}
(\textbf{0.50})} & {\footnotesize \textbf{0.90} (\textbf{0.50})} &  & {\footnotesize \textbf{1.00} (\textbf{0.10})} & 
{\footnotesize 0.80 (\textbf{0.40})} \\ 
& \multicolumn{1}{|l}{\footnotesize \texttt{rejectionABC} 95\%(50\%)c} & {\footnotesize \textbf{1.00}
(0.60)} & {\footnotesize 0.20 (0.10)} &  & {\footnotesize \textbf{1.00} (0.00)} & 
{\footnotesize \textbf{0.90} (\textbf{0.40})} \\ \hline
{\footnotesize 50} & \multicolumn{1}{|l}{\footnotesize \texttt{copulaABCdrf} mean} & 
{\footnotesize \textbf{0.04} (\textbf{0.00})} & {\footnotesize \textbf{0.06} (\textbf{0.01})} & {\footnotesize %
300} & {\footnotesize \textbf{0.03} (\textbf{0.00})} & {\footnotesize \textbf{0.06} (\textbf{0.01})} \\ 
{\footnotesize} & \multicolumn{1}{|l}{\footnotesize \texttt{rejectionABC} mean} & 
{\footnotesize 0.06 (\textbf{0.00})} & {\footnotesize 0.22 (0.08)} & {\footnotesize} & {\footnotesize 0.06 (\textbf{0.00})} & {\footnotesize 0.24 (0.10)} \\ 
& \multicolumn{1}{|l}{\footnotesize \texttt{copulaABCdrf} median} & {\footnotesize \textbf{0.05}
(\textbf{0.00})} & {\footnotesize \textbf{0.05} (\textbf{0.00})} &  & {\footnotesize \textbf{0.03} (\textbf{0.00})} & 
{\footnotesize \textbf{0.05} (\textbf{0.00})} \\ 
& \multicolumn{1}{|l}{\footnotesize \texttt{rejectionABC} median} & {\footnotesize 0.06
(\textbf{0.00})} & {\footnotesize 0.22 (0.08)} &  & {\footnotesize 0.07 (\textbf{0.00})} & 
{\footnotesize 0.24 (0.10)} \\ 
& \multicolumn{1}{|l}{\footnotesize \texttt{copulaABCdrf} modeMLE} & {\footnotesize 0.06
(0.01)} & {\footnotesize \textbf{0.08} (\textbf{0.01})} &  & {\footnotesize \textbf{0.06} (\textbf{0.00})} & 
{\footnotesize \textbf{0.06} (\textbf{0.01})} \\ 
& \multicolumn{1}{|l}{\footnotesize \texttt{rejectionABCkern} modeMLE} & {\footnotesize \textbf{0.05}
(\textbf{0.00})} & {\footnotesize 0.22 (0.10)} &  & {\footnotesize 0.09 (0.01)} & 
{\footnotesize 0.30 (0.17)} \\ 
& \multicolumn{1}{|l}{\footnotesize \texttt{copulaABCdrf} 95\%(50\%)c} & {\footnotesize \textbf{1.00}
(\textbf{0.60})} & {\footnotesize \textbf{1.00} (\textbf{0.50})} &  & {\footnotesize \textbf{1.00} (\textbf{0.80})} & 
{\footnotesize \textbf{1.00} (\textbf{0.30})} \\ 
& \multicolumn{1}{|l}{\footnotesize \texttt{rejectionABC} 95\%(50\%)c} & {\footnotesize \textbf{1.00}
(0.20)} & {\footnotesize \textbf{0.90} (0.10)} &  & {\footnotesize \textbf{1.00} (0.10)} & 
{\footnotesize 0.60 (\textbf{0.30})} \\ \hline\hline
\multicolumn{7}{l}{\footnotesize \textit{Note:} \textbf{Bold} indicates the more accurate ABC method for the given estimator.}
\end{tabular}
\end{center}

\begin{changemargin}{-.7in}{-.7in}
\begin{center}
\begin{tabular}{c|ccc|c|cc}
\multicolumn{7}{l}{\footnotesize \textbf{Table 13.} MAE (MSE), mean credible interval coverage (95\%(50\%)c) for DMR model over 10 replicas, varying $n$ and ${n}_{\text{sim}}$.}\\\hline\hline
&  & \multicolumn{2}{c|}{{\footnotesize \textit{n} = 50}} &  & \multicolumn{2}{|c}{%
{\footnotesize \textit{n} = 300}} \\ 
\footnotesize{\textit{n}}$_{\text{sim}}$ & {\footnotesize Posterior} & \footnotesize${q}_{\text{del}}{ \ }$ & \footnotesize${q}_{\text{new}}$ & {\footnotesize \textit{n}}$_{\text{sim}}$ & \footnotesize${q}_{\text{del}}{\ }$ & \footnotesize${q}_{\text{new}}$ \\ \hline
{\footnotesize 5} & \multicolumn{1}{|l}{\footnotesize \texttt{copulaABCdrf} mean} & {\footnotesize %
0.06 (\textbf{0.00})} & {\footnotesize 0.49 (0.25)} & {\footnotesize 30} & 
{\footnotesize 0.06 (\textbf{0.00})} & {\footnotesize 0.21 (0.05)} \\ 
{\footnotesize } & \multicolumn{1}{|l}{\footnotesize \texttt{rejectionABCselect} mean} & 
{\footnotesize \textbf{0.05}\textsubscript{-.01}(\textbf{0.00})} & {\footnotesize \textbf{0.43}\textsubscript{-.06}(\textbf{0.18}\textsubscript{-.06})} & {\footnotesize} & {\footnotesize \textbf{0.05}\textsubscript{-.01}(\textbf{0.00})} & {\footnotesize \textbf{0.17}\textsubscript{-.01}(\textbf{0.03})} \\ 
& \multicolumn{1}{|l}{\footnotesize \texttt{copulaABCdrf} median} & {\footnotesize 0.06
(\textbf{0.00})} & {\footnotesize 0.50 (0.26)} &  & {\footnotesize 0.06 (\textbf{0.00})} & 
{\footnotesize 0.19 (0.04)} \\ 
& \multicolumn{1}{|l}{\footnotesize \texttt{rejectionABCselect}  median} & {\footnotesize \textbf{0.05}\textsubscript{-.02}(\textbf{0.00})} & {\footnotesize \textbf{0.42}\textsubscript{-.07}(\textbf{0.18}\textsubscript{-.07})} &  & {\footnotesize \textbf{0.05}\textsubscript{-.01}(\textbf{0.00})} & {\footnotesize \textbf{0.14}\textsubscript{-.02}(\textbf{0.03})} \\ 
& \multicolumn{1}{|l}{\footnotesize \texttt{copulaABCdrf} modeMLE} & {\footnotesize \textbf{0.07}
(\textbf{0.01})} & {\footnotesize 0.53 (0.32)} &  & {\footnotesize 0.06 (0.01)} & 
{\footnotesize 0.23 (0.07)} \\ 
& \multicolumn{1}{|l}{\footnotesize \texttt{rejectionABCkern.select} modeMLE} & {\footnotesize 0.08\textsubscript{-.02}(\textbf{0.01})} & {\footnotesize \textbf{0.34}\textsubscript{-.20}(\textbf{0.12}\textsubscript{-.18})} &  & {\footnotesize \textbf{0.04}\textsubscript{-.02}(\textbf{0.00})} & {\footnotesize \textbf{0.09}\textsubscript{-.04}(\textbf{0.02})} \\ 
& \multicolumn{1}{|l}{\footnotesize \texttt{copulaABCdrf}95\%(50\%)c} & {\footnotesize \textbf{1.00} (0.00)} & {\footnotesize 0.00 (\textbf{0.00})} &  & \multicolumn{1}{|c}{\footnotesize \textbf{1.00} (0.00)} & {\footnotesize 0.70 (0.00)} \\ 
& \multicolumn{1}{|l}{\footnotesize \texttt{rejectionABCselect} 95\%(50\%)c} & {\footnotesize \textbf{1.00}(\textbf{0.10}\textsubscript{+.10})} & {\footnotesize \textbf{0.60}\textsubscript{+.60}(\textbf{0.00})} &  & {\footnotesize \textbf{1.00} (\textbf{0.20}\textsubscript{+.10})} & {\footnotesize \textbf{0.90} (\textbf{0.30}\textsubscript{+.30})} \\ \hline
{\footnotesize 16} & \multicolumn{1}{|l}{\footnotesize \texttt{copulaABCdrf} mean} & {\footnotesize \textbf{0.05} (\textbf{0.00})} & {\footnotesize 0.29 (0.10)} & {\footnotesize 100} & {\footnotesize \textbf{0.05} (\textbf{0.00})} & {\footnotesize \textbf{0.12} (\textbf{0.02})} \\ 
{\footnotesize } & \multicolumn{1}{|l}{\footnotesize \texttt{rejectionABCselect} mean} & 
{\footnotesize 0.06\textsubscript{+.01}(0.00)} & {\footnotesize \textbf{0.25}\textsubscript{-.03}(\textbf{0.07}\textsubscript{-.02})} & {\footnotesize } & {\footnotesize \textbf{0.05} (\textbf{0.00})} & {\footnotesize \textbf{0.12}\textsubscript{+.02}(\textbf{0.02}\textsubscript{+.01})} \\ 
& \multicolumn{1}{|l}{\footnotesize \texttt{copulaABCdrf} median} & {\footnotesize \textbf{0.06}
(\textbf{0.00})} & {\footnotesize 0.27 (0.09)} &  & {\footnotesize \textbf{0.05} (\textbf{0.00})} & 
{\footnotesize \textbf{0.10} (\textbf{0.02})} \\ 
& \multicolumn{1}{|l}{\footnotesize \texttt{rejectionABCselect} median} & {\footnotesize \textbf{0.06} (\textbf{0.00})} & {\footnotesize \textbf{0.23}\textsubscript{-.03}(\textbf{0.06}\textsubscript{-.02})} &  & {\footnotesize 0.06\textsubscript{+.01}(\textbf{0.00})} & 
{\footnotesize \textbf{0.10}\textsubscript{+.02}(\textbf{0.02}\textsubscript{+.01})} \\ 
& \multicolumn{1}{|l}{\footnotesize \texttt{copulaABCdrf} modeMLE} & {\footnotesize \textbf{0.05}
(\textbf{0.00})} & {\footnotesize 0.32 (0.17)} &  & {\footnotesize \textbf{0.04} (\textbf{0.00})} & 
{\footnotesize \textbf{0.10} (\textbf{0.02})} \\ 
& \multicolumn{1}{|l}{\footnotesize \texttt{rejectionABCkern.select} modeMLE} & {\footnotesize 0.06\textsubscript{-.02}(0.01)} & 
{\footnotesize \textbf{0.17}\textsubscript{-.06}(\textbf{0.06}\textsubscript{-.02})} &  & {\footnotesize 0.05\textsubscript{-.01}(\textbf{0.00}\textsubscript{-.01})} & 
{\footnotesize 0.11\textsubscript{+.05}(\textbf{0.02}\textsubscript{+.01})} \\ 
& \multicolumn{1}{|l}{\footnotesize \texttt{copulaABCdrf}95\%(50\%)c} & {\footnotesize \textbf{1.00}
(\textbf{0.30})} & {\footnotesize 0.60 (\textbf{0.00})} &  & \multicolumn{1}{|c}{\footnotesize \textbf{1.00} (0.30)} & {\footnotesize \textbf{0.80} (\textbf{0.40})} \\ 
& \multicolumn{1}{|l}{\footnotesize \texttt{rejectionABCselect} 95\%(50\%)c} & {\footnotesize \textbf{1.00} (0.00\textsubscript{-.20})} & {\footnotesize \textbf{0.80} (\textbf{0.00})} &  & {\footnotesize \textbf{1.00} (\textbf{0.40}\textsubscript{-.10})} & {\footnotesize \textbf{0.80} (\textbf{0.40})} \\ \hline
{\footnotesize 25} & \multicolumn{1}{|l}{\footnotesize \texttt{copulaABCdrf} mean} & {\footnotesize \textbf{0.05} (\textbf{0.00})} & {\footnotesize 0.17 (0.04)} & {\footnotesize 150} & 
{\footnotesize 0.06 (\textbf{0.00})} & {\footnotesize 0.06 (\textbf{0.01})} \\ 
{\footnotesize } & \multicolumn{1}{|l}{\footnotesize \texttt{rejectionABCselect} mean} & 
{\footnotesize \textbf{0.05} (\textbf{0.00})} & {\footnotesize \textbf{0.16}\textsubscript{-.02}(\textbf{0.03}\textsubscript{-.01})} & {\footnotesize} & {\footnotesize \textbf{0.05}\textsubscript{-.01}(\textbf{0.00})} & {\footnotesize \textbf{0.05}\textsubscript{-.02}(\textbf{0.01})} \\ 
& \multicolumn{1}{|l}{\footnotesize \texttt{copulaABCdrf} median} & {\footnotesize \textbf{0.05}
(\textbf{0.00})} & {\footnotesize \textbf{0.14} (\textbf{0.03})} &  & {\footnotesize \textbf{0.06} (\textbf{0.00})} & 
{\footnotesize 0.06 (\textbf{0.01})} \\ 
& \multicolumn{1}{|l}{\footnotesize \texttt{rejectionABCselect} median} & {\footnotesize 0.06 (\textbf{0.00})} & {\footnotesize \textbf{0.14}\textsubscript{-.02}(\textbf{0.03}\textsubscript{-.01})} &  & {\footnotesize \textbf{0.06} (\textbf{0.00})} & 
{\footnotesize \textbf{0.06}\textsubscript{+.01}(\textbf{0.01})} \\ 
& \multicolumn{1}{|l}{\footnotesize \texttt{copulaABCdrf} modeMLE} & {\footnotesize \textbf{0.07}
(\textbf{0.01})} & {\footnotesize 0.17 (0.08)} &  & {\footnotesize \textbf{0.05} (\textbf{0.00})} & 
{\footnotesize 0.11 (0.03)} \\ 
& \multicolumn{1}{|l}{\footnotesize \texttt{rejectionABCkern.select} modeMLE} & {\footnotesize \textbf{0.07}\textsubscript{+.01}(\textbf{0.01})} & {\footnotesize \textbf{0.11}\textsubscript{-.02}(\textbf{0.04})} &  & {\footnotesize 0.06\textsubscript{-.01}(\textbf{0.00}\textsubscript{-.01})} & {\footnotesize \textbf{0.09}\textsubscript{+.03}(\textbf{0.01})} \\ 
& \multicolumn{1}{|l}{\footnotesize \texttt{copulaABCdrf}95\%(50\%)c} & {\footnotesize \textbf{1.00}
(\textbf{0.30})} & {\footnotesize \textbf{0.90} (\textbf{0.10})} &  & {\footnotesize \textbf{1.00} (\textbf{0.30})} & 
{\footnotesize \textbf{0.90} (\textbf{0.70})} \\ 
& \multicolumn{1}{|l}{\footnotesize \texttt{rejectionABCselect} 95\%(50\%)c} & {\footnotesize \textbf{1.00} (0.00\textsubscript{-.40})} & {\footnotesize \textbf{1.00} (0.00\textsubscript{-.10})} &  & {\footnotesize \textbf{1.00} (0.00)} & {\footnotesize \textbf{0.90} (0.80\textsubscript{-.10})} \\ \hline
{\footnotesize 37} & \multicolumn{1}{|l}{\footnotesize \texttt{copulaABCdrf} mean} & {\footnotesize %
\textbf{0.05} (\textbf{0.00})} & {\footnotesize \textbf{0.10} (\textbf{0.01})} & {\footnotesize 225} & 
{\footnotesize \textbf{0.05} (\textbf{0.00})} & {\footnotesize \textbf{0.04} (\textbf{0.00})} \\ 
{\footnotesize }{\footnotesize } & \multicolumn{1}{|l}{\footnotesize \texttt{rejectionABCselect} mean} & 
{\footnotesize \textbf{0.05} (\textbf{0.00})} & {\footnotesize 0.11\textsubscript{+.01}(0.02\textsubscript{+.01})} & {\footnotesize } & {\footnotesize \textbf{0.05} (\textbf{0.00})} & {\footnotesize 0.06\textsubscript{+.02}(\textbf{0.00})} \\ 
& \multicolumn{1}{|l}{\footnotesize \texttt{copulaABCdrf} median} & {\footnotesize \textbf{0.05}
(\textbf{0.00})} & {\footnotesize \textbf{0.08} (\textbf{0.01})} &  & {\footnotesize \textbf{0.05} (\textbf{0.00})} & 
{\footnotesize \textbf{0.04} (\textbf{0.00})} \\ 
& \multicolumn{1}{|l}{\footnotesize \texttt{rejectionABCselect} median} & {\footnotesize 0.06\textsubscript{+.01}(\textbf{0.00})} & {\footnotesize 0.09\textsubscript{+.02}(\textbf{0.01})} &  & {\footnotesize \textbf{0.05} (\textbf{0.00})} & 
{\footnotesize 0.06\textsubscript{+.03}(\textbf{0.00})} \\ 
& \multicolumn{1}{|l}{\footnotesize \texttt{copulaABCdrf} modeMLE} & {\footnotesize \textbf{0.06}
(\textbf{0.00})} & {\footnotesize 0.10 (0.02)} &  & {\footnotesize 0.08 (0.01)} & 
{\footnotesize 0.11 (0.02)} \\ 
& \multicolumn{1}{|l}{\footnotesize \texttt{rejectionABCkern.select} modeMLE} & {\footnotesize 0.07\textsubscript{+.04}(0.01\textsubscript{+.01})} & {\footnotesize \textbf{0.06}\textsubscript{+.01}(\textbf{0.01}\textsubscript{+.01})} &  & {\footnotesize \textbf{0.06} (\textbf{0.00}\textsubscript{-.01})} & 
{\footnotesize \textbf{0.06}\textsubscript{+.01}(\textbf{0.00})} \\ 
& \multicolumn{1}{|l}{\footnotesize \texttt{copulaABCdrf}95\%(50\%)c} & {\footnotesize \textbf{1.00}
(\textbf{0.60})} & {\footnotesize \textbf{1.00} (\textbf{0.40})} &  & {\footnotesize \textbf{1.00} (\textbf{0.40})} & 
{\footnotesize \textbf{1.00} (0.80)} \\ 
& \multicolumn{1}{|l}{\footnotesize \texttt{rejectionABCselect} 95\%(50\%)c} & {\footnotesize \textbf{1.00} (0.00\textsubscript{-.40})} & {\footnotesize \textbf{0.90}\textsubscript{-.10}(0.8\textsubscript{+.50})} &  & {\footnotesize \textbf{1.00} (0.00\textsubscript{-.40})} & 
{\footnotesize \textbf{1.00} (\textbf{0.50}\textsubscript{-.40})} \\ \hline
{\footnotesize 50} & \multicolumn{1}{|l}{\footnotesize \texttt{copulaABCdrf} mean} & {\footnotesize \textbf{0.05} (\textbf{0.00})} & {\footnotesize \textbf{0.14} (\textbf{0.05})} & {\footnotesize 300} & {\footnotesize \textbf{0.05} (\textbf{0.00})} & {\footnotesize \textbf{0.04} (\textbf{0.00})} \\ 
{\footnotesize } & \multicolumn{1}{|l}{\footnotesize \texttt{rejectionABCselect} mean} & 
{\footnotesize \textbf{0.05} (\textbf{0.00})} & {\footnotesize 0.17\textsubscript{+.04}(0.07\textsubscript{+.03})} & {\footnotesize } & {\footnotesize 0.06\textsubscript{+.01}(0.00)} & {\footnotesize 0.07\textsubscript{+.03}(0.01\textsubscript{+.01})} \\ 
& \multicolumn{1}{|l}{\footnotesize \texttt{copulaABCdrf} median} & {\footnotesize \textbf{0.05}
(\textbf{0.00})} & {\footnotesize \textbf{0.13} (\textbf{0.06})} &  & {\footnotesize \textbf{0.05} (\textbf{0.00})} & 
{\footnotesize \textbf{0.05} (\textbf{0.00})} \\ 
& \multicolumn{1}{|l}{\footnotesize \texttt{rejectionABCselect} median} & {\footnotesize \textbf{0.05} (\textbf{0.00})} & {\footnotesize 0.16\textsubscript{+.04}(0.08\textsubscript{+.04})} &  & {\footnotesize 0.06\textsubscript{+.01}(\textbf{0.00})} & {\footnotesize 0.07\textsubscript{+.04}(0.01\textsubscript{+.01})} \\ 
& \multicolumn{1}{|l}{\footnotesize \texttt{copulaABCdrf} modeMLE} & {\footnotesize \textbf{0.05}
(\textbf{0.00})} & {\footnotesize 0.19 (\textbf{0.07})} &  & {\footnotesize \textbf{0.06} (\textbf{0.01})} & 
{\footnotesize 0.08 (\textbf{0.01})} \\ 
& \multicolumn{1}{|l}{\footnotesize \texttt{rejectionABCkern.select} modeMLE} & {\footnotesize 0.08\textsubscript{+.02}(0.01\textsubscript{+.01})} & {\footnotesize \textbf{0.14}\textsubscript{+.01}(0.09\textsubscript{+.02})} &  & {\footnotesize 0.07\textsubscript{+.02}(\textbf{0.01}\textsubscript{+.01})} & 
{\footnotesize \textbf{0.07}\textsubscript{+.04}(\textbf{0.01}\textsubscript{+.01})} \\ 
& \multicolumn{1}{|l}{\footnotesize \texttt{copulaABCdrf}95\%(50\%)c} & {\footnotesize \textbf{1.00}
(\textbf{0.50})} & {\footnotesize \textbf{0.90} (\textbf{0.50})} &  & {\footnotesize \textbf{1.00} (\textbf{0.40})} & {\footnotesize \textbf{1.00} (\textbf{0.60})} \\ 
& \multicolumn{1}{|l}{\footnotesize \texttt{rejectionABCselect} 95\%(50\%)c} & {\footnotesize \textbf{1.00} (0.20\textsubscript{-.20})} & {\footnotesize 0.80\textsubscript{-.10}(0.70\textsubscript{+.20})} &  & {\footnotesize \textbf{1.00} (0.10\textsubscript{-.50})}
& {\footnotesize \textbf{1.00} (0.30\textsubscript{-.70})} \\ \hline\hline
\multicolumn{7}{l}{\footnotesize \textit{Note:}  Subscript is change in MAE (MSE) based on \texttt{drf} selecting the 2 best of the 3 summaries.}\\
\multicolumn{7}{l}{\footnotesize \textbf{Bold} indicates the more accurate ABC method for the given estimator.}
\end{tabular}
\end{center}
\end{changemargin}

\subsection{Real Citation Network Analysis}

With Algorithm 1 validated by simulation studies, we apply this algorithm to analyze real-life datasets. Specifically, we now analyze a large binary directed citation \texttt{HepPh} network dataset, a arXiv High Energy Physics paper citation network of $n=28,093$ papers (nodes) and $3,148,447$ citations (directed edges) among them \citep{RossiAhmed15}. (This dataset was obtained from https://networkrepository.com/cit.php). This dataset is analyzed by the ERGM model and by the Price model, each model using a vector of summary statistics $s(\cdot)$ being the network size invariant offset MPLEs (resp.) of geometrically weighted in-degree distribution (\texttt{gwidegree}), its decay (degree weighting) parameter estimate (\texttt{gwidegree.decay}), and \texttt{triangle} count. The ERGM, based on network sufficient statistics \texttt{gwidegree}, \texttt{gwidegree.decay}, and \texttt{triangle}, was assigned a trivariate normal prior $\theta\sim$ $\mathcal{N}(0,10I_{3})$. The Price model was assigned priors $k_{0}\sim$ $\mathcal{U}_{(0.9,1.1)}$ and $p\sim\mathcal{U}_{(0,0.20)}$.

\medskip
\begin{center}
\begin{tabular}{lccccc}
\multicolumn{6}{l}{\footnotesize \textbf{Table 14.} ABC posterior estimates for the Price model and for the ERGM, obtained from } \\ 
\multicolumn{6}{l}{\footnotesize the arXiv High Energy Physics paper citation network dataset.} \\ \hline\hline
& \multicolumn{2}{c}{\footnotesize Price model} & \multicolumn{1}{|c}{} & {\footnotesize %
ERGM} &  \\ \hline
& \footnotesize${k}_{0}$ & \footnotesize${p}$ & \multicolumn{1}{|c}{\footnotesize $\theta$ {\footnotesize gwidegree}} & {\footnotesize $\theta$ {\footnotesize \ decay}} & {\footnotesize $\theta$ {\footnotesize \ triangles}} \\\hline
{\footnotesize \texttt{copulaABCdrf} mean} & {\footnotesize 1.01} & {\footnotesize 0.01} & \multicolumn{1}{|c}{\footnotesize -4.46} & {\footnotesize 3.34} & {\footnotesize -1.07} \\
{\footnotesize \texttt{rejectionABC} mean} & {\footnotesize 1.01} & {\footnotesize 0.01} & \multicolumn{1}{|c}{\footnotesize -1.49} & {\footnotesize -0.72} & {\footnotesize -0.58} \\
{\footnotesize \texttt{rejectionABCselect} mean} & {\footnotesize 1.01} & {\footnotesize 0.01} & \multicolumn{1}{|c}{\footnotesize  } & {\footnotesize  } & {\footnotesize }\\\hline
{\footnotesize \texttt{copulaABCdrf} median} & {\footnotesize 1.02} & {\footnotesize 0.01} & \multicolumn{1}{|c}{\footnotesize -4.96} & {\footnotesize 3.66} & {\footnotesize -0.58} \\ 
{\footnotesize \texttt{rejectionABC} median} & {\footnotesize 1.02} & {\footnotesize 0.01} & \multicolumn{1}{|c}{\footnotesize -1.86} & {\footnotesize -1.45} & {\footnotesize -0.53} \\
{\footnotesize \texttt{rejectionABCselect} median} & {\footnotesize 1.02} & {\footnotesize 0.01} & \multicolumn{1}{|c}{\footnotesize } & {\footnotesize } & {\footnotesize }\\\hline
{\footnotesize \texttt{copulaABCdrf} mode} & {\footnotesize 1.08} & {\footnotesize 0.01} & \multicolumn{1}{|c}{\footnotesize -1.76} & {\footnotesize -0.80} & {\footnotesize -4.59} \\ 
{\footnotesize \texttt{rejectionABC} mode} & {\footnotesize 1.02} & {\footnotesize 0.01} & \multicolumn{1}{|c}{\footnotesize -2.18} & {\footnotesize 4.49} & {\footnotesize -0.51} \\
{\footnotesize \texttt{rejectionABCkern.select} mode} & {\footnotesize 1.02} & {\footnotesize 0.01} & \multicolumn{1}{|c}{\footnotesize } & {\footnotesize } & {\footnotesize }\\\hline
{\footnotesize \texttt{copulaABCdrf} MLE} & {\footnotesize 1.08} & {\footnotesize 0.01} & \multicolumn{1}{|c}{\footnotesize -4.82} & {\footnotesize 6.00} & {\footnotesize -0.58} \\ 
{\footnotesize \texttt{rejectionABC} MLE} & {\footnotesize 1.02} & {\footnotesize 0.01} & \multicolumn{1}{|c}{\footnotesize 5.39} & {\footnotesize -5.86} & {\footnotesize -7.30} \\
{\footnotesize \texttt{rejectionABCkern.select} MLE} & {\footnotesize 1.02} & {\footnotesize 0.01} & \multicolumn{1}{|c}{\footnotesize } & {\footnotesize  } & {\footnotesize }\\\hline
{\footnotesize \texttt{copulaABCdrf} s.d.} & {\footnotesize 0.05} & {\footnotesize 0.00} & \multicolumn{1}{|c}{\footnotesize 2.41} & {\footnotesize 2.62} & {\footnotesize 2.06} \\ 
{\footnotesize \texttt{rejectionABC} s.d.} & {\footnotesize 0.06} & {\footnotesize 0.003} & \multicolumn{1}{|c}{\footnotesize 2.81} & {\footnotesize 3.29} & {\footnotesize 2.85} \\
{\footnotesize \texttt{rejectionABCselect} s.d.} & {\footnotesize 0.06} & {\footnotesize 0.003} & \multicolumn{1}{|c}{\footnotesize} & {\footnotesize} & {\footnotesize} \\\hline
{\footnotesize \texttt{copulaABCdrf} 50\%} & {\footnotesize (0.97, 1.06)} & {\footnotesize (0.01,\ 0.01)} & \multicolumn{1}{|c}{\footnotesize (-6.09, -2.81)} & {\footnotesize (2.23, 4.96)} & {\footnotesize (-1.17, -0.23)} \\ 
{\footnotesize \texttt{rejectionABC} 50\%} & {\footnotesize (0.97, 1.05)} & {\footnotesize (0.01, 0.01)} & \multicolumn{1}{|c}{\footnotesize (-2.62, 0.51)} & {\footnotesize (-3.21, 1.92)} & {\footnotesize (-1.99, 1.07)} \\
{\footnotesize \texttt{rejectionABCselect} 50\%} & {\footnotesize (0.97, 1.05)} & {\footnotesize (0.01, 0.01)} & \multicolumn{1}{|c}{\footnotesize} & {\footnotesize} & {\footnotesize} \\\hline
{\footnotesize \texttt{copulaABCdrf} 95\%} & {\footnotesize (0.91, 1.09)} & {\footnotesize (0.00,\ 0.02)} & \multicolumn{1}{|c}{\footnotesize (-9.12, 0.73)} & {\footnotesize (-2.08, 6.80)} & {\footnotesize (-6.09, 3.22)} \\
{\footnotesize \texttt{rejectionABC} 95\%} & {\footnotesize (0.90, 1.09)} & {\footnotesize (0.003, 0.01)} & \multicolumn{1}{|c}{\footnotesize (-6.60, 4.00) } & {\footnotesize (-5.44, 5.86)} & {\footnotesize (-6.12, 4.22)} \\
{\footnotesize \texttt{rejectionABCselect} 95\%} & {\footnotesize (0.90, 1.09)} & {\footnotesize (0.003, 0.01)} & \multicolumn{1}{|c}{\footnotesize} & {\footnotesize} & {\footnotesize} \\\hline
{\footnotesize \texttt{copulaABCdrf}} & {\footnotesize d.f.} & {\footnotesize scale} & \multicolumn{1}{|c}{\footnotesize d.f.} & \multicolumn{2}{c}{\footnotesize scale matrix} \\ 
{\footnotesize Copula d.f. and scale} & {\footnotesize 50.42} & {\footnotesize -0.20} & 
\multicolumn{1}{|c}{\footnotesize 6.31} & {\footnotesize $\theta $\ decay} & {\footnotesize $\theta $\ triangles} \\ 
&  &  & \multicolumn{1}{|c}{{\footnotesize $\theta $\ gwidegree}} & {\footnotesize -0.60} & {\footnotesize 0.30} \\ 
&  &  & \multicolumn{1}{|c}{{\footnotesize $\theta $\ decay}} &  & {\footnotesize -0.28}
\\ \hline\hline
&  &  &  &  & 
\end{tabular}
\end{center}

The MPLEs for the three network summaries (resp.) could not be computed directly on this large network; these computations were prohibitively long. Therefore, these 3 MPLEs were estimated by their geometric median over 100,000 subsamples of the 28,093 papers (nodes), each subsample inducing a subgraph of size $\lfloor\sqrt{28,093}\rfloor=167$, providing an outlier-robust divide-and-conquer (DAC) subsampling MPLE estimator \citep{Minsker19}, resulting in the following MPLEs (resp.): for \texttt{gwidegree}, -0.8418461; for \texttt{gwidegree.decay}, 0.2365466; for \texttt{triangle}: 1.6904300. Recall from \S 3.2 that the MPLE is the MLE for logistic regression, and they are both consistent over a growing number of networks, observed from the same set of fixed nodes (while the MPLE assumes dyadic independence of the edge observations). These three DAC MPLEs then specified the vector of summary statistics $s(x)$ for the real observed citation \texttt{HepPh} network dataset, analyzed by each of the ERGM and Price model using Algorithm 1, with each model simulating a network of size $n_{\text{sim}}=$ $167$ in each algorithm iteration.

Table 14 presents the posterior estimates and MLEs obtained from \texttt{copulaABCdrf}, this ABC algorithm. Running Step 2 of this algorithm, for model selection, output a posterior probability estimate of $0.99$ for the Price model, compared to the ERGM. This table also presents the results of \texttt{rejectionABC}, for ERGM, and for the Price model based on all three network summary statistics. In addition, results are presented for the Price model after pre-selecting the most important summary statistics (variable) of the three total summaries, based on a \texttt{drf} regression fit to the Reference table, regressing the prior parameter samples on the samples of all three summary statistics. It was found that among all three network statistics, being the network size (\textit{n}) invariant MPLEs of \texttt{gwidegree} (importance measure of 0.14), \texttt{gwidegree.decay} (importance 0.07), and \texttt{triangle} count (importance 0.79), the first and third of these statistics were most important and thus pre-selected. Table 14 shows that for the Price model, the posterior-based parameter estimates were similar across all \texttt{copulaABCdrf} and \texttt{rejectionABC} methods, while for ERGM they were less similar.

\subsection{Real Multilayer Network Analysis}

The multilayer, weighted, and directed \texttt{BostonBomb2013} Twitter network dataset consists of $n =4,377,184$ persons (nodes), and $9,480,331$ count-weighted edges, across $L=3$ layers: retweets, mentions and replies, occurring between dates 4-15-2013 and 4-22-2013 of the Boston Bombing Attacks of 2013 \citep{DeDomenicoAltmann20}. (The dataset was obtained from \\ https://manliodedomenico.com/data.php). Using Algorithm 1, the \texttt{BostonBomb2013} network dataset was analyzed by a three-Layer ERGM with Poisson reference measure $b$,
\[
f(X=x\mid\theta)=
{\displaystyle\prod\limits_{l=1}^{L=3}}
f(X_{l}=x_{l}\mid\theta)=\dfrac{b(x_{l})\exp\left\{
{\textstyle\sum\nolimits_{l=1}^{L=3}}
\theta_{l}^{\intercal}h_{l}(x_{l})\right\}  }{%
{\textstyle\prod\limits_{l=1}^{L=3}}
{\textstyle\sum\limits_{x^{\prime}\in\mathcal{X}}}
b(x^{\prime})\exp\left\{
{\textstyle\sum\nolimits_{l=1}^{L=3}}
\theta_{l}^{\intercal}h_{l}(x^{\prime})\right\}  },
\]
which was assigned a multivariate normal prior $\theta\sim$
$\mathcal{N}(0,10I_{18})$ for 18 parameters $\theta$ of six network sufficient statistics (vector $h_{l}$) specified for each of the 3 layers, namely: (1) Indicator of edge weight equal to 1 in the given layer (\texttt{equalto(1)}); (2) Indicator of edge weight greater than 1 in the given layer; (3) minimum mutuality (\texttt{mutual.min}) which is analogous to mutuality for binary networks; (4) triad closure represented by transitive weights with twopath = min, combine = sum, and affect = min(\texttt{transitiveweights.min.sum.min}), which is analogous to triangle count from a binary network; (5) triad closure represented by transitive weights with twopath = min, combine = max, and affect = min(\texttt{transitiveweights.min.max.min}), which is analogous to a count of transitive ties from a binary network; and finally (6), the term \texttt{CMP}, which specifies the reference measure $b$ by the Conway-Maxwell-Poisson (CMP) distribution for the count-weighted edges, with corresponding coefficient parameter $\theta_{\text{CMP}}$, which controls the degree of dispersion relative to Poisson distribution. In particular, $\theta_{\text{CMP}}=0$ defines the Poisson distribution; $\ \theta_{\text{CMP}}$$<0$ defines a more underdispersed distribution; the value $\theta_{\text{CMP}}$ $\rightarrow$$-\infty$ leads to the Bernoulli distribution; $\theta_{\text{CMP}}$ $>0$ defines a more overdispersed distribution; and $\theta_{\text{CMP}}=1$ corresponds to the geometric distribution, being most overdispersed. More details about these network statistics of valued networks and the CMP distribution for ERGM are provided elsewhere \citep{Krivitsky12}.

These 18 ERGM sufficient statistics could not be easily computed on the \texttt{BostonBomb2013} network dataset, due to its massive size. Therefore, these sufficient statistics were approximated by computing 21 (mostly) UMP summary statistics $s(x)$, each of which were computable on the full network, for each of the 3 network layers. The values of the summaries $s(x)$ of the \texttt{BostonBomb2013} network dataset are shown in Table 15. 
\begin{center}
\begin{tabular}{lccc}
\multicolumn{4}{l}{\footnotesize \textbf{Table 15.} Summary statistics $s(x)$ of the \texttt{%
BostonBomb2013} network dataset.} \\ \hline\hline
{\footnotesize Summary Statistic} & {\footnotesize Layer 1} & {\footnotesize Layer 2} & {\footnotesize %
Layer 3} \\ \hline
{\footnotesize \texttt{meanIndegree}} & {\footnotesize 1.35253692785} & {\footnotesize 0.6141674647} & 
{\footnotesize 0.1590769317} \\ 
{\footnotesize \texttt{varIndegree}} & {\footnotesize 8474.74052782986} & {\footnotesize 1804.0675198631}
& {\footnotesize 31.5665679809} \\ 
{\footnotesize \texttt{meanOutdegree}} & {\footnotesize 1.35253692785} & {\footnotesize 0.6141674647} & 
{\footnotesize 0.1590769317} \\ 
{\footnotesize \texttt{varOutdegree}} & {\footnotesize 7.61861428910} & {\footnotesize 4.5875615155} & 
{\footnotesize 0.8667280838} \\ 
{\footnotesize \texttt{wClusteringCoef}} & {\footnotesize 0.00008622621} & {\footnotesize 0.0004401573} & 
{\footnotesize 0.0001867074} \\ 
{\footnotesize \texttt{assortativityDegree}} & {\footnotesize -0.06229008973} & {\footnotesize %
-0.0233535797} & {\footnotesize -0.0165706996} \\ 
{\footnotesize \texttt{reciprocity}} & {\footnotesize 0.00379811641} & {\footnotesize 0.0379874337} & 
{\footnotesize 0.0625188215} \\ \hline\hline
\end{tabular}
\end{center}

The \texttt{copulaABCdrf} (Algorithm 1) was run on this model, based on these network data summaries $s(x)$, and based on summaries $s(y)$ of each network $y$ of $n_{\text{sim}}=500$ nodes simulated from the ERGM, conditionally on given proposed model parameters, in each iteration of this algorithm. The posterior distribution and MLE estimates delivered by the algorithm are shown in Table 16. This table also shows the posterior parameter estimates of \texttt{rejectionABC} using the same summary statistics and $n_{\text{sim}}$. The posterior estimates noticeably differed between these two ABC methods.

To provide a further investigation, a simulation study of \texttt{copulaABCdrf} and \texttt{rejectionABC} was conducted using the posterior mean estimates shown in Table 16 as the true data-generating parameters of the same multilayer ERGM, and based on $n_{\text{sim}}=100$ nodes and $n_{\text{sim}}=$ 10 and 20, while noting that even these analyses were very computationally expensive. Tables 17, 18, and 19 present the results of this simulation study. Table 17 shows the results show MSEs and MAEs for the posterior mean, median and mode estimates. Table 18 presents the MLE estimates compared to MCMLEs. Table 19 presents the mean 95\% and 50\% interval coverage. According to these three tables, both ABC methods were competitive, while \texttt{copulaABCdrf} tended to produce superior results.

\bigskip
\begin{center}
\begin{tabular}{lcccccccccc}
\multicolumn{11}{l}{\footnotesize \textbf{Table 16.} ABC posterior estimates, 3-layer valued ERGM parameters, from \texttt{BostonBomb2013} network dataset.} \\\hline\hline
{\footnotesize Layer 1} & {\footnotesize Method} & {\footnotesize Mean} & {\footnotesize Median} & {\footnotesize Mode} & {\footnotesize MLE} & {\footnotesize S.D.}
& {\footnotesize 2.5\%} & {\footnotesize 25\%} & {\footnotesize 75\%} & 
{\footnotesize 97.5\%} \\\hline 
\footnotesize\texttt{equalto.1.pm.0} & {\footnotesize \texttt{copulaABCdrf}} & {\footnotesize -5.65} & 
{\footnotesize -5.42} & {\footnotesize -4.07} & {\footnotesize -4.69} & 
{\footnotesize 1.38} & {\footnotesize -9.07} & {\footnotesize -6.30} & 
{\footnotesize -4.69} & {\footnotesize -3.59} \\ 
& {\footnotesize \texttt{rejectionABC}} & {\footnotesize -3.40} & {\footnotesize -3.73} & 
{\footnotesize -3.65} & {\footnotesize -5.55} & {\footnotesize 2.59} & 
{\footnotesize -7.45} & {\footnotesize -4.56} & {\footnotesize -2.96} & 
{\footnotesize 3.16} \\ 
\footnotesize\texttt{greaterthan.1} & {\footnotesize \texttt{copulaABCdrf}} & {\footnotesize -2.85} & 
{\footnotesize -3.18} & {\footnotesize -0.67} & {\footnotesize 1.33} & 
{\footnotesize 2.74} & {\footnotesize -7.72} & {\footnotesize -4.77} & 
{\footnotesize -1.06} & {\footnotesize 3.06} \\ 
& {\footnotesize \texttt{rejectionABC}} & {\footnotesize -1.47} & {\footnotesize -1.56} & 
{\footnotesize -1.95} & {\footnotesize -13.22} & {\footnotesize 3.37} & 
{\footnotesize -7.51} & {\footnotesize -3.33} & {\footnotesize 0.63} & 
{\footnotesize 4.54} \\ 
\footnotesize\texttt{mutual.min} & {\footnotesize \texttt{copulaABCdrf}} & {\footnotesize -3.85} & 
{\footnotesize -3.63} & {\footnotesize -2.43} & {\footnotesize 0.12} & 
{\footnotesize 2.20} & {\footnotesize -8.72} & {\footnotesize -5.27} & 
{\footnotesize -2.42} & {\footnotesize 0.14} \\ 
& {\footnotesize \texttt{rejectionABC}} & {\footnotesize -0.17} & {\footnotesize -0.13} & 
{\footnotesize 0.13} & {\footnotesize 4.18} & {\footnotesize 2.71} & 
{\footnotesize -5.56} & {\footnotesize -1.86} & {\footnotesize 1.65} & 
{\footnotesize 5.07} \\ 
\footnotesize\texttt{tw.min.sum.min} & {\footnotesize \texttt{copulaABCdrf}} & {\footnotesize 0.05} & 
{\footnotesize 0.18} & {\footnotesize -2.04} & {\footnotesize 1.24} & 
{\footnotesize 3.25} & {\footnotesize -6.71} & {\footnotesize -2.30} & 
{\footnotesize 2.38} & {\footnotesize 5.74} \\ 
& {\footnotesize \texttt{rejectionABC}} & {\footnotesize 3.06} & {\footnotesize 2.84} & 
{\footnotesize 0.12} & {\footnotesize 1.80} & {\footnotesize 2.31} & 
{\footnotesize -1.59 } & {\footnotesize 1.76} & {\footnotesize 4.67} & 
{\footnotesize 7.54} \\ 
\footnotesize\texttt{tw.min.max.min} & {\footnotesize \texttt{copulaABCdrf}} & {\footnotesize -1.88} & 
{\footnotesize -1.90} & {\footnotesize -0.33} & {\footnotesize -1.30} & 
{\footnotesize 2.89} & {\footnotesize -7.29} & {\footnotesize -3.84} & 
{\footnotesize 0.06} & {\footnotesize 3.78} \\ 
& {\footnotesize \texttt{rejectionABC}} & {\footnotesize 2.98} & {\footnotesize 3.01} & 
{\footnotesize 2.32} & {\footnotesize 5.96} & {\footnotesize 2.36} & 
{\footnotesize -1.05} & {\footnotesize 1.48} & {\footnotesize 4.31} & 
{\footnotesize 7.24} \\ 
\footnotesize\texttt{CMP} & {\footnotesize \texttt{copulaABCdrf}} & {\footnotesize -2.15} & 
{\footnotesize -2.24} & {\footnotesize 3.29} & {\footnotesize 5.45} & 
{\footnotesize 2.66} & {\footnotesize -7.33} & {\footnotesize -3.89} & 
{\footnotesize -0.38} & {\footnotesize 3.04} \\ 
& {\footnotesize \texttt{rejectionABC}} & {\footnotesize -0.11} & {\footnotesize -0.19} & 
{\footnotesize 1.79} & {\footnotesize 5.17} & {\footnotesize 3.65} & 
{\footnotesize -7.69} & {\footnotesize -2.08} & {\footnotesize 2.36} & 
{\footnotesize 7.23} \\ \hline
{\footnotesize Layer 2} & {\footnotesize Method} & {\footnotesize Mean} & {\footnotesize Median} & {\footnotesize Mode} & {\footnotesize MLE} & {\footnotesize S.D.}
& {\footnotesize 2.5\%} & {\footnotesize 25\%} & {\footnotesize 75\%} & 
{\footnotesize 97.5\%} \\ \hline
\footnotesize\footnotesize\texttt{equalto.1.pm.0} & {\footnotesize \texttt{copulaABCdrf}} & {\footnotesize -6.39} & 
{\footnotesize -6.33} & {\footnotesize -3.81} & {\footnotesize -5.26} & 
{\footnotesize 1.44} & {\footnotesize -9.76} & {\footnotesize -7.26} & 
{\footnotesize -5.39} & {\footnotesize -3.91} \\ 
& {\footnotesize \texttt{rejectionABC}} & {\footnotesize -0.08} & {\footnotesize 0.06} & 
{\footnotesize 0.19} & {\footnotesize -0.06} & {\footnotesize 3.17} & 
{\footnotesize -5.75} & {\footnotesize -2.44} & {\footnotesize 1.72} & 
{\footnotesize 6.60} \\ 
\footnotesize\texttt{greaterthan.1} & {\footnotesize \texttt{copulaABCdrf}} & {\footnotesize -2.76} & 
{\footnotesize -2.66} & {\footnotesize 1.68} & {\footnotesize 3.49} & 
{\footnotesize 2.80} & {\footnotesize -7.83} & {\footnotesize -4.71} & 
{\footnotesize -1.02} & {\footnotesize 3.13} \\ 
& {\footnotesize \texttt{rejectionABC}} & {\footnotesize -0.44} & {\footnotesize -0.32} & 
{\footnotesize -4.59} & {\footnotesize -2.16} & {\footnotesize 3.38} & 
{\footnotesize -6.38} & {\footnotesize -2.88} & {\footnotesize 1.97} & 
{\footnotesize 6.20} \\ 
\footnotesize\texttt{mutual.min} & {\footnotesize \texttt{copulaABCdrf}} & {\footnotesize 1.70} & 
{\footnotesize 1.59} & {\footnotesize 0.57} & {\footnotesize 0.44} & 
{\footnotesize 1.25} & {\footnotesize -0.32} & {\footnotesize 0.78} & 
{\footnotesize 2.57} & {\footnotesize 3.83} \\ 
& {\footnotesize \texttt{rejectionABC}} & {\footnotesize 0.16} & {\footnotesize 0.50} & 
{\footnotesize 0.64} & {\footnotesize -2.14} & {\footnotesize 2.99} & 
{\footnotesize -5.79} & {\footnotesize -1.67} & {\footnotesize 1.70} & 
{\footnotesize 5.91} \\ 
\footnotesize\texttt{tw.min.sum.min} & {\footnotesize \texttt{copulaABCdrf}} & {\footnotesize 0.23} & 
{\footnotesize 0.27} & {\footnotesize -2.83} & {\footnotesize 3.39} & 
{\footnotesize 2.98} & {\footnotesize -5.81} & {\footnotesize -1.64} & 
{\footnotesize 2.33} & {\footnotesize 5.87} \\ 
& {\footnotesize \texttt{rejectionABC}} & {\footnotesize 1.98} & {\footnotesize 1.42} & 
{\footnotesize 0.60} & {\footnotesize 8.29} & {\footnotesize 3.14} & 
{\footnotesize -3.33} & {\footnotesize -0.10} & {\footnotesize 4.11} & 
{\footnotesize 8.10} \\ 
\footnotesize\texttt{tw.min.max.min} & {\footnotesize \texttt{copulaABCdrf}} & {\footnotesize -2.48} & 
{\footnotesize -2.45} & {\footnotesize 1.65} & {\footnotesize 1.59} & 
{\footnotesize 2.79} & {\footnotesize -8.02} & {\footnotesize -4.36} & 
{\footnotesize -0.64} & {\footnotesize 3.30} \\ 
& {\footnotesize \texttt{rejectionABC}} & {\footnotesize 1.37} & {\footnotesize 1.41} & 
{\footnotesize 2.10} & {\footnotesize 5.36} & {\footnotesize 2.79} & 
{\footnotesize -3.48} & {\footnotesize -0.60} & {\footnotesize 3.13} & 
{\footnotesize 6.96} \\ 
\texttt{CMP} & {\footnotesize \texttt{copulaABCdrf}} & {\footnotesize -2.04} & 
{\footnotesize -2.10} & {\footnotesize -2.82} & {\footnotesize 4.43} & 
{\footnotesize 2.93} & {\footnotesize -7.55} & {\footnotesize -3.95} & 
{\footnotesize -0.01} & {\footnotesize 3.84} \\ 
& {\footnotesize \texttt{rejectionABC}} & {\footnotesize 0.18} & {\footnotesize 0.32} & 
{\footnotesize 0.83} & {\footnotesize -2.18} & {\footnotesize 3.27} & 
{\footnotesize -5.74} & {\footnotesize -2.00} & {\footnotesize 2.54} & 
{\footnotesize 5.99} \\ \hline
{\footnotesize Layer 3} & {\footnotesize Method} & {\footnotesize Mean} & {\footnotesize Median} & {\footnotesize Mode} & {\footnotesize MLE} & {\footnotesize S.D.}
& {\footnotesize 2.5\%} & {\footnotesize 25\%} & {\footnotesize 75\%} & {\footnotesize 97.5\%} \\ \hline
\footnotesize\texttt{equalto.1.pm.0} & {\footnotesize \texttt{copulaABCdrf}} & {\footnotesize -5.67} & 
{\footnotesize -5.62} & {\footnotesize -4.63} & {\footnotesize -4.28} & 
{\footnotesize 1.25} & {\footnotesize -8.44} & {\footnotesize -6.52} & 
{\footnotesize -4.78} & {\footnotesize -3.71} \\ 
& {\footnotesize \texttt{rejectionABC}} & {\footnotesize 0.41} & {\footnotesize 0.31} & 
{\footnotesize 0.24} & {\footnotesize 4.40} & {\footnotesize 3.22} & 
{\footnotesize -5.31} & {\footnotesize -1.63} & {\footnotesize 2.47} & 
{\footnotesize 7.00} \\ 
\footnotesize\texttt{greaterthan.1} & {\footnotesize \texttt{copulaABCdrf}} & {\footnotesize -3.22} & 
{\footnotesize -3.17} & {\footnotesize 0.82} & {\footnotesize -0.24} & 
{\footnotesize 2.76} & {\footnotesize -8.22} & {\footnotesize -5.20} & 
{\footnotesize -1.45} & {\footnotesize 2.37} \\ 
& {\footnotesize \texttt{rejectionABC}} & {\footnotesize 0.21} & {\footnotesize -0.21} & 
{\footnotesize -1.93} & {\footnotesize 3.16} & {\footnotesize 2.73} & 
{\footnotesize -4.26} & {\footnotesize -1.80} & {\footnotesize 1.79} & 
{\footnotesize 6.05} \\ 
\footnotesize\texttt{mutual.min} & {\footnotesize \texttt{copulaABCdrf}} & {\footnotesize 2.62} & 
{\footnotesize 2.41} & {\footnotesize 4.05} & {\footnotesize 0.78} & 
{\footnotesize 1.21} & {\footnotesize 0.65} & {\footnotesize 1.86} & 
{\footnotesize 3.33} & {\footnotesize 5.16} \\ 
& {\footnotesize \texttt{rejectionABC}} & {\footnotesize -0.07} & {\footnotesize 0.03} & 
{\footnotesize 0.77} & {\footnotesize -2.10} & {\footnotesize 3.23} & 
{\footnotesize -5.58} & {\footnotesize -2.57} & {\footnotesize 2.22} & 
{\footnotesize 5.60} \\ 
\footnotesize\texttt{tw.min.sum.min} & {\footnotesize \texttt{copulaABCdrf}} & {\footnotesize -1.14} & 
{\footnotesize -1.23} & {\footnotesize -0.15} & {\footnotesize -6.25} & 
{\footnotesize 2.82} & {\footnotesize -6.86} & {\footnotesize -2.79} & 
{\footnotesize 0.68} & {\footnotesize 4.23} \\ 
& {\footnotesize \texttt{rejectionABC}} & {\footnotesize 0.22} & {\footnotesize 0.35} & 
{\footnotesize 0.18} & {\footnotesize 2.70} & {\footnotesize 3.26} & 
{\footnotesize -5.67} & {\footnotesize -1.81} & {\footnotesize 1.98} & 
{\footnotesize 5.74} \\ 
\footnotesize\texttt{tw.min.max.min} & {\footnotesize \texttt{copulaABCdrf}} & {\footnotesize -2.15} & 
{\footnotesize -2.28} & {\footnotesize -3.66} & {\footnotesize -0.16} & 
{\footnotesize 2.85} & {\footnotesize -7.68} & {\footnotesize -4.03} & 
{\footnotesize -0.33} & {\footnotesize 3.92} \\ 
& {\footnotesize \texttt{rejectionABC}} & {\footnotesize 0.31} & {\footnotesize 0.46} & 
{\footnotesize 1.22} & {\footnotesize -0.54} & {\footnotesize 2.67} & 
{\footnotesize -5.07} & {\footnotesize -1.24} & {\footnotesize 1.84} & 
{\footnotesize 6.21} \\ 
\footnotesize\texttt{CMP} & {\footnotesize \texttt{copulaABCdrf}} & {\footnotesize -1.34} & 
{\footnotesize -1.20} & {\footnotesize -1.99} & {\footnotesize 2.29} & 
{\footnotesize 2.72} & {\footnotesize -6.82} & {\footnotesize -3.20} & 
{\footnotesize 0.47} & {\footnotesize 4.11} \\ 
& {\footnotesize \texttt{rejectionABC}} & {\footnotesize -0.32} & {\footnotesize -0.47} & 
{\footnotesize -1.09} & {\footnotesize -1.36} & {\footnotesize 2.98} & 
{\footnotesize -5.98} & {\footnotesize -2.00} & {\footnotesize 1.72} & 
{\footnotesize 5.59} \\ \hline\hline
\multicolumn{11}{l}{\footnotesize Note:\ \texttt{tw} refers to \texttt{transitiveweights}.}\\
\multicolumn{11}{l}{\footnotesize }
\end{tabular}
\end{center}

\bigskip

\begin{changemargin}{-.7in}{-.7in}
\begin{center}
{\footnotesize
\begin{tabular}{lccccccc}
\multicolumn{8}{l}{\textbf{Table 17.} ABC posterior mean, median, and mode estimates of the 3-layer valued ERGM parameters from network datasets} \\ 
\multicolumn{8}{l}{simulated to reflect the \texttt{BostonBomb2013} network dataset, over 10 replications, for $n=100$ nodes and for ${n}_{\text{sim}}$ = 10 and 20.}\\\hline\hline
               & $n_{\text{sim}} =$ & 10           & 20           & 10           & 20           & 10           & 20           \\\hline
Layer 1        & Method   & Mean         & Mean         & Median       & Median       & Mode         & Mode         \\\hline
\texttt{equalto.1.pm.0} & \texttt{copulaABCdrf}    & \textbf{3.15} (\textbf{9.96})  & \textbf{2.28} (\textbf{5.19})  & \textbf{3.34} (\textbf{11.20}) & \textbf{2.43} (\textbf{5.93})  & 4.04 (17.43) & \textbf{3.38} (\textbf{11.95}) \\
               & \texttt{rejectionABC}     & 4.02 (16.28) & 3.91 (15.37) & 4.08 (16.97) & 4.00 (16.3)     & \textbf{3.82} (\textbf{16.49}) & 3.76 (16.02) \\
\texttt{greaterthan.1}  & \texttt{copulaABCdrf}    & 1.92 (3.71)  & \textbf{1.00} (\textbf{1.10})  & 1.98 (4.00)  & \textbf{1.03} (\textbf{1.17})  & \textbf{1.81} (4.17)  & \textbf{1.95} (\textbf{6.06})  \\
               & \texttt{rejectionABC}     & \textbf{1.72} (\textbf{3.59})  & 1.52 (2.67)  & \textbf{1.48} (\textbf{2.79})  & 1.18 (1.87)  & 1.91 (5.31)  & 2.40 (6.89)   \\
\texttt{mutual.min}     & \texttt{copulaABCdrf}    & \textbf{1.06} (\textbf{1.23})  & \textbf{1.14} (\textbf{1.36})  & \textbf{1.25} (\textbf{1.64})  & \textbf{1.31} (\textbf{1.80})  & \textbf{2.15} (\textbf{6.03})  & 3.20 (13.16)  \\
               & \texttt{rejectionABC}     & 1.89 (3.98)  & 2.19 (5.37)  & 1.83 (4.04)  & 2.11 (4.98)  & 2.82 (8.81)  & \textbf{2.91} (\textbf{11.09}) \\
\texttt{tw.min.sum.min} & \texttt{copulaABCdrf}    & 2.25 (\textbf{5.08})  & 2.27 (5.19)  & \textbf{2.17} (\textbf{4.71})  & 2.15 (4.69)  & \textbf{0.87} (\textbf{1.13})  & \textbf{1.22} (\textbf{2.26})  \\
               & \texttt{rejectionABC}     & \textbf{2.22} (5.29)  & \textbf{1.97} (\textbf{4.04})  & 2.32 (5.71)  & \textbf{1.81} (\textbf{3.60})   & 2.69 (10.21) & 1.80 (4.43)   \\
\texttt{tw.min.max.min} & \texttt{copulaABCdrf}    & \textbf{0.09} (\textbf{0.01})  & \textbf{0.30} (\textbf{0.14})  & \textbf{0.07} (\textbf{0.01})  & \textbf{0.25} (\textbf{0.10})  & 1.64 (3.62)  & 1.96 (6.51)  \\
               & \texttt{rejectionABC}     & 0.48 (0.31)  & 0.76 (0.90)   & 0.45 (0.36)  & 0.82 (1.00)     & \textbf{1.38} (\textbf{3.29})  & \textbf{1.39} (\textbf{2.09})  \\
\texttt{CMP}            & \texttt{copulaABCdrf}    & \textbf{0.13} (\textbf{0.03})  & \textbf{0.26} (\textbf{0.08})  & \textbf{0.27} (\textbf{0.12})  & \textbf{0.15} (\textbf{0.02})  & 1.96 (4.60)   & 2.11 (5.07)  \\
               & \texttt{rejectionABC}     & 0.41 (0.33)  & 0.34 (0.18)  & 0.66 (0.70)   & 0.43 (0.23)  & \textbf{1.31} (\textbf{3.26})  & \textbf{1.03} (\textbf{1.50})   \\\hline
Layer 2        & Method   & Mean         & Mean         & Median       & Median       & Mode         & Mode         \\\hline
\texttt{equalto.1.pm.0} & \texttt{copulaABCdrf}    & \textbf{3.96} (\textbf{15.73}) & \textbf{3.11} (\textbf{9.69})  & \textbf{4.13} (\textbf{17.11}) & \textbf{3.34} (\textbf{11.16}) & \textbf{4.15} (\textbf{18.40})  & \textbf{4.67} (\textbf{22.17}) \\
               & \texttt{rejectionABC}     & 4.77 (23.18) & 4.35 (19.03) & 4.92 (24.50)  & 4.34 (19.00)    & 5.41 (33.17) & 4.96 (26.54) \\
\texttt{greaterthan.1}  & \texttt{copulaABCdrf}    & \textbf{1.78} (\textbf{3.21})  & \textbf{0.94} (\textbf{0.92})  & \textbf{1.79} (\textbf{3.30})  & \textbf{1.02} (\textbf{1.09})  & 3.71 (17.44) & 2.34 (7.59)  \\
               & \texttt{rejectionABC}     & 2.03 (4.84)  & 1.31 (1.88)  & 2.06 (5.32)  & 1.11 (1.35)  & \textbf{2.06} (\textbf{6.73})  & \textbf{1.78} (\textbf{4.80})   \\
\texttt{mutual.min}     & \texttt{copulaABCdrf}    & \textbf{3.04} (\textbf{9.83})  & \textbf{2.36} (\textbf{6.27})  & \textbf{2.91} (\textbf{9.06})  & \textbf{2.28} (\textbf{5.83})  & \textbf{1.92} (\textbf{4.17})  & 2.91 (9.06)  \\
               & \texttt{rejectionABC}     & 3.72 (14.03) & 3.65 (13.78) & 3.80 (14.86)  & 3.39 (12.10)  & 3.65 (15.88) & \textbf{1.86} (\textbf{4.75})  \\
\texttt{tw.min.sum.min} & \texttt{copulaABCdrf}    & 2.36 (5.60)  & 2.56 (6.57)  & 2.26 (\textbf{5.15})  & 2.57 (6.62)  & \textbf{1.43} (\textbf{3.07})  & 1.68 (4.81)  \\
               & \texttt{rejectionABC}     & \textbf{2.33} (\textbf{5.80})   & \textbf{2.18} (\textbf{5.16})  & \textbf{2.22} (5.60)   & \textbf{2.18} (\textbf{5.30})   & 3.00 (15.47)    & \textbf{1.49} (\textbf{3.51})  \\
\texttt{tw.min.max.min} & \texttt{copulaABCdrf}    & \textbf{0.56} (\textbf{0.33})  & \textbf{0.41} (\textbf{0.20})  & \textbf{0.61} (\textbf{0.41})  & \textbf{0.47} (\textbf{0.27})  & 1.45 (2.88)  & 1.83 (7.65)  \\
               & \texttt{rejectionABC}     & 0.61 (0.43)  & 0.81 (0.81)  & 0.67 (0.59)  & 1.10 (1.38)   & \textbf{1.33} (\textbf{2.62})  & \textbf{1.20} (\textbf{1.89})   \\
\texttt{CMP}            & \texttt{copulaABCdrf}    & \textbf{0.12} (\textbf{0.02})  & \textbf{0.32} (\textbf{0.14})  & \textbf{0.17} (\textbf{0.04})  & \textbf{0.18} (\textbf{0.06})  & 1.39 (2.79)  & 2.21 (6.37)  \\
               & \texttt{rejectionABC}     & 0.59 (0.61)  & 0.75 (0.92)  & 0.73 (0.85)  & 0.58 (0.78)  & \textbf{1.19} (\textbf{2.24})  & \textbf{1.68} (\textbf{3.68})  \\\hline
Layer 3        & Method   & Mean         & Mean         & Median       & Median       & Mode         & Mode         \\\hline
\texttt{equalto.1.pm.0} & \texttt{copulaABCdrf}    & \textbf{3.94} (\textbf{15.57}) & \textbf{2.61} (\textbf{6.91})  & \textbf{4.20} (\textbf{17.72}) & \textbf{2.88} (\textbf{8.37})  & 4.92 (25.87) & 2.91 (9.88)  \\
               & \texttt{rejectionABC}     & 4.56 (21.06) & 3.73 (14.11) & 4.59 (21.38) & 3.75 (14.41) & \textbf{4.68} (\textbf{24.47}) & \textbf{2.84} (\textbf{9.46})  \\
\texttt{greaterthan.1}  & \texttt{copulaABCdrf}    & 2.45 (6.03)  & \textbf{1.45} (\textbf{2.19})  & 2.52 (6.40)  & \textbf{1.48} (\textbf{2.25})  & 3.32 (12.91) & \textbf{2.54} (\textbf{9.59})  \\
               & \texttt{rejectionABC}     & \textbf{2.04} (\textbf{4.34})  & 2.38 (6.05)  & \textbf{2.05} (\textbf{4.67})  & 2.26 (5.35)  & \textbf{1.68} (\textbf{3.75})  & 3.66 (16.5)  \\
\texttt{mutual.min}     & \texttt{copulaABCdrf}    & \textbf{2.79} (\textbf{7.81})  & \textbf{2.12} (\textbf{4.64})  & \textbf{2.70} (\textbf{7.30})  & \textbf{2.07} (\textbf{4.37})  & \textbf{3.57} (\textbf{13.85}) & \textbf{2.40} (\textbf{6.56})   \\
               & \texttt{rejectionABC}     & 4.81 (23.87) & 4.74 (22.61) & 4.77 (23.84) & 4.57 (21.18) & 4.05 (21.73) & 4.55 (23.11) \\
\texttt{tw.min.sum.min} & \texttt{copulaABCdrf}    & 0.97 (\textbf{0.96})  & 1.08 (\textbf{1.19})  & \textbf{0.88}  (\textbf{0.80})  & 1.01 (1.05)  & \textbf{0.93} (\textbf{1.15})  & \textbf{0.86} (\textbf{1.07})  \\
               & \texttt{rejectionABC}     & \textbf{0.96} (1.02)  & \textbf{0.98} (1.22)  & 0.99 (1.16)  & \textbf{0.74} (\textbf{0.85})  & 1.46 (2.61)  & 2.64 (9.04)  \\
\texttt{tw.min.max.min} & \texttt{copulaABCdrf}    & \textbf{0.21} (\textbf{0.05})  & \textbf{0.19} (\textbf{0.05})  & \textbf{0.28} (\textbf{0.10})  & 0.30 (0.12)  & 2.21 (\textbf{6.86})  & 1.98 (6.30)   \\
               & \texttt{rejectionABC}     & 0.61 (0.51)  & 0.26 (0.10)   & 0.67 (0.59)  & \textbf{0.24} (\textbf{0.07})  & \textbf{2.16} (6.89)  & \textbf{1.49} (\textbf{3.71})  \\
\texttt{CMP}            & \texttt{copulaABCdrf}    & 0.89 (0.83)  & 1.06 (1.22)  & 0.65 (\textbf{0.48})  & 0.80 (0.76)  & \textbf{1.12} (\textbf{2.14})  & 1.53 (3.00)     \\
               & \texttt{rejectionABC}     & \textbf{0.50} (\textbf{0.33})   & \textbf{0.67} (\textbf{0.67})  & \textbf{0.64} (0.56)  & \textbf{0.69} (\textbf{0.61})  & 1.77 (5.75)  & \textbf{0.67} (\textbf{0.68})  \\\hline\hline
\multicolumn{8}{l}{\footnotesize \textit{Note:} \textbf{Bold} indicates the more accurate ABC method for the given estimator. \texttt{tw} refers to \texttt{transitiveweights}.}
\end{tabular}}
\end{center}
\end{changemargin}

\begin{center}
{\footnotesize
\begin{tabular}{lccccccc}
\multicolumn{6}{l}{\textbf{Table 18.} ABC MLEs and MCMLEs of the 3-layer valued ERGM parameters from network} \\ 
\multicolumn{6}{l}{datasets simulated to reflect the BostonBomb2013 network dataset, over 10 replications,}\\
\multicolumn{6}{l}{for $n=100$ nodes and for ${n}_{\text{sim}}$ = 10 and 20.}\\
\hline\hline
               & $n_{\text{sim}} =$ & 10           & 20           & 10           & 20           \\\hline
Layer 1        & Method   & MLE          & MLE          & MCMLE        & MCMLE        \\\hline
\texttt{equalto.1.pm.0} & \texttt{copulaABCdrf}    & \textbf{3.63} (\textbf{15.07}) & \textbf{3.27} (\textbf{11.11}) & 0.19 (0.05)  & 0.21 (0.06)  \\
               & \texttt{rejectionABCkern}     & 4.20 (18.15)  & 4.34 (19.74) &              &              \\
\texttt{greaterthan.1}  & \texttt{copulaABCdrf}    & 3.15 (13.25) & 2.90 (\textbf{8.75})   & 1.31 (5.35)  & 0.42 (0.24)  \\
               & \texttt{rejectionABCkern}     & \textbf{1.79} (\textbf{6.89})  & \textbf{2.61} (8.85)  &              &              \\
\texttt{mutual.min}     & \texttt{copulaABCdrf}    & \textbf{1.72} (\textbf{4.33})  & 2.19 (6.15)  & 4.96 (43.20) & 6.79 (138.42)\\
               & \texttt{rejectionABCkern}     & 2.30 (6.95)   & \textbf{1.68} (\textbf{3.74})  &              &              \\
\texttt{tw.min.sum.min} & \texttt{copulaABCdrf}    & \textbf{2.16} (\textbf{5.91})  & 3.10 (16.61)  & $\infty$ ($\infty$) & $\infty$ ($\infty$)    \\
               & \texttt{rejectionABCkern}     & 3.05 (15.01) & \textbf{1.30} (\textbf{1.96})   &              &              \\
\texttt{tw.min.max.min} & \texttt{copulaABCdrf}    & \textbf{2.23} (\textbf{9.67})  & 2.82 (8.40)   & $\infty$ ($\infty$) & $\infty$ ($\infty$)    \\
               & \texttt{rejectionABCkern}     & 3.00 (16.48) & \textbf{2.27} (\textbf{6.40})   &              &  \\
\texttt{CMP}           & \texttt{copulaABCdrf}    & \textbf{1.48} (\textbf{2.83})  & \textbf{1.78} (\textbf{4.38})  & 1.71 (9.56)  & 0.36 (0.18)  \\
               & \texttt{rejectionABCkern}     & 2.00 (5.89)  & 3.32 (20.56) &              &              \\\hline
Layer 2        & Method   & MLE          & MLE          & MCMLE        & MCMLE        \\\hline
\texttt{equalto.1.pm.0} & \texttt{copulaABCdrf}    & \textbf{4.36} (\textbf{20.49}) & \textbf{3.62} (\textbf{13.42}) & 0.39 (0.20)  & 0.35 (0.15)  \\
               & \texttt{rejectionABCkern}     & 5.00 (31.16)    & 4.79 (26.93) &              &              \\
\texttt{greaterthan.1}  & \texttt{copulaABCdrf}    & 2.88 (8.89)  & 2.93 (11.84) & 0.33 (0.16)  & 0.50 (0.39)  \\
               & \texttt{rejectionABCkern}     & \textbf{2.48} (\textbf{8.05})  & \textbf{2.17} (\textbf{5.73})  &              &              \\
\texttt{mutual.min}     & \texttt{copulaABCdrf}    & 4.24 (26.38) & \textbf{1.24} (\textbf{1.61})  & 0.38 (0.22)  & 0.48 (0.35)  \\
               & \texttt{rejectionABCkern}     & \textbf{3.29} (\textbf{18.69}) & 4.63 (39.55) &              &              \\
\texttt{tw.min.sum.min} & \texttt{copulaABCdrf}    & \textbf{2.25} (\textbf{7.67})  & 5.12 (28.20)  & $\infty$ ($\infty$)    & $\infty$ ($\infty$)    \\
               & \texttt{rejectionABCkern}     & 3.36 (17.19) & \textbf{2.24} (\textbf{8.13})  &              &              \\
\texttt{tw.min.max.min} & \texttt{copulaABCdrf}    & 2.35 (8.18)  & \textbf{3.11} (14.38) & $\infty$ ($\infty$)    & $\infty$ ($\infty$)           \\
               & \texttt{rejectionABCkern}     & \textbf{1.62} (\textbf{3.58})  & 3.31 (\textbf{14.09}) &  &   \\
\texttt{CMP}            & \texttt{copulaABCdrf}    & \textbf{1.61} (\textbf{3.46})  & 1.38 (2.83)  & 0.30 (0.12)  & 0.49 (0.40)  \\
               & \texttt{rejectionABCkern}     & 2.05 (5.21)  & \textbf{1.04} (\textbf{1.76})  &     &    \\\hline
Layer 3        & Method   & MLE          & MLE          & MCMLE        & MCMLE        \\\hline
\texttt{equalto.1.pm.0} & \texttt{copulaABCdrf}    & 4.39 (\textbf{21.34}) & 3.56 (\textbf{13.41}) & 0.24 (0.07)  & 0.23 (0.06)  \\
               & \texttt{rejectionABCkern}     & \textbf{4.21} (21.71) & \textbf{3.29} (17.39) &              &              \\
\texttt{greaterthan.1}  & \texttt{copulaABCdrf}    & 4.67 (30.13) & 3.36 (15.32) & 0.24 (0.09)  & 0.23 (0.08) \\
               & \texttt{rejectionABCkern}     & \textbf{2.13} (\textbf{5.91})  & \textbf{3.14} (\textbf{10.99})& & \\
\texttt{mutual.min}     & \texttt{copulaABCdrf}    & \textbf{2.79} (\textbf{9.52})&\textbf{2.85} (\textbf{8.74})&0.58 (0.36)  & 0.68 (0.47)\\
               & \texttt{rejectionABCkern}     & 5.45 (35.26) & 6.51 (46.49) &              &              \\
\texttt{tw.min.sum.min} & \texttt{copulaABCdrf}    & 2.14 (5.98)  & 3.23 (15.80)  & $\infty$ ($\infty$)    & $\infty$ ($\infty$)    \\
               & \texttt{rejectionABCkern}     & \textbf{2.03} (\textbf{5.16})  & \textbf{2.97} (\textbf{11.92}) &&\\
\texttt{tw.min.max.min} & \texttt{copulaABCdrf}    & \textbf{1.85} (\textbf{4.07})  & \textbf{1.39} (\textbf{2.41})  & $\infty$ ($\infty$)    & $\infty$ ($\infty$)    \\
               & \texttt{rejectionABCkern}     & 2.43 (11.27) & 1.58 (3.61)  &              &              \\
\texttt{CMP}            & \texttt{copulaABCdrf}    & \textbf{1.32} (\textbf{2.44})  & \textbf{0.93} (\textbf{1.29})  & 0.20 (0.08)  & 0.31 (0.12)  \\
               & \texttt{rejectionABCkern}     & 1.62 (4.22)  & 3.34 (16.36) &              &              \\\hline\hline
\multicolumn{6}{l}{\footnotesize \textit{Note:} \textbf{Bold} indicates the more accurate ABC method for the given estimator.}\\
\multicolumn{6}{l}{\footnotesize \texttt{tw} refers to \texttt{transitiveweights}.}
\end{tabular}}
\end{center}

\begin{center}
{\footnotesize
\begin{tabular}{lccccc}

\multicolumn{6}{l}{\textbf{Table 19.} ABC and MCMLE mean 95\% and 50\% coverage of the 3-layer valued ERGM} \\ 
\multicolumn{6}{l}{parameters from network datasets simulated to reflect the BostonBomb2013 network dataset, }\\
\multicolumn{6}{l}{over 10 replications, for $n=100$ nodes and for ${n}_{\text{sim}}$ = 10 and 20.}\\\hline\hline
               & $n_{\text{sim}} =$ & 10          & 20          & 10             & 20             \\\hline
Layer 1        & Method   & 95\%(50\%)c  & 95\%(50\%)c& MCMLE95\%c & MCMLE95\%c \\\hline
\texttt{equalto.1.pm.0} & \texttt{copulaABCdrf}    & \textbf{0.90} (\textbf{0.00}) & \textbf{1.00} (\textbf{0.00}) & 0.90           & 0.80           \\
               & \texttt{rejectionABC}     & 0.30 (\textbf{0.00}) & 0.20 (\textbf{0.00}) &     &         \\
\texttt{greaterthan.1}  & \texttt{copulaABCdrf}    & \textbf{1.00} (0.30) & \textbf{1.00} (1.00) & 1.00  & 1.00  \\
               & \texttt{rejectionABC}     & \textbf{1.00} (\textbf{0.40}) & \textbf{1.00} (\textbf{0.60}) &  &   \\
\texttt{mutual.min}     & \texttt{copulaABCdrf}    & \textbf{1.00} (\textbf{0.70}) & \textbf{1.00} (\textbf{0.60}) & 1.00   & 1.00 \\
               & \texttt{rejectionABC}     & \textbf{1.00} (0.20) & \textbf{1.00} (0.20) &      &        \\
\texttt{tw.min.sum.min} & \texttt{copulaABCdrf}    & \textbf{1.00} (0.00) & \textbf{1.00} (0.00) & 0.10  & 0.00  \\
               & \texttt{rejectionABC}     & \textbf{1.00} (\textbf{0.10}) & \textbf{1.00} (\textbf{0.20}) &  &\\
\texttt{tw.min.max.min} & \texttt{copulaABCdrf}    & \textbf{1.00} (\textbf{1.00}) & \textbf{1.00} (\textbf{1.00}) & 0.10 & 0.00   \\
               & \texttt{rejectionABC}     & \textbf{1.00} (\textbf{1.00}) & \textbf{1.00} (\textbf{1.00}) &   &           \\
\texttt{CMP}            & \texttt{copulaABCdrf}    & \textbf{1.00} (\textbf{1.00}) & \textbf{1.00} (\textbf{1.00}) & 1.00  & 1.00  \\
               & \texttt{rejectionABC}     & \textbf{1.00} (\textbf{1.00}) & \textbf{1.00} (\textbf{1.00}) & &\\\hline
Layer 2        & Method   & 95\%(50\%)c & 95\%(50\%)c  & MCMLE95\%c & MCMLE95\%c \\\hline
\texttt{equalto.1.pm.0} & \texttt{copulaABCdrf}    & \textbf{0.50} (\textbf{0.00}) & \textbf{0.80} (\textbf{0.00}) & 0.70           & 0.80  \\
               & \texttt{rejectionABC}     & 0.30 (\textbf{0.00}) & 0.00 (\textbf{0.00}) & &    \\
\texttt{greaterthan.1}  & \texttt{copulaABCdrf}    & \textbf{1.00} (\textbf{0.60}) & \textbf{1.00} (1.00) & 1.00   & 1.00 \\
               & \texttt{rejectionABC}     & \textbf{1.00} (\textbf{0.40}) & \textbf{1.00} (\textbf{0.80})  &    &         \\
\texttt{mutual.min}     & \texttt{copulaABCdrf}    & 0.80 (\textbf{0.00}) & \textbf{1.00} (\textbf{0.00}) & 0.60 & 0.40 \\
               & \texttt{rejectionABC}     & \textbf{1.00} (\textbf{0.00}) & 0.60 (\textbf{0.00}) &   &         \\
\texttt{tw.min.sum.min} & \texttt{copulaABCdrf}    & \textbf{1.00} (0.00) & \textbf{1.00 }(\textbf{0.00}) & 0.00 & 0.00 \\
               & \texttt{rejectionABC}     & \textbf{1.00} (\textbf{0.10}) & \textbf{1.00} (\textbf{0.00}) &     &         \\
\texttt{tw.min.max.min} & \texttt{copulaABCdrf}    & \textbf{1.00} (1.00) & \textbf{1.00} (1.00) & 0.00  & 0.00  \\
               & \texttt{rejectionABC}     & \textbf{1.00} (\textbf{1.00}) & \textbf{1.00} (\textbf{0.60})  &      &  \\
\texttt{CMP}            & \texttt{copulaABCdrf}    & \textbf{1.00} (\textbf{1.00}) & \textbf{1.00} (1.00) & 1.00  & 1.00 \\
               & \texttt{rejectionABC}     & \textbf{1.00} (\textbf{0.90}) & \textbf{1.00} (\textbf{0.80}) &       &\\\hline
Layer 3        & Method   & 95\%(50\%)c  & 95\%(50\%)c& MCMLE95\%c & MCMLE95\%c \\\hline
\texttt{equalto.1.pm.0} & \texttt{copulaABCdrf}    & \textbf{0.90} (\textbf{0.00}) & \textbf{1.00} (\textbf{0.00}) & 0.60  & 0.80\\
               & \texttt{rejectionABC}     & 0.00 (\textbf{0.00}) & 0.60 (\textbf{0.00}) &     &         \\
\texttt{greaterthan.1}  & \texttt{copulaABCdrf}    & \textbf{1.00} (0.00) & \textbf{1.00} (\textbf{0.60}) & 0.90  & 1.00 \\
               & \texttt{rejectionABC}     & \textbf{1.00} (\textbf{0.10}) & \textbf{1.00} (0.00) & & \\
\texttt{mutual.min}     & \texttt{copulaABCdrf}    & \textbf{0.50} (\textbf{0.00}) & \textbf{0.60} (\textbf{0.00}) & 0.00 & 0.00  \\
               & \texttt{rejectionABC}     & 0.30 (\textbf{0.00}) & 0.20 (\textbf{0.00}) &        &      \\
\texttt{tw.min.sum.min} & \texttt{copulaABCdrf}    & \textbf{1.00} (1.00) & \textbf{1.00} (1.00) & 0.00 & 0.00   \\
               & \texttt{rejectionABC}     & \textbf{1.00} (\textbf{0.90}) & \textbf{1.00} (\textbf{0.80}) &      &        \\
\texttt{tw.min.max.min} & \texttt{copulaABCdrf}    & \textbf{1.00} (\textbf{1.00}) & \textbf{1.00} (\textbf{1.00}) & 0.00 & 0.00   \\
               & \texttt{rejectionABC}     & \textbf{1.00} (\textbf{1.00}) & \textbf{1.00} (\textbf{1.00}) &      &        \\
\texttt{CMP}            & \texttt{copulaABCdrf}    & \textbf{1.00} (\textbf{1.00)} & \textbf{1.00} (1.00) & 0.90 & 0.80 \\
               & \texttt{rejectionABC}     & \textbf{1.00} (\textbf{1.00}) & \textbf{1.00} (\textbf{0.80}) &      & \\\hline\hline
\multicolumn{6}{l}{\footnotesize \textit{Note:} \textbf{Bold} indicates the more accurate ABC method for the given estimator.}
\end{tabular}}
\end{center}

\subsection{Real Social Network Analysis}

We now apply Algorithm 1 to analyze the massive \texttt{friendster} social network of $n=65,608,366$ persons (nodes) and 1,806,067,135 undirected edges \citep{YangLeskovec12} (This dataset was obtained from https://snap.stanford.edu/data/com-Friendster.html), using the NLPA model, assigned uniform priors $(\alpha,p)\sim\mathcal{U}_{(0,3)}\mathcal{U}_{(0,0.20)}$, and using summary statistics $s(x)=$ (Average clustering coefficient, Diameter, Density) $=(0.1623,32,8.39161\times10^{-7})$. Also, these network summaries also computed from a network of size $n_{\text{sim}}=1,000$ simulated from the model (given a set of proposed parameters) in each of the $N=10,000$ iterations of Algorithm 1. The algorithm routinely delivered posterior distribution and MLE estimates, shown in Table 20, despite the sheer size of the dataset. Also, this table reports the results of \texttt{rejectionABC} using all the three summary statistics, and the results of \texttt{rejectionABC}. For the \texttt{rejectionABC} analysis, the first two summary statistics were pre-selected as the most important, among the three total summary statistics, namely, average clustering coefficient (with importance measure of 0.66), diameter (importance 0.22), and density (importance of 0.12). The importance of each summary was measured from \texttt{drf} regression performed on the Reference Table, as before, regressing the prior parameter samples on all three summaries. The preselection of summaries led \texttt{rejectionABC} to produce posterior mean and median estimates that were similar to those of \texttt{copulaABCdrf}.

\begin{center}
\begin{tabular}{lcc}
\multicolumn{3}{l}{\footnotesize \textbf{Table 20.} Posterior estimates of the NLPA model from} \\ 
\multicolumn{3}{l}{\footnotesize the \texttt{friendster} network dataset.} \\ 
\hline\hline
& \footnotesize${\alpha }$ & \footnotesize${p}$ \\ \hline

{\footnotesize \texttt{copulaABCdrf} mean} & {\footnotesize 1.21} & {\footnotesize 0.001} \\ 
{\footnotesize \texttt{rejectionABC} mean} & {\footnotesize 0.69} & {\footnotesize 0.002}\\
{\footnotesize \texttt{rejectionABCselect} mean} & {\footnotesize 1.21} & {\footnotesize 0.02}\\\hline

{\footnotesize \texttt{copulaABCdrf} median} & {\footnotesize 1.25} & {\footnotesize 0.001} \\ 
{\footnotesize \texttt{rejectionABC} median} & {\footnotesize 0.64} & {\footnotesize 0.002}\\
{\footnotesize \texttt{rejectionABCselect} median} & {\footnotesize 1.20} & {\footnotesize 0.02}\\\hline

{\footnotesize \texttt{copulaABCdrf} mode} & {\footnotesize 0.24} & {\footnotesize 0.0002} \\ 
{\footnotesize \texttt{rejectionABC} mode} & {\footnotesize 0.42} & {\footnotesize 0.001}\\
{\footnotesize \texttt{rejectionABCkern.select} mode} & {\footnotesize 1.25} & {\footnotesize 0.01}\\\hline

{\footnotesize \texttt{copulaABCdrf} MLE} & {\footnotesize 0.24} & {\footnotesize 0.0002} \\ 
{\footnotesize \texttt{rejectionABC} MLE} & {\footnotesize 0.42} & {\footnotesize 0.001}\\
{\footnotesize \texttt{rejectionABCkern.select} MLE} & {\footnotesize 1.25} & {\footnotesize 0.01}\\\hline

{\footnotesize \texttt{copulaABCdrf} s.d.} & {\footnotesize 0.28} & {\footnotesize 0.002} \\ 
{\footnotesize \texttt{rejectionABC} s.d.} & {\footnotesize 0.44} & {\footnotesize 0.002}\\
{\footnotesize \texttt{rejectionABCselect} s.d.} & {\footnotesize 0.12} & {\footnotesize 0.01}\\\hline

{\footnotesize \texttt{copulaABCdrf} 50\%} & {\footnotesize (1.18, 1.31)} & {\footnotesize (0.0003, 0.001)} \\ 
{\footnotesize \texttt{rejectionABC} 50\%} & {\footnotesize (0.33, 1.01)} & {\footnotesize (0.001, 0.004)}\\
{\footnotesize \texttt{rejectionABCselect} 50\%} & {\footnotesize (1.14, 1.26)} & {\footnotesize (0.01, 0.02)}\\\hline

{\footnotesize \texttt{copulaABCdrf} 95\%} & {\footnotesize (0.37, 2.05)} & {\footnotesize (0.0001, 0.005)}\\
{\footnotesize \texttt{rejectionABC} 95\%} & {\footnotesize (0.04, 1.44)} & {\footnotesize (0.0002, 0.005)}\\
{\footnotesize \texttt{rejectionABCselect} 95\%} & {\footnotesize (1.04, 1.37)} & {\footnotesize (0.002, 0.03)}\\\hline

{\footnotesize \texttt{copulaABCdrf} copula} & {\footnotesize d.f. 28.91} & {\footnotesize scale 0.32} \\ \hline\hline
\end{tabular}
\end{center}

\section{Conclusions and Discussion}

This article introduced \texttt{copulaABCdrf} as a framework that unifies previous methods of \texttt{copulaABC} and \texttt{abcrf}, which were introduced separately in previous articles. This unified method aimed to provide, in a single unified framework, a wide range of inferences from the approximate posterior distribution (including posterior means, medians, mode, and the MLE) for models that may be defined by intractable likelihoods and many parameters, and model selection. All of this was done while being able to automatically select the subset of the most relevant summary statistics, from a potentially high number of summary statistics, without requiring the user to identify all the important summary statistics by hand before data analysis. Further, \texttt{copulaABCdrf} avoids the need to use tolerance and distance measure tuning parameters of \texttt{rejectionABC} methods.

This paper also proposed a new solution to the simulation problem in ABC, which is based on simulating datasets from the exact model likelihood of size smaller than the potentially-vary large size of the datasets being analyzed by the given model. This strategy could potentially be useful for models from which it is computationally-costly to simulate datasets from, including models for large networks such as ERGM and mechanistic network models, using UNC summary statistics calculated on simulated network datasets that were smaller than the potentially large network dataset being analyzed.

Based on the results of many simulation studies performed in this paper, evaluating and comparing \texttt{copulaABCdrf}, \texttt{rejectionABC}, and semiautomatic ABC \texttt{rejectionABCselect} methods based on \texttt{drf}, the following general conclusions can be made:

\begin{itemize}
    \item For the low dimensional parametric models ($d\leq5$) considered in this paper, \texttt{copulaABcdrf} and \texttt{rejectionABC} methods (including \texttt{rejectionABCselect} which uses pre-selection of summary statistics using \texttt{drf}) were competitive in terms of MAE and MSE of estimation of marginal posterior mean and median, and of the posterior mode and MLE of the model parameters. Also, \texttt{rejectionABCselect} often outperformed \texttt{rejectionABC} without pre-selection of the subset of summaries, and often produced posterior estimates that were similar than those of \texttt{copulaABcdrf}.    
    \item The \texttt{copulaABcdrf} framework tended to outperform \texttt{rejectionABC} methods in the estimation accuracy of univariate marginal posterior distributions, according to KS distance to the exact univariate marginal posterior distributions. However, all KS tests of all ABC methods indicated statistically significant departures.
    \item For high-dimensional parametric models ($d\ge5$) with posterior that was not highly multimodal, \texttt{copulaABcdrf} far outperformed \texttt{rejectionABC} methods.
    \item For all ABC methods, posterior estimation accuracy tended to be best when $n_{\text sim}/n = 1$, and accuracy decreased as $n_{\text sim}/n$ decreased below 1, which is not surprising because then the summaries become less informative about the data.
    \item For all ABC methods, estimation from the posterior distribution can be challenging when the true posterior is highly-multimodal. 
\end{itemize}

Further, it was found in the simulations that the network size-offset MPLEs were useful, but too computationally costly, compared to the other UNC network summary statistics, which were far more rapidly computable because they did not require optimization algorithms. For future research, computational speed can be drastically improved by using parallel computing methods to compute each of the summary statistics while constructing the Reference Table. 

The results suggest that \texttt{drf} is the main driving force of \texttt{copulaABCdrf}, as suggested, for example, by the fact that \texttt{copulaABCdrf} often produced results similar to that of \texttt{rejectionABCselect}. However, according to the KS tests, \texttt{drf} always produced univariate marginal posterior estimates that significantly departed from the corresponding exact posterior densities. On a related note, after the initial writing of this manuscript in early January 2024 (see Acknowledgements), \texttt{drf} was proposed as a tool for recursively estimating weights for sequential Monte Carlo ABC Algorithms \citep{DinhEtAl24}, for models defined by intractable likelihoods with less than five parameters. The \texttt{copulaABCdrf} framework, seemingly because it is based on the meta-$t$ posterior distribution approximation, was limited (like the \texttt{rejectionABC} methods) to the accurate estimation of more symmetric or skewed unimodal posterior distributions. The framework is partially motivated by the fact that the posterior given $n$ i.i.d. observations converge to a multivariate normal distribution asymptotically as $n \rightarrow \infty$ of i.i.d. observations, according to the Bernstein von Mises theorem, under mild regularity conditions \citep{KleijnVanderVaart12}.

In other words, \texttt{copulaABCdrf} based on the meta-$t$ is limited in estimating highly-multimodal posteriors, at least according to one simulation study. For future research, this issue can potentially be addressed by using more flexible copula density functions which can handle nonlinear relationships among parameters, perhaps while taking advantage of recent developments in the continually-active field of copula-based multivariate density estimation. However, such more flexible nonlinear copulas can be very computationally costly to estimate, especially for high dimensional parameters. Still, finding a method to efficiently compute estimates of such a more flexible copula density in high dimensional settings seems worth pursuing in future research.

\section*{Acknowledgments}{
The author thanks anonymous reviewers for providing helpful comments on a previous version of this manuscript, which helped improve its presentation. This manuscript was presented at a Biostatistics Seminar at the Medical College of Wisconsin during March 19, 2024, and at the conferences COMPSTAT 2024 (University of Giessen, 27-30 August 2024) and CFE-CMStatistics 2024 (King's College London, 14-16 December 2024). The initial version of this manuscript, including the results of simulation studies and analyses of real networks, obtained using Algorithm 1 implementing Distribution Random Forests and copula modeling, was presented as a report in a grant proposal submitted to the U.S. National Science Foundation on February 14, 2024. The authors declare no conflicts of interest. The real datasets can be obtained through sources cited within the paper. The R software code files, used to simulate the data and to analyze the real and simulated datasets, can be obtained from the author via \url{https://github.com/GeorgeKarabatsos/copulaABCdrf}.}

\newpage
\bibliographystyle{rss.bst}
\bibliography{Karabatsos.bib}

\end{document}